\documentclass[acmsmall]{acmart}
\pdfoutput=1

\renewcommand\footnotetextcopyrightpermission[1]{} % removes footnote with conference information in first column
\AtBeginDocument{%
  \providecommand\BibTeX{{%
    \normalfont B\kern-0.5em{\scshape i\kern-0.25em b}\kern-0.8em\TeX}}}

\setcopyright{acmcopyright}
\copyrightyear{2021}
\acmYear{2021}
\usepackage{graphicx}
\usepackage{multirow}
\usepackage[figuresright]{rotating}
\usepackage{amsmath}
\usepackage{graphicx}
\usepackage{float}
\usepackage{subfig}
\usepackage{threeparttable}
\usepackage{longtable}
\usepackage{color}
\usepackage{xcolor}
\usepackage{array}
\usepackage{longtable}
\begin{document}

%%
%% The "title" command has an optional parameter,
%% allowing the author to define a "short title" to be used in page headers.
\title{Epidemic Model-based Network Influential Node Ranking Methods: A Ranking Rationality Perspective}

%%
%% The "author" command and its associated commands are used to define
%% the authors and their affiliations.
%% Of note is the shared affiliation of the first two authors, and the
%% "authornote" and "authornotemark" commands
%% used to denote shared contribution to the research.
\author{Bing Zhang}
\author{Xuyang Zhao}
\email{bingzhang@ysu.edu.cn}
\orcid{0000-0002-9867-8439}
\affiliation{%
  \institution{School of information science and engineering, Yanshan University}
  \streetaddress{438W Hebei Avenue}
  \city{Qin Huangdao, Hebei Province}
  \country{China}
  \postcode{066004}
}

\author{Jiangtian Nie}
\affiliation{%
  \institution{School of Computer Science and Engineering, Nanyang Technological University}
  \country{Singapore}}
\email{jnie001@e.ntu.edu.sg}

\author{Jianhang Tang}
\authornote{Corresponding author}
\author{Yuling Chen}
\affiliation{%
  \institution{The State Key Laboratory of Public Big Data, Guizhou University}
  \city{Guiyang}
  \country{China}
   \postcode{550025}}
\email{tangjh91@ysu.edu.cn}
%\orcid{0000-0002-7958-391X}

\author{Yang Zhang}
\affiliation{%
  \institution{College of Computer Science and Technology, Nanjing University of Aeronautics and Astronautics}
  \country{China}}
\email{yangzhang@nuaa.edu.cn}

\author{Dusit Niyato}
\affiliation{%
  \institution{School of Computer Science and Engineering, Nanyang Technological University}
  \country{Singapore}}
\email{dniyato@ntu.edu.sg}

\renewcommand{\shortauthors}{Bing Zhang et al.}

\begin{abstract}
 % Since influential nodes in real-world networks, e.g., communication and social networks, become more and more important in diverse applications, many Influential Node Ranking Methods (INRMs) have been invented to detect them in the past 20 years. 
Most recent surveys and reviews on Influential Node Ranking Methods (INRMs) hightlight discussions on the methods’ technical details, but there still lacks in-depth research on the fundamental issue of how to verify the considerable influence of these nodes in a network. Compared to conventional verification models such as cascade failure and linear threshold, the epidemic model is more widely used. Accordingly, we conducted a survey of INRM based on epidemic model on 81 primary studies and analyzed their Capability and Correctness which we defined in our work. Our study categorized 4 types of networks used by INRM, classified 7 categories of INRMs for analyzing the networks and defined 2 evaluation metrics set of Capability and Correctness for evaluating INRM from Ranking Rationality Perspective. We also discussed particular real-world networks that were used to evaluate INRM and the Capability and Correctness of different INRMs on ranking nodes in specific networks. This is, as far as we know, the first survey aimed at systematically summarizing the Capability and Correctness of INRM. Our findings can assist practitioners and researchers in choosing and comparing INRMs and identifying research gaps.
\end{abstract}




\maketitle

\section{Introduction}
Due to the scale-free nature of complex networks, most degrees of nodes are small, but few nodes can lead to the power-law distribution of degrees in the networks. As nodes play different roles in information dissemination within the networks, their degrees are heterogeneous. Although nodes that influence the structure and information transmission of the network more than other nodes are clearly considered as influential ones, the effective identification of the most important nodes in the network has long been a hot research area. As is well-known, 
% as shown in Fig.\ref{fig_networks}, 
finding Internet celebrities in social networks can promote products or information, tracing the high-risk groups in human communication network can reduce widespread transmission of coronavirus, and identifying the essential proteins in biological networks can help people understand what is necessary to sustain life. Likewise, identifying important functions that are prone to vulnerabilities in \textit{Log4j} software network or locating nodes that are prone to cause grid failure in power grid network can provide targets for maintainers. 

% \begin{figure}[htbp]
% \centering
% \includegraphics[width=0.6\textwidth]{networks.pdf}
% \caption{\textsc{The applications of influential nodes}}
% \label{fig_networks}
% \end{figure}

\textcolor{black}{The influence of a node in a network is usually measured by the node’s ability to disseminate information. Assuming that information starts to spread from a certain node, if it spreads more quickly over a larger area of the network, the node will be more influential. \cite{Borgatti200501}. Once such nodes fail, they would greatly affect the flow of traffic in the network. A minimum of influential nodes that are capable of maximizing the total amount of information is called the Influence Maximization problem  
 \cite{LiYuchenandFan2018,ZHONG201878,HuangDawenYu2017,WangXiaojieandSu2016,ZhangJian-XiongandChen2016,HeJialinandFuYanandChen2015,ChenDongmingandPanpanDuandFang2020,KumarSanjayandPanda2020,AhmadAmreenandAhmad201911,YuEnyuandFuYanandTang202001,BerahmandKamalandBouyer201803,GaoChaoandZhongLuandLi201510}. Ranking these nodes in the network by calculation indicators like the \textcolor{black}{propagation capability \cite{LiuYingandTangMingandDo2017,WangJunyiandHou201702,theroleofclustering201310}and the network topology \cite{WenTaoandJiang201902,Ibnoulouafi201809,YangPingleandLiu201806,SheikhahmadiAmir2017}} has been become one of the common ways for locating the most influential (\textit{top-k}) nodes, which is also known as the influential node ranking method (INRM). \textcolor{black}{Unlike the Influence Maximization problem that considers the combination of the set of nodes, which is unacceptable in large-scale networks, the INRM is dedicated to determining the impact of individual nodes, and it has less complexity.}}

\textcolor{black}{Currently, there are two serious challenges that have never been extensively reviewed for INRMs in different types of networks. The first one is that almost no studies focus on how to verify the capability of the ranking results, mainly the ranking granularity and credibility of INRMs. The second one is that most studies highlight the applications of INRMs and the INRMs themseleves but ignore how to validate the correctness of INRMs, that is, how the INRMs approach benchmarks. It is noteworthy that capability and correctness are collectively known as Ranking Rationality of INRMs. For the first challenge, fewer approaches \textcolor{black}{ \cite{MA2016205,QiuLiqing202107,7979310,BAO2017391,8351889,BIAN2017422}} attempt to evaluate and improve the capability through \textcolor{black}{distinct metric \cite{MaQianandMaJun201608,ZENG20131031}, complementary cumulative distribution function \cite{ZAREIE2017485}, monotonicity function \cite{Ibnoulouafi201809,ShaoZengzhenandLiu201911}, etc.} But these methods and evaluation metrics are relatively scattered and not deeply reviewed from the perspectives of ranking granularity and credibility. For the second challenge, the correctness of most INRMs can be validated by different information spreading or diffusion models, including the epidemic model \cite{NathDilipandDas201704}, the cascade failure model \cite{MotterAdilsonandLai200301}, the independent cascade model \cite{GoldenbergJacobandLibaiBarakandMuller200108} and the linear threshold model \cite{MohapatraDebasisandPradhan201812}. These models are used to simulate the actual information propagation process, and the ranking results of nodes are often regarded as benchmarks compared with those of proposed influential node identification methods to validate the correctness. Nevertheless, no existing studies systematically reviewed and evaluated the correctness of INRMs using these models. Compared to other models, the epidemic model, which is the most widely used in different types of networks, can simulate the entire information propagation process in a network in a near-realistic way, and therefore can differentiate well the propagation abilities of nodes in the network. Thus, in this paper, our study mainly discussed INRMs validated by the epidemic model.}

\textcolor{black}{In recent two decades, a large variety of influential node ranking methods have sprout out in various networks. A few studies surveyed or reviewed them from different viewpoints. Approaches proposed in references \cite{10.1145/3301286,7854364,MAJI2020113681,LU20161} involved the validation metrics, but they focused on the classification of methods \cite{HAFIENE2020113642,TULU2020102768,2021ScChE..64..451L,LALOU201892}, and did not systematically analyzing and summarize the evaluation metrics and their fitness for different type of networks. References \cite{10.1145/3301286,HAFIENE2020113642,TULU2020102768} summarized and categorized influential node detection methods designed just for social networks, and reference \cite{7854364} surveyed the evaluation methods for node importance in world city networks. Reference \cite{MAJI2020113681} systematically surveyed INRMs based on 
\textit{k-shell} decomposition with part capability metrics. Despite the fact that existing surveys and reviews have provided an overview of INRMs from various angles, only a few of them investigated the capability and correctness of INRMs when the called influential, critical or important nodes were identified.} To perform a network information mining task, a summary of the empirical assessment of INRMs can help network researchers to select the optimal method. Specifically, a comprehensive understanding of the advantages and disadvantages of various approaches in capability and correctness can enable researchers to identify the gaps in the research and to address the deficiencies of these methods.

To summarize INRMs and their empirical evaluation metrics, we systematically reviewed articles published over the last 20 years. We found that networks mentioned in these papers covered most real-world networks, and there are huge differences in information flow among these networks. Moreover, there are clear differences in the concept of information dissemination in these networks and computational process among different INRMs. Third, the evaluation and verification metrics for different identification results on different networks are complex and various. Based on these findings, we attempts to address the following research issues.

RQ1: What types of networks have been analyzed by INRMs based on the epidemic model?

RQ2: What are the existing INRMs and how they are classified on different types of networks?

RQ3: How to evaluate the result of these INRMs?

\textcolor{black}{We combined the thematic analysis and the simple statistical analysis to answer these questions, with reference to the extensive literature (81 primary studies) in this field.} The results and the main contributions of our study are shown below.

% \begin{figure*}[htbp!]
% \centering
% \includegraphics[width=0.85\textwidth]{figure2_original5.pdf}
% \caption{\textsc{Layout and structure of this survey}}
% \label{fig_Layout}
% \end{figure*}

(1) Based on the application scenarios and the exploration of how the traffic (information) is disseminated in networks, existing real-world networks used frequently in INRMs are divided into four categories, which are the social network, knowledge network, technological network, and biological network. INRM-focused companies can use these networks to benchmark different tools. Furthermore, the lack of complete information on undirected, unweighted and low-quality networks can encourage collaboration between researchers from both industry and academia to deliver optimized test suites and networks and thus better serve the complex network community.
% And undirected and unweighted, low-quality networks without full information can drive the industry practitioners and researchers to work collaboratively to provide more accomplished test suites and networks to benefit the complex network community.

\textcolor{black}{(2) Based on different computation of feature influence for nodes in the networks, INRMs can be grouped into four main categories, which are the structural centrality method, the iterative refinement method, the MADM method and the machine learning method. The structural centrality method can be further classified into four categories of meta-methods, including local centrality methods, semi-local centrality methods, global centrality methods and hybrid centrality methods. Through results analysis, the semi-local, global, hybrid and MADM technique methods are widely applied in four kinds of networks. The local centrality and iterative refinement method are more widely used in social networks and technical networks. These conclusions would help engineers and researchers narrow the choosing range of INRMs.}

\textcolor{black}{(3) Specific evaluation metrics to verify INRMs are defined as Capability and Correctness. Capability verifies the ability of the method in differentiating influential nodes and assessing ranking credibility, including \textit{Distinct Metric, Complementary Cumulative Distribution Function, Monotonicity Function} and \textit{Ranking similarity}, while Correctness verifies the validity and consistency to approach
the benchmarks in different scenarios, including \textit{Kendall’s tau Correlation Coefficient, Spearman Coefficient, Spearman Coefficient, Imprecision function, Pearson Coefficient}, \textit{Robustness}. Moreover, five selection strategies \textit{I, II, III, IV, V} are summarized to provide the data to the convincing comparison of INRMs under the epidemic model. Through results analysis, \textit{Ranking similarity} and \textit{Kendall’s tau Correlation Coefficient} metric in Capability and Correctness are used frequently, and \textit{Selection Strategy II}, which is the most frequently used among all methods, can assist engineers and researchers in prioritizing and selecting proper tools on a better basis.}

The rest of this paper is arranged as follows. Section 2 introduced the preliminary concepts. Section 3 classified different networks used to evaluate the primary studies. Section 4 summarized the INRMs. Section 5 defined Capability and Correctness evaluation metrics. Section 6 presented our survey results. Section 7 compared our work with related studies. Section 8 discussed the results, followed by conclusions and suggestions for future work in Section 9. 
% The layout and structure of this survey is shown as Fig. \ref{fig_Layout}}.

\section{PRELIMINARIES}

\textcolor{black}{In this section, we introduced the preliminary concepts of traditional Node Ranking, Influence Maximization problem and Epidemic model to locate \textit{top-k} influential nodes. We defined the capability and correctness of influential node ranking. Finally, relationships among them were discussed. }

\subsection{Node Ranking}
Ranking nodes in a network by their “importance” is a classic problem \cite{Ahierarchicalapproachforinfluential}. The importance of a node is also called “centrality”, which is equivalent to either the significance of the node's connections to other nodes \cite{Borgatti200501} or the information diffusion capability of the node in the entire network \cite{IyerSwamiandKillingback}. \textcolor{black}{Taking degree centrality \cite{FreemanLinton197901} as an example, a toy network is shown in Fig. \ref{fig_ToyNetwork}, nodes $a, i, j$ are assigned 
“centrality” value 5, nodes $b, e, f, g, k$ are assigned “centrality” value 4, nodes $m, n, o, h, c, d$ are assigned “centrality” value 3, node $y$ are assigned “centrality” value 2, nodes $u, q, r, p, t, s, v$ are assigned “centrality” value 1. Then the importance of nodes is ranked based on their “centrality” values. Here, the \textit{top-3} nodes such as $a, i, j$ may be the most influential nodes.}

\subsection{Influence Maximization problem}
\begin{figure*}[htbp!]
\centering
\begin{minipage}{0.45\linewidth}
\includegraphics[width=1\textwidth]{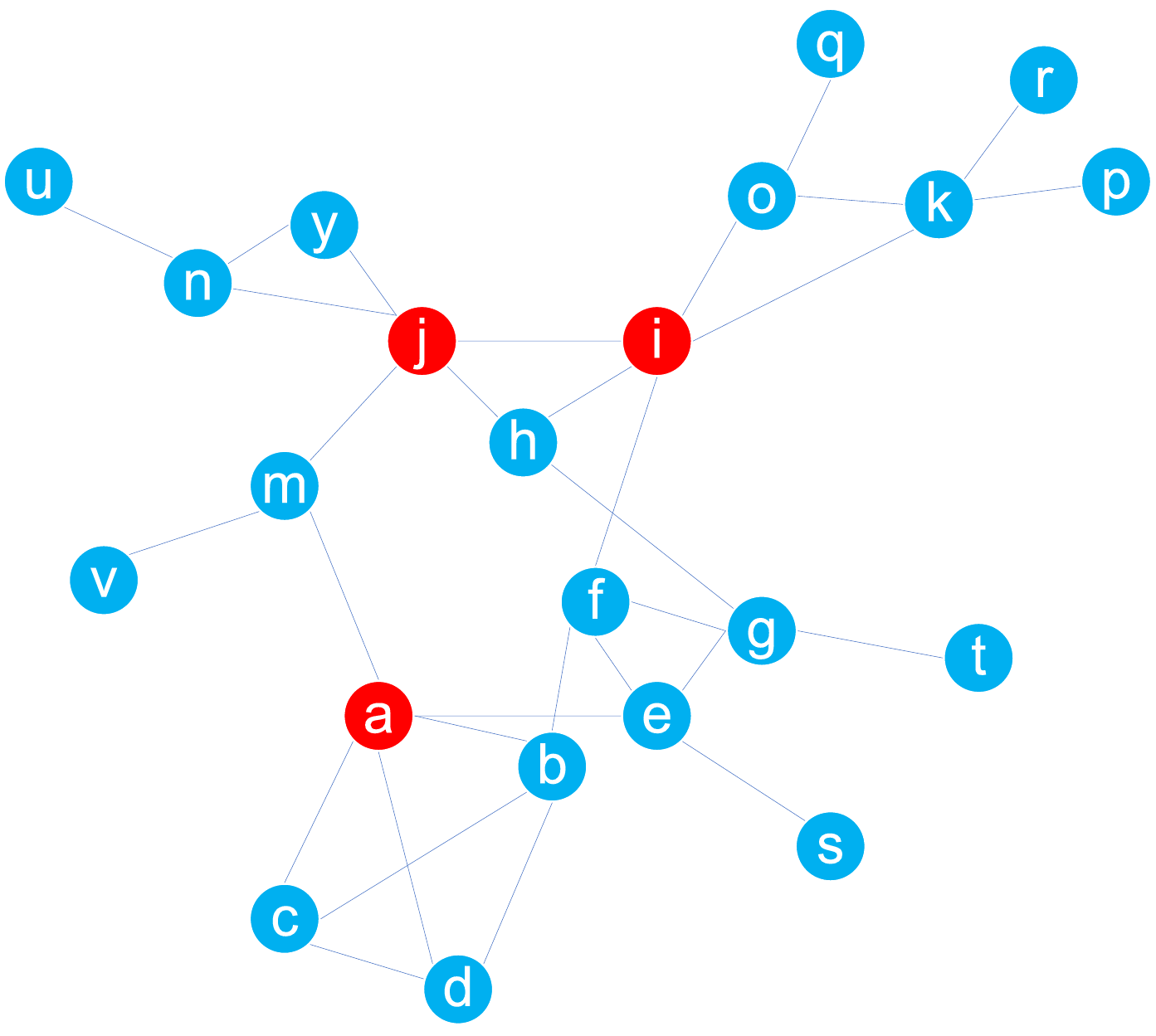}
\caption{\textsc{An example of Toy Network}}
\label{fig_ToyNetwork}
\end{minipage}
\qquad\qquad
\begin{minipage}{0.4\linewidth}
\includegraphics[width=0.82\textwidth]{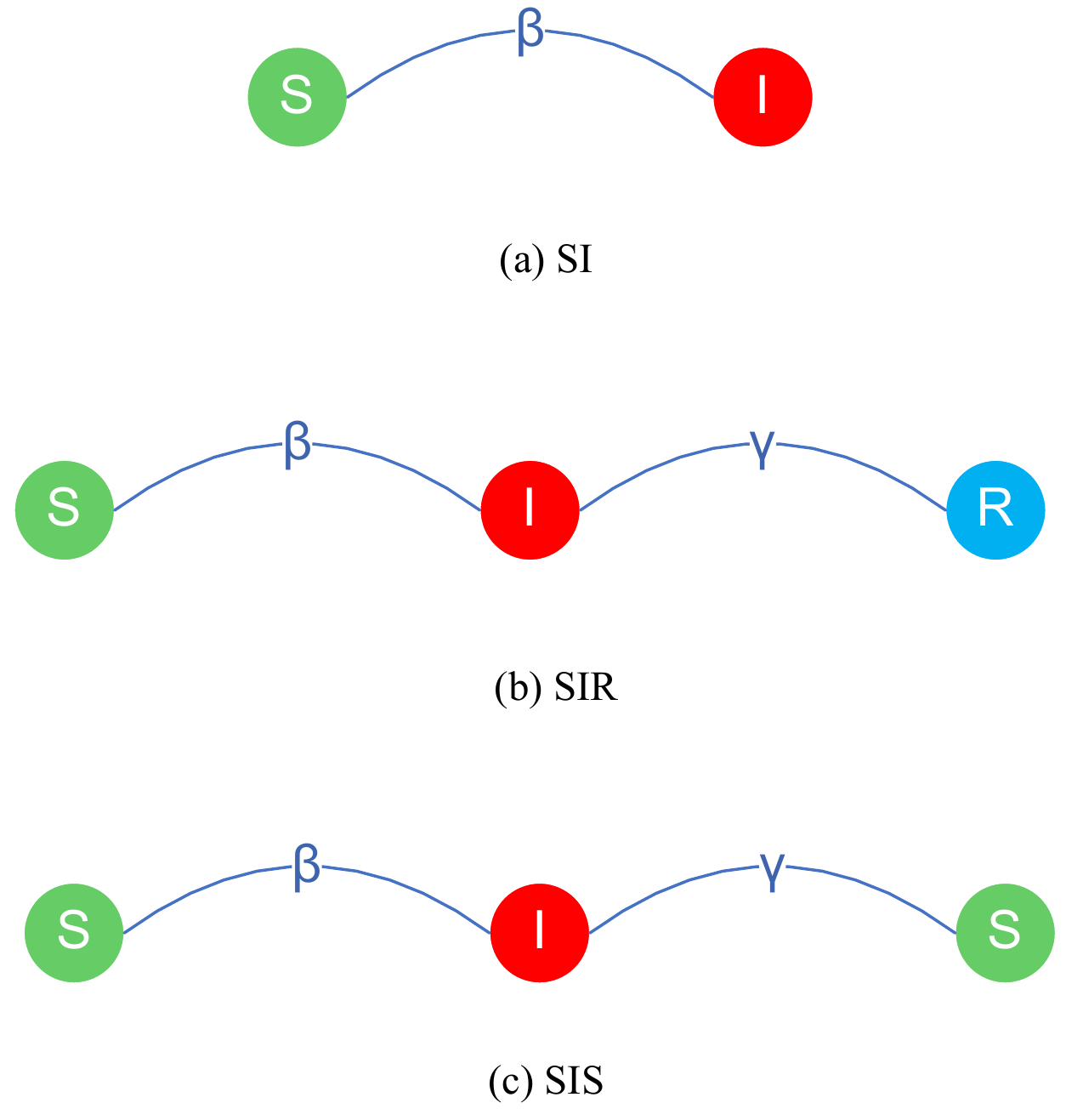}
\vspace{23pt}
\caption{\textsc{Epidemic Model}}
\label{fig_DynamicOfEpidemicModel}
\end{minipage}
\end{figure*}

Maximum Influence initially refers to the task of selecting $k$ seed nodes in social networks to maximize the transmission of the seed's influence \cite{LiYuchenandFan,KumarSanjayandPanda,ZhangZufanandLiXieliang,IRIEScalableandRobustInfluence}. It models viral marketing scenarios and can be applied to other scenarios like cascade monitoring and rumor control. Since its proposal in 2003, Maximum Influence and its variants have been extensively studied, and the domain is still actively developing. Maximum Influence is also a good demonstration of how classical algorithms fit into the social networking environment, such as the greedy algorithm \cite{IMRankInfluenceMaximization} and the Dijkstra’s shortest path algorithm \cite{FuzzyDijkstraalgorithm}. \textcolor{black}{Contrary to the Node Ranking problem, the Influence Maximization problem endeavors to clarify the importance of a set of nodes. For example, in Fig. \ref{fig_ToyNetwork}, if we aim to find \textit{top-5} influential nodes, nodes $j, a, f, i, k$ can be an alternative node set, or nodes $y, m, b, g ,i$ can be another alternative node set. At this point, the solution of the influence maximization problem requires finding five nodes that together have the greatest influence in the network compared to all other five possible nodes.}

\subsection{Epidemic Model}
% The epidemic model \cite{KeelingMattandEames} examines the transmission speed, the spatial scope, the transmission route, the dynamic mechanism and other issues of infectious diseases in order to guide the effective prevention and control of infectious diseases. It focuses on the transmission mechanism and coping strategies of infectious diseases, computer viruses, rumors, etc. in the population by establishing mathematical models based on Markov process. The classical epidemic models are the Susceptible-Infected (SI) model, the Susceptible-Infected-Recovered (SIR) model and the Susceptible-Infected-Susceptible (SIS) model \cite{Epidemicthresholdofnodeweighted}. The SIR model of great epidemiological signifance is considered as the most basic model of infectious disease dynamics. 

The epidemic model \cite{KeelingMattandEames} studies the spread rate, the space range, the transmission path and the dynamic mechanism to direct the prevention and control of infectious diseases. Its emphasis is placed on the transmission mechanism and the coping strategy of infectious diseases or rumors in the population by establishing mathematical models based on Markov process. Classic epidemic models are the Susceptible-Infected (SI) model, the Susceptible-Infected-Recovered (SIR) model and the Susceptible-Infected-Susceptible (SIS) model \cite{Epidemicthresholdofnodeweighted}. It is worth noting that the SIR model, which has epidemiological significance, is a fundamental model for the dynamics of infection. Therefore, \textcolor{black}{most studies \cite{BjornstadOttarandShea2020,LiuZuhanandTian2020,JiShenggongand2017,SunHangandShengYuhong2021}} on epidemic dynamics of network are mainly based on classic algorithms of the epidemic model, and each node in the network is in one of the three states at any given moment: Susceptible, Infected, and Recovered.

The SI model has only \textit{"susceptible" (S)} and \textit{"infected" (I)} individuals. There are multiple rounds of infection in the network. Susceptible individuals are infected by their infected neighbors with the transmission rate of $\beta$ in each round. $S(t)$ and $I(t)$ are the ratios of susceptible and infected individuals at moment $t$. We have $S(t)+I(t)=1$. Once the susceptible individuals are infected, they cannot be cured. After multiple rounds of infection, all the individuals are at the infected state. The contagion rule of the SI model is shown in Fig. \ref{fig_DynamicOfEpidemicModel} (a). The differential equation for the SI model is shown in equation (1).

\begin{equation}
\begin{cases}
\frac{\mathrm{d}S(t)}{\mathrm{d}t}=-\beta S(t)I(t),\\
\frac{\mathrm{d}I(t)}{\mathrm{d}t}=\beta S(t)I(t).
\end{cases} 
\end{equation}

The SIR model categorizes individuals by three states: \textit{“Susceptible”(S), “Infected”(I)} and \textit{“Recovered”(R)}. Nodes with state \textit{R} indicates that the individual has been cured and cannot be infected again, or it is no longer infectious. In the SIR model, the contagion path is $S \to I \to R$, as shown in Fig. \ref{fig_DynamicOfEpidemicModel} (b). The susceptible individuals ($S$ state) is infected by infected individuals ($I$ state) with probability $\beta$ in each round. At the meantime, infected individuals ($I$ state) recover to $R$ state with probability $\gamma$. At the end of the propagation process, only state $S$ and state $R$ exist in the network. The differential equation for the SIR model is as follows.

\begin{equation}
\begin{cases}
\frac{\mathrm{d}S(t)}{\mathrm{d}t}=-\beta S(t)I(t),\\
\frac{\mathrm{d}I(t)}{\mathrm{d}t}=\beta S(t)I(t)-\gamma I(t),\\
\frac{\mathrm{d}R(t)}{\mathrm{d}t}=-\gamma I(t).
\end{cases} 
\end{equation}

The SIS model differs from the SIR model in that infected individuals are still susceptible. The SIS model describes the epidemic transmission behavior of individuals that can be repeatedly infected and susceptible. Its contagion rule is shown in Fig. \ref{fig_DynamicOfEpidemicModel} (c). The differential equation for SIS model is below.

\begin{equation}
\begin{cases}
\frac{\mathrm{d}S(t)}{\mathrm{d}t}=-\beta S(t)I(t)+\gamma I(t),\\
\frac{\mathrm{d}I(t)}{\mathrm{d}t}=\beta S(t)I(t)-\gamma I(t).
\end{cases} 
\end{equation}

\subsection{The Capability of Influential node ranking}
A node's importance depends on its "importance" value (also called centrality value) assigned by these methods, while the capability of INRMs is reflected by its ability to differentiate the importance of various nodes in the network. If two or more nodes are assigned the same importance in an INRM, the capability of the method is poor because their importance in complex networks is usually different. Moreover, classic or popular methods have been validated by the industry, and the similarity of the ranking results with these methods confirms the credibility of INRMs. The more similar they are, the more credible the INRMs are, the better the capability of INRMs is.

\subsection{The Correctness of influential node ranking}
An INRM can produce a node importance ranking list according to the measured node importance or influence value. The correctness analyses of the methods can be verified by the correlations between the ranking list and the benchmark importance ranking list of nodes generated by the epidemic model. The more positively correlated to it, the higher the correctness of the method. Analyzing its consistency with the ranking list generated by the impact from changes in indicators (ranking values under epidemic model, centrality values) before and after removing the node, which can be any node in a network, can also verify the correctness of INRMs. For the removed node, greater changes indicate greater impact on the network and greater importance of the node. For the correctness, more consistency of the ranking list with the one generated by the impact value of each node reveals better correctness of INRMs. 

\subsection{Summary}
Influence maximization and influential node ranking problems are both important directions for influential node mining in complex networks \cite{YangXandHuangDC}, and their difference lies in that solving the influence maximization problem requires comparing the influence of a group of nodes in the network, while solving the influential node ranking problem requires comparing the influence of each individual node \cite{ZhangJianXiongandChenDuanbingandDong}. Capability and correctness are two evaluation methods of INRMs. Capability tests whether the ranking results can distinguish the similarities and differences of importance of nodes, while correctness tests the rationality of the ranking results. Further, the epidemic model simulates the information propagation process of nodes in the network and denotes the approximation of the actual importance of the node as the size of the spreading range. A proof of correctness requires the consistency of the ranking results of INRMs with the results of the epidemic model.

\section{The Classification of the networks}
To answer RQ1, we first collected basic statistics of real-world networks used to evaluate the important nodes ranking methods based on epidemic Model. We then categorized these networks according to \textcolor{black}{their application scenarios and how the traffic (information) is disseminated in the network \cite{Borgatti200501,doi10.1137S003614450342480}}.

\subsection{The statistics of real-world networks}

We surveyed 58 different type of networks in various fields such as communication, security, medicine, and society from the 81 primary studies. We collected the statistic information (numbers of nodes and edges, node degree, etc.) from the network data sets in Table \ref{tb_NetworkStatistics}, which are critical to investigating the structure, function and traffic flow of these networks. Among these real-world networks, only 
34.5\% of them showed their basic information and listed corresponding accessible URL.

\begin{table*}[hbtp!]
\centering

\caption{The Statistic Information of Real-world Networks}
\label{tb_NetworkStatistics}
\resizebox{\linewidth}{!}{
\begin{threeparttable}
\setlength\tabcolsep{5pt}
\begin{tabular}{c|m{3.5cm}|c|c|c|c|c|c|c|c|m{4.9cm}|l}
\hline
\textbf{ID} & \textbf{The name of network}                                                                    & \textbf{Type} & $\left | V \right | $   & $\left | E \right | $    & $\left \langle k \right \rangle $    & $k_{max} $    & \textbf{Assortativity} & \textbf{Directed} & \textbf{Weighted} & \textbf{The refs that use networks}                                                                                                                                                                                                                                                                  & \textbf{Url}                                                                              \\ \hline
1  & Blogs                                                                      & SN   & 1.2K  & 33K    & 55     & 702  & -0.2213       &          &          & \cite{CHEN20121777,khop201502,MaQianandMaJun201608,RePEc:eee:phsmap:v:403:y:2014:i:c:p:130-147,spreadingdynamics201505,BAE2014549,LiuDongandNie201810,LiuYingandTang201705,ZhangRuishengandYang201608,MA2016205,CaiBiaoandTuoXianGuo2014,QingChengHuanfYanShen201301,HU201673,ZhaoJieandWang202004,ZhaoJieandSong202008,ZhaoJieandSongYutong202003}                                                                                                                                                                                                   & http://konect.cc/networks/dimacs10-polblogs/        \\
\hline
\end{tabular}
 \begin{tablenotes}
        \footnotesize
        \item $\left | V \right | $: number of nodes. $\left | E \right | $: number of edges. $\left \langle k \right \rangle =2\left | E \right |/\left | V \right |  $: average nodal degree. $k_{max}$: maximum degree. $k_{min}$: minimum degree. 
        \item Assortativity: the propensity for large-degree nodes to connect to other large-degree nodes and low-degree to low-degree\cite{ZHOU202047}.
      \end{tablenotes}
  \end{threeparttable}
}
\end{table*}

\subsection{Classification of real-world networks}
A set of nodes connected by edges is only the simplest type of network, while real-world networks are much more complicated than this \cite{doi10.1137S003614450342480}. In real-world networks, there may be one or more type of nodes and edges. For instance, in Internet network, the vertices may represent computers, mobile phones, smartwatches, or a variety of local area networks, and edges between nodes may represent electric cable, optical cable, wireless signal and so on, which can carry weight. The weight of an edge in the Internet can be the network transmission rate or the distance between two hosts. In logistics network, it can be the volume or weight of a parcel, or the distance between two locations. Moreover, edges can be directed. In email network, a directed edge means an email sent from one person to another.

According to reference \cite{Borgatti200501}, many INRMs first assume the characteristics of traffic flow (information flow) in the network, and then design algorithms to score the importance of nodes based on their contribution to information dissemination. However, there are various characteristics of traffic flow in different types of networks. At this time, if the characteristics of traffic flow assumed by an INRM do not conform with the network, then the results are likely to be not accurate. \textcolor{black}{For example, on Twitter platform, the posts posted by a person are usually passed on to all its subscribers. Likely, if the INRM assumes that the posts of this person will only be notified to a small number of subscribers, then the impact of dissemination of the person will be underestimated.}

\begin{figure}[htbp]
\includegraphics[width=0.58\textwidth]{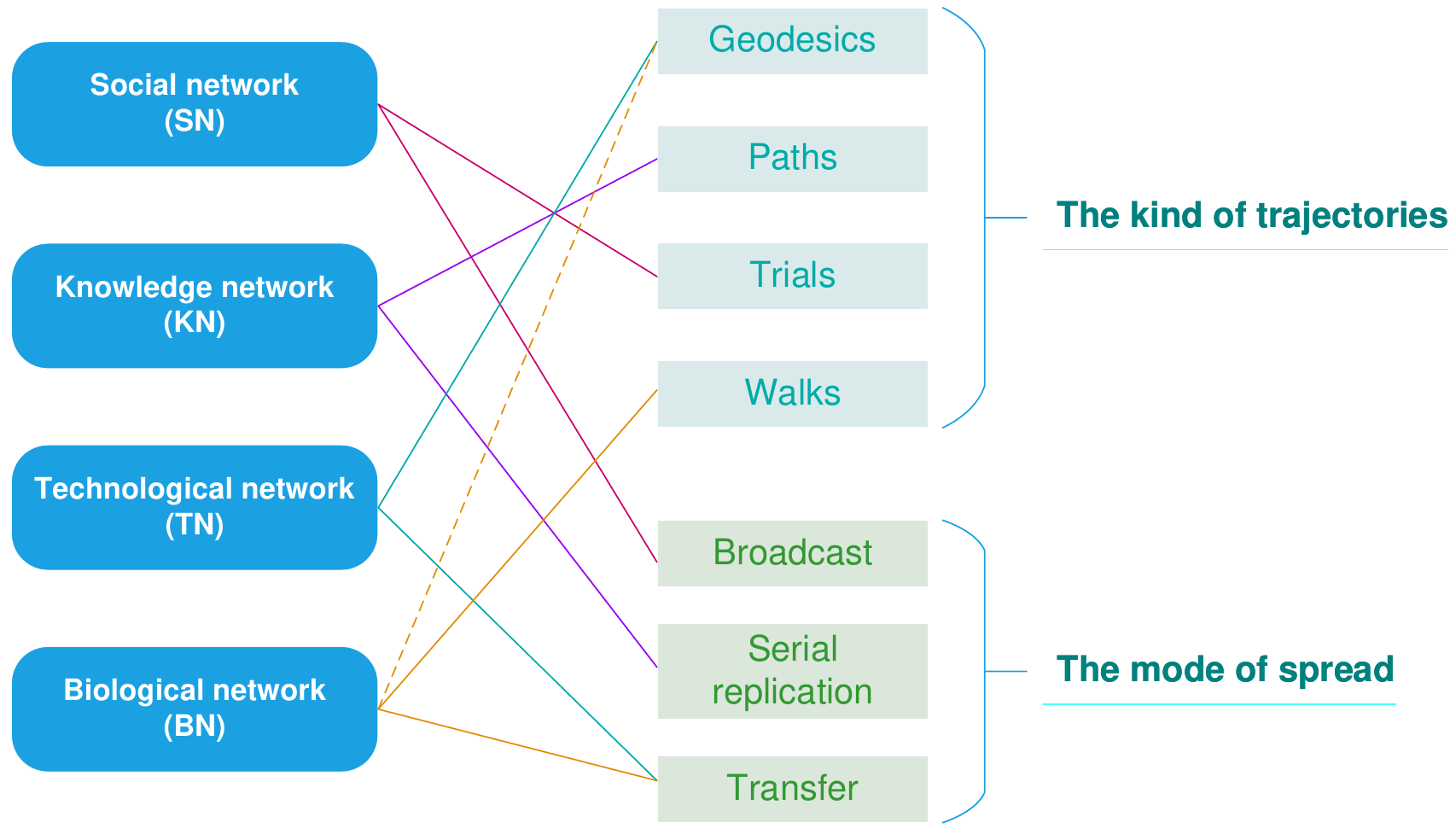}
\caption{\textsc{Categories of Networks and Features}}
\label{fig_MappingAndFeatures}
\end{figure}

\textcolor{black}{The pattern of traffic flow is usually determined by two factors \cite{Borgatti200501,doi10.1137S003614450342480}, namely the trajectories possibly followed by traffic and the mode of spread. The former can be geodesics, paths, trails, or walks, and the latter includes broadcast, serial replication, or transfer. For the trajectories that traffic follows, geodesics are the shortest paths between nodes, and they are all the possible paths connecting nodes. The nodes in a path cannot be repeated, while the nodes in a trail can be repeated many times. The walks consist of a series of random contacts. Each node in a walk will randomly select a neighbouring node to disseminate information. As for the mode of spread, in broadcast (parallel duplication) and serial replication, the traffic will be copied at every step of dissemination, while in serial replication, because the traffic will only be passed on a specific path, so the traffic will be copied once at a time. On the contrary, when traffic flow in transfer mode, the traffic will not be replicated throughout the dissemination process and the intermediate nodes are only responsible for the transfer of traffic.} Based on these factors and network's typical application scenarios, we classified the real-world networks in Table \ref{tb_NetworkStatistics}  into four categories, which are Social network, Knowledge network, Technological network and Biological network. The mapping of networks and features is shown in Fig. \ref{fig_MappingAndFeatures}.

\subsubsection{\textbf{Social network (SN)}}
\textcolor{black}{A social network refers to a group of individuals (mostly people) having some interactions or some pattern of contact between them. The patterns of contact are face to face or via social websites or emails, anonymous chat via the Internet, academic discussions among professors. The trajectories in social networks mainly follows trails and the mode of spread is usually broadcast. For the reason that the information propagation in social networks usually starts from a single node, replicates along all paths and starts a new propagation process from new nodes. These nodes can be repeated in a path. \textcolor{black}{For example, gossip and attitude are the most common traffic in social networks, which can be in several places and can be propagated by all nodes that hold them to all neighboring nodes. Eventually, all nodes in the network will retain this gossip and attitude. Similar scenarios include random academic communications between professors in \textit{Netscience (NS)} network \cite{Garas_2012,HuPingandMei201709,ZAREIE2017485,hindex201601,CHEN20121777,WEI2015277,ZENG20131031,BAE2014549,10.1155/2020/5903798,ZareieAhmad201811,LiuYingandTang201705,MA2016205,CaiBiaoandTuoXianGuo2014,LvZhiweiandZhao201902,WangShashaandDu201611,QiuLiqing202107,ZhaoJieandWang202004,FuYuHsiangandHuang201504,WangJunyiandHou201702,QingChengHuanfYanShen201301,ZhaoJieandSong202008,YangPingleandLiu201806,7979310,ZhaoJieandSongYutong202003}, publishers and subscribers on \textit{Twitter} network \cite{RePEc:eee:phsmap:v:403:y:2014:i:c:p:130-147,10.1155/2020/5903798} and community chats in \textit{HighSchool} network \cite{7978752}. In addition to human social networks, some other social networks have the same characteristics. \textit{Dolphins} will share route and food information with all their companions, or warn of danger \cite{Ibnoulouafi201809,HuPingandMei201709,ZAREIE2017485,e22080848,ShaoZengzhenandLiu201911,ZENG20131031,BAE2014549,ZareieAhmad201811,ZhangRuishengandYang201608,8322554,8259501,CaiBiaoandTuoXianGuo2014,QiuLiqing202107,FuYuHsiangandHuang201504,WangJunyiandHou201702,YangYuanzhi201911}. The production of a commodity will involve close cooperation between many companies \cite{7925142,Garas_2012}, and the reputation of a company in \textit{Corporate ownership network (CON)} will be widely spread in the industry. }} 

\subsubsection{\textbf{Knowledge network (KN)}}
\textcolor{black}{Knowledge network can also be called Information network. The nodes in Knowledge networks usually store knowledge or so-called information, and the information between two nodes connected by one edge are related.} \textcolor{black}{The trajectories in knowledge networks follows paths, and the mode of spread is serial duplication. This is because information dissemination in knowledge networks often requires preconditions. As a result, the information in the knowledge network can only be spread along a path that meets specific conditions, and each node in a path will learn this information on the condition that these nodes are not repeated. \textcolor{black}{For example, the existence of a match between two teams in \textit{American football} network presupposes that no match has been played between these two teams, and the team must have played against all other teams \cite{8259501,ZhaoZijuan201912,YangYuanzhi201911,DU201457}. It is impossible to have the same two teams in each match. Unlike the random communication between professors, collaborative networks of professors like \textit{Arxiv HEP-PH (High Energy Physics-Phenomenology)/CA-Hep} are limited to specific areas, e.g., communication is restricted to those professors majoring in complex networks \cite{HuPingandMei201709}. In \textit{PGP web of trust(2004)} network \cite{Ibnoulouafi201809,ZAREIE2017485,BAO2017391,LIU2016289,e22080848,RePEc:eee:phsmap:v:403:y:2014:i:c:p:130-147,spreadingdynamics201505,LIU20134154,ZENG20131031,BAE2014549,ZareieAhmad201811,LiuYingandTang201705,ZhangRuishengandYang201608,MA2016205,CaiBiaoandTuoXianGuo2014,QiuLiqing202107,WangJunyiandHou201702,ZHONG20152272,HU201673}, 
if individuals want to send or receive encrypted emails or encrypted files, they need to first gain the trust of others. Therefore, the transmission of information will only take place in a circle of mutual trust.} Knowledge network is different from social network. In social networks, network nodes disseminate the information they hold to all nodes connected to them, 
while in knowledge networks, network nodes only select specific different neighbors in a path for dissemination. In social networks, information may arrive at a node many times, whereas in knowledge networks, information reaches a node only once.} 

\subsubsection{\textbf{Technological network (TN)}}
\textcolor{black}{A technological network is a network used to distribute or deliver certain type of goods or resources.} \textcolor{black}{The trajectories in technological networks follow geodesics, and the mode of spread is transfer. This is because the number goods or resources in technological networks are usually limited and the traffic flow process in technological networks has a feature of fixed destination or target. \textcolor{black}{For example, in \textit{Airport/Airline} networks \cite{ZhaoJieandSong202008,BIAN2017422,FEI2017257,ZhaoJieandSongYutong202003,FeiLiguoandmo201708,8259501,ZhongLinFengandLiu201807,ZhongLinFengandShang201806}, airplanes usually travel along the shortest route to transfer people from one place to another. In \textit{Powergrid} networks \cite{DaiJinYingandWang201909,ZAREIE2017485,e22080848,ShaoZengzhenandLiu201911,ZENG20131031,BAE2014549,ZareieAhmad201811,LiuDongandNie201810,ZhaoZhiyingandWang201504,QiuLiqing202107,ShengJinfang201910,FeiLiguoandZhang201808}, researchers try to minimise losses in the transport of electricity and design the shortest possible transmission path. Likely, in \textit{E-road} network \cite{WenTaoandJiang201902,WangShashaandDu201611,ZhaoJieandWang202004,ZhaoJieandSong202008,BIAN2017422,7979310,ZhaoJieandSongYutong202003,ZAREIE2017485,khop201502,e22080848,ZareieAhmad201811,ZhangRuishengandYang201608,QiuLiqing202107} or \textit{Router/Routers} network \cite{hindex201601,CHEN20121777,BAO2017391,LIU2016289,spreadingdynamics201505,WEI2015277,LiuDongandNie201810,LiuYingandTang201705,MA2016205,ZhaoZhiyingandWang201504,QingChengHuanfYanShen201301}, roads between cities or network packets are usually aligned along the shortest paths to save cost and overhead.} Since “the information” in technological networks is usually something valuable, the information in the technical network is only relayed between nodes, and the transmission of 'this information" has cost. Therefore, in order to control the transmission cost, "the information" is usually transmitted along the shortest path.} 

\subsubsection{\textbf{Biological network (BN)}}
A biological network is marked by the graphical and mathematical modeling of interactions or relationships between cells or proteins \cite{inbookKim201905}. \textcolor{black}{The trajectories in biological networks follow geodesics or walks, and the mode of spread is transfer. The flow of information in biological networks is usually the transfer of certain substances. \textcolor{black}{In a few cases, information may be passed along the geodesics \cite{Computational_network_biology}, like sensory signal transmission in \textit{C. Elegans metabolic network (Elegans)} network \cite{ZAREIE2017485,e22080848,LIN20143279,WEI2015277,ZENG20131031,Garas_2012,BAE2014549,ZareieAhmad201811,ZhaoZhiyingandWang201504,LvZhiweiandZhao201902,FuYuHsiangandHuang201504}. Leukocytes move along the shortest path to destroy germs in \textit{Bio-dmela} network. But in most cases, information will propagate randomly between multiple nodes. A classical example is in \textit{Yeast} network, the transmission of neurotransmitters that randomly select their receptors \cite{Ibnoulouafi201809,e22080848,ZENG20131031,MA2016205,ZhongLinFengandLiu201807,ZhongLinFengandShang201806,BIAN2017422,HU201673,YangPingleandLiu201806,BAO2017391,ShengJinfang201910,ZhangJunkai201910,ZHAO202018}. Other scenarios 
like mutations in \textit{Protein-Protein Interactions} or \textit{Human protein} networks are random. When they infect other proteins or cells, the nutrients or molecules would not be copied during the transmission process \cite{ShengJinfang201910,ZhangJunkai201910}.} Therefore, similar to the case in technological networks, the information in biological networks is usually some biological substance, so its mode of spread is transfer, and there are usually many "receptors" in each information transmission round in the biological network. As mentioned above, when the information is transmitted to the next node, it may be a random selection of a "receptor", or it may be a competition between multiple "receptors", so its kind of trajectories may be geodesics or random walks, Which is shown in Fig. \ref{fig_MappingAndFeatures}. The dotted line in Fig. \ref{fig_MappingAndFeatures} represents the situation of few cases.} 

There also exist few synthetic (artificial) networks such as Lancichinetti–Fortunato–Radicchi \cite{ZAREIE2017485,ZhaoZijuan201912}, Watts–Strogatz \cite{WattsDuncan201112}, Barabási–Albert \cite{BAO2017391,BAE2014549,MA2016205,8259501,WangJunyiandHou201702,YuEnYuandWang202004}, and Factional Preferential Attachment (FPA) network \cite{RakRafalandRak202004}, and they can be combined with the above four types of real-world networks to test the performance of INRMs. These synthetic networks are also classified as one of the four types of network accordingly.

\subsection{Summary}
After extensive investigation, we classified 58 real-world networks used by the important nodes ranking methods based on the epidemic model into four categories according to the seven indicators shown in Fig. \ref{fig_MappingAndFeatures} and the network’s typical application scenario. We found that social networks are most studied, and only 34.5\% networks give their detail information and the accessible URLs. Based on the features of the seven indicators, researchers and practitioners can classify a real-world network accordingly.

\section{Existing influential node ranking methods}
\label{SV}

\label{sec2.2}
To present the results of RQ2, 81 INRMs based on the epidemic model were extracted in our survey. According to different computation of feature influence for nodes in networks, we divided INRMs into four major categories, which are structural centrality methods, iterative refinement centrality methods, MADM technique methods and machine learning methods. Structural centrality methods use the topological information of the network to directly obtain the centrality value of nodes. Iterative refinement centrality methods, on the other hand, apply random wandering to iteratively update the centrality value of each node until those of all nodes in the network converge to stable values, and they are mostly designed for directed networks. MADM technique refers to the integration and ranking of the criterion values of multiple schemes under multiple criteria. Accordingly, MADM technique methods treat the ranking results under multiple classical methods as criteria in MADM technique and reorder. Machine learning methods transform the influential node ranking problem into a regression or classification problem, and result values in structural centrality methods are used as the input to the machine learning model.

% The classification results and their representative methods is shown in Fig. \ref{fig_ClassificationOfMethods}.
% \begin{figure*}[htbp!]
% \centering
% \tiny
% \includegraphics[width=1\textwidth]{methodsClassification_original2.pdf}
% \caption{\textsc{Classification of Methods and Representative Methods}}
% \label{fig_ClassificationOfMethods}
% \end{figure*}

\subsection{Structural centrality method}
Depending on the scale of network topology considered, structural centrality methods are further divided into local centrality methods, semi-local centrality methods, global centrality methods, and hybrid centrality methods. Their difference in influence measures and the major metrics or methods related to node importance used in INRMs are summarized in Table \ref{tab:structural_centrality_method}.

\begin{table*}[htbp!]
\caption{The influence measures and major metrics or methods to compute node importance}
\label{tab:structural_centrality_method}
\resizebox{\linewidth}{!}{
\begin{tabular}{l|l|l|l|l|m{10.5cm}}
\hline
\textbf{ Meta-methods} & \multicolumn{1}{l|}{{\textbf{PNI}}} & \multicolumn{1}{l|}{{\textbf{INMN}}} &\multicolumn{1}{l|}{{\textbf{IMN}}} & {\textbf{IAN}} & \textbf{The major metrics or methods related to node importance used in INRMs} \\ \hline
Local & \multicolumn{1}{l|}{$\bullet$} & \multicolumn{1}{l|}{$\bullet$}  & &  & {Degree Centrality \cite{FreemanLinton197901}, Temporal Degree Deviation Centrality \cite{7978752}, Hybrid degree centrality \cite{MaQianandMaJun201608}, \textit{H-index} \cite{hindex201601}, KED method \cite{Chen_2013}, CSE method \cite{Mekonnen202003}, Local dimension method \cite{PuJunandChenXiaowu201407}, FLD method \cite{WenTaoandJiang201902}, \textit{K-hop} method \cite{khop201502}, Density centrality \cite{Ibnoulouafi201809}, Local neighbor contribution centrality \cite{DaiJinYingandWang201909}} \\ \hline
Semi-local & \multicolumn{1}{l|}{$\circ$} & \multicolumn{1}{l|}{$\circ$} & \multicolumn{1}{l|}{$\bullet$} &  & {ERM \cite{ZAREIE2017485}, LISH \cite{HuiyuandZun201301}, ISH \cite{YuHuiandCao201705}, E-Burt \cite{HuPingandMei201709}, Bao et al. \cite{BAO2017391}, Semi-local centrality method \cite{8351889}, Dong et al. \cite{8351889}, Weighted Semi-local Centrality Method \cite{7925142}, Shao et al. \cite{ShaoZengzhenandLiu201911}, Cluster Rank \cite{theroleofclustering201310}, Chen et al. \cite{theroleofclustering201310}, Gao et al. \cite{RePEc:eee:phsmap:v:403:y:2014:i:c:p:130-147}} \\ \hline
Global & \multicolumn{1}{l|}{$\circ$} & \multicolumn{1}{l|}{$\circ$} &\multicolumn{1}{l|}{$\circ$} & $\bullet$ & {Closeness centrality\cite{Sabidussi196602}, Betweeness centrality \cite{FreemanLinton197901}, Quasi-local structure\cite{WEN2020549}, The \textit{k-shell} decomposition method and its variants \cite{KitsakMaksim201001,ZENG20131031,Garas_2012,BAE2014549,LiuYingandTang201705,ZhangRuishengandYang201608}, Community partitioning algorithms \cite{8322554,8259501,Piccardi201111,ZhaoZijuan201912,londel200804,ZhaoZhiyingandWang201504,Duch200509,CaiBiaoandTuoXianGuo2014}, Node deletion \cite{LvZhiweiandZhao201902,BoccalettiStefano200602,WangShashaandDu201611}} \\ \hline
Hybrid & $\circ$ & $\circ$  & $\circ$ & $\circ$  & {Neighborhood Centrality \cite{LIU2016289}, TsallisRank \cite{e22080848}, Liu et al. \cite{LIU2016289}, Lin et al. \cite{LIN20143279}, Liu et al. \cite{spreadingdynamics201505}, Wei et al. \cite{WEI2015277}, Liu et al. \cite{LIU20134154}, Basaras et al. \cite{6477051}, Improved \textit{k-shell} decomposition algorithms \cite{10.1155/2020/5903798,LiuDongandNie201810,YangYuanzhi202005,ShengJinfang201910,ZhongLinFengandLiu201807}, Zareie et al. \cite{ZareieAhmad201811}, Gravity formula \cite{MA2016205}, Liu et al. \cite{8529839}, Qiu et al.\cite{QiuLiqing202107}, Zhao et al. \cite{ZhaoJieandWang202004}, Zhong et al. \cite{ZhongLinFengandLiu201807}, Percolation theory \cite{KarrerBrian201405}, Sheng et al. \cite{ShengJinfang201910}, Zhang et al. \cite{ZhangJunkai201910}, Zhong et al. \cite{ZhongLinFengandShang201806}, Fu et al. \cite{FuYuHsiangandHuang201504}, Qing et al. \cite{QingChengHuanfYanShen201301}} \\ \hline
\end{tabular}
}

 \begin{tablenotes}
        \footnotesize
        \item Influence measures (PNI: The properties of the target node itself, INMN: The influence of the number of multi-hop \\neighbors, IMN: The influence among multi-hop neighbors, IAN: The influence of all nodes in the network)
        \item $\bullet$ means the methods prefer to use the corresponding influence measures. 
        \item $\circ$ shows that the methods may use this type of influence measures, or does not use.
      \end{tablenotes}

\end{table*}

\subsubsection{\textbf{Local}} \textcolor{black}{Local centrality methods use only the properties of the target node itself. As shown in Table \ref{tab:structural_centrality_method}, the properties can be Degree Centrality \cite{FreemanLinton197901}, Temporal Degree Deviation Centrality \cite{7978752} 
and Hybrid degree centrality \cite{MaQianandMaJun201608}, or the influence of multi-hop neighbors on the node, such as \textit{H-index}\cite{hindex201601}, KED method \cite{Chen_2013}, Content spreading
efficiency (CSE) method \cite{Mekonnen202003}, local dimension method \cite{PuJunandChenXiaowu201407},  Fuzzy local dimension (FLD) method \cite{WenTaoandJiang201902}, \textit{K-hop} method \cite{khop201502}, Density centrality \cite{Ibnoulouafi201809} and local neighbor contribution
centrality \cite{DaiJinYingandWang201909}.} For example, Degree Centrality \cite{FreemanLinton197901} assumes that influential nodes are nodes with many connections. A greater degree of a node indicates that its influence is greater. The KED method \cite{Chen_2013} argues that only calculating the number of paths which contain multi-hop neighbor nodes is not sufficient when ranking the influence of nodes. Path diversity is usually overlooked. Path diversity is often evaluated by \textcolor{black}{Information entropy. If a target node has better degree evenness, then its path diversity is greater, and the information through the node have more ways to spread outward along its neighbors.} The KED method also used a normalization approach for the information entropy path diversity, which shields the effect of the degree difference of the central node.

\subsubsection{\textbf{Semi-local}} \textcolor{black}{Semi-local centrality methods also prefer to consider the effect among multi-hop neighbors, namely the relationship among multi-hop neighbors (Table \ref{tab:structural_centrality_method}), which  
% The local centrality method either calculates the number of direct neighbor nodes or paths, or considers the influence of the number of multi-hop neighbors in order to improve performance. However, the semi-local centrality method believes that it is insufficient to only calculate the number when considering the influence of multi-hop neighbors. For a Semi-local centrality method, the relationship between multi-step neighbor nodes 
may have either positive or negative effect on the importance of the target nodes.} \textcolor{black}{For example, when there are more common neighbors among neighbors, the higher the degree of coupling in the influence propagation range between neighbors, which has positive impact on the importance of nodes. If there is no direct connection between neighbor nodes, it means that the channels of information dissemination are independent of each other and also have negative impact on the importance of nodes.} The semi-local centrality method believes that it is insufficient to only calculate the number when considering the influence of multi-hop neighbors. Entropy-Based Ranking Measure (ERM) \cite{ZAREIE2017485} holds that a node has strong propagation ability if the degree of neighbor nodes is both high and uniform, and information entropy is used to quantify the high and uniform metrics. \textcolor{black}{When information spreads through networks, a few nodes in networks bridge the information transmission between neighbor nodes, which is called structural hole. Identification of the node can better evaluate its importance. However,} classic structural hole method considers only the nearest neighbors. Local Improved Structural Holes (LISH) \cite{HuiyuandZun201301} and Improved Structural Holes (ISH) \cite{YuHuiandCao201705} take \textit{2-hop} neighbors into consideration. E-Burt \cite{HuPingandMei201709} proposed the entropic degree concept and modified structural hole method. Moreover, \textcolor{black}{local cluster coefficient is a coefficient used to describe the agglomeration between nodes in a network. If the local cluster coefficient is larger, it means that the connection between nodes is closer.} 
% Cluster Rank \cite{theroleofclustering201310} are designed for directed network, which holds that local clustering usually has negative effect on information dissemination. Since the neighbors of the target node closely interact with each other rather than other nodes in the network, the information originated from the target node will more likely confined in a local area. Based on this, 
Chen et al. \cite{theroleofclustering201310} proposed a simple exponential function to quantify the local Cluster coefficient.
% They hold that local clustering confine information into a local area, while Gao et al. \cite{RePEc:eee:phsmap:v:403:y:2014:i:c:p:130-147} investigated that local clustering can have positive effect on information dissemination. By considering the number of \textit{2-hop} neighbors, 
\textcolor{black}{Gao et al. \cite{RePEc:eee:phsmap:v:403:y:2014:i:c:p:130-147} calculated a cluster coefficient for each node by calculating the connections between the node and within two-step neighbor nodes. When calculating the importance of the node, the node cluster coefficient is taken as an important factor, and the larger the cluster coefficient, the higher the importance of the node.}

\subsubsection{\textbf{Global}} \textcolor{black}{Both local centrality methods and semi-local centrality methods measure the importance of a node by considering the node itself or its neighbor nodes. However, global centrality methods hold that when measuring the importance of a node, it is more accurate to consider all network nodes (Table \ref{tab:structural_centrality_method}). For example, when considering the importance of a node, local centrality methods may involve the degree of the node itself, semi-local centrality methods consider the degree relationship between the node and the one within the two-step neighbor range, and global centrality methods take into account the degree relationship of all nodes in the network. Therefore, global centrality methods usually have \textit{O(N)} and above algorithm complexity.} To assess relationship of all nodes in the network, closeness centrality \cite{Sabidussi196602}, betweeness centrality \cite{FreemanLinton197901}, and quasi-local structure \cite{WEN2020549} are used to compute the shortest path between each pair of nodes in the network. The \textit{k-shell} decomposition method and its variants are used to  \cite{KitsakMaksim201001,ZENG20131031,Garas_2012,BAE2014549,LiuYingandTang201705,ZhangRuishengandYang201608} peel the network layer by layer from the outside to the inside. Other methods apply community partitioning algorithms \cite{8322554,8259501,Piccardi201111,ZhaoZijuan201912,londel200804,ZhaoZhiyingandWang201504,Duch200509,CaiBiaoandTuoXianGuo2014}, or measure the importance of nodes by deleting nodes and observing network changes \cite{LvZhiweiandZhao201902,BoccalettiStefano200602,WangShashaandDu201611}.

\subsubsection{\textbf{Hybrid}} Hybrid centrality methods can be fusion methods of previous three kinds of structural centrality methods (Table \ref{tab:structural_centrality_method}) or fusion methods of custom metrics for the network topology. The hybrid centrality method believes that each structural centrality method has its own advantages and disadvantages, so the hybrid centrality method usually integrates multiple structural centrality methods in some way. For example, Neighborhood Centrality \cite{LIU2016289} calculates the sum of Degree Centrality and Coreness Centrality of multi-hop neighbor nodes to quantify the spreading influence of the target node, TsallisRank \cite{e22080848} used Closeness Centrality as parameter of entropy. The TsallisRank \cite{e22080848} method used the closeness centrality as the parameter of Tsallis entropy, and then combined the entropy of nearest with the next nearest neighbors to calculate the spreading influence. There are a number of other integration forms for different Hybrid centrality methods. Lin et al. \cite{LIN20143279} evaluated the propagation capability of the target node by referring to the sum of \textit{k-shell} values of its neighbors. Qiu et al.\cite{QiuLiqing202107} and Zhao et al. \cite{ZhaoJieandWang202004} proposed two indexes like local index and global index to evaluate the node.

\subsection{Iterative refinement centrality method}
\textcolor{black}{In this method, each node in the network is given an initial score, and then several iterations are performed. In each round of iteration, the score of each node accumulates by specific rules \cite{Moler_Cleve}. The algorithm converges after the score of each node has been stabilized. At this point, the final scores of nodes serve as the basis of judgment for their importance. Classical iterative refinement centrality methods include eigenvector centrality \cite{Estrada200506}, PageRank \cite{PageL199901} method and its variants \cite{ZhaoJieandSong202008,leaders201112,LI201447,PageL199901,Ren_2014,ZHONG20152272}.} Among them, the main idea of eigenvector centrality \cite{Estrada200506} is that the importance of a node is determined not only by the number of its neighbors, but also by their importance. In the computation of eigenvector centrality, the weight of each node is diffused to its neighbors. After enough rounds of diffusion, the weights of all the nodes will reach stability, and the importance of a node is determined by its final weight. PageRank \cite{PageL199901} was originally applied in Google search engine to rank web ranges. The PageRank method uses hyperlinks in web pages to jump pages to simulate random walking. The random walking process in the PageRank method is like the diffusion process in eigenvector centrality. Finally, the score of each web page reaches stability and is used as the basis for page ranking. In a standard PageRank method, the probability of jumping from one node to another is the same, and the correlation between these nodes is neglected. Based on Kullback-Leibler divergence, Zhao et al. \cite{ZhaoJieandSong202008} proposed a structural similarity calculation method in place of the original transition module in the standard PageRank method. Because the PageRank method is computationally intensive and not applicable to dynamic networks, Lü et al. \cite{leaders201112} put forward a LeaderRank method for the rapidly-changing characteristics of social network. LeaderRank introduced a ground node which is connected to every node in the network by bidirectional links and makes the network  strongly connected. This ensures that the scores of all the nodes converge to unique steady state. Compared to PageRank, LeaderRank is parameter-freeness, which can be applied to any type of network, and is independent of the initial conditions.

% Considering the number of fans in social network is an important indicator for a people’ influence, Li et al. \cite{LI201447} proposed a biased random walk, which allowed nodes with greater in-degree gets more scores in each round of iteration. In references \cite{PageL199901,leaders201112,LI201447}, each node is assigned the same score. To incorporate classic centrality methods with iterative refinement centrality method, Ren et al. \cite{Ren_2014} proposed IRA method and used centrality value under some certain centrality method as the initial score of nodes. The references \cite{Ren_2014,ZHONG20152272} considered that the spreading rate may affect the process of resource allocation.

\subsection{MADM technique}
\textcolor{black}{Different from Hybrid centrality methods that accumulate the index values of each INRM based on
 structural centrality, Multiple Attribute Decision Making (MADM) technique deals with the assessment of a set of alternatives through multi-criteria decision making \cite{Hwang1981MultipleAD}. MADM technique can integrate various types of INRMs rather than only integrate structural centrality methods like Hybrid methods. \textcolor{black}{There are three basic elements in multi-attribute decision-making, namely schemes, evaluation attributes and weight allocation. Multi-attribute decision-making first selects multiple schemes, then evaluates the performance of evaluation attributes under the schemes, and finally assigns weight to attributes. In the application of important node ranking, schemes and attributes that are often determined by traditional ranking methods are used to calculate the weight of these ranking methods, and the weight can be further applied to obtain the values of importance of these nodes.} MADM techniques include the Analytic Hierarchy Process (AHP)-based method \cite{BIAN2017422}, the TOPSIS method \cite{DU201457,HU201673,FEI2017257,YangYuanzhi201911,YangPingleandLiu201806}, and the Dempster–Shafer (DS) evidence theory method \cite{7979310,WEI20132564,ZhaoJieandSongYutong202003}.}

\subsection{Machine learning}
The machine learning method usually divides the node ranking problem into regression and classification problems, and then builds models for machine learning or neural networks to address these problems. The inputs of these models are usually node features in the network. The simulation results of the epidemic model are often used as ground truth to train the model and evaluate model performance \cite{ZHAO202018,YuEnYuandWang202004}. Machine learning models for ranking important nodes include Graph Convolutional Network (GCN), Naive Bayes (NB), Decision Tree (DT), Random Forest(RF), Support Vector Machine (SVM), K-Nearest Neighbor (KNN), Logistic Regression (LR), and Multi-Layer Perceptron (MLP) \cite{ZhaoGouhengandJia202003}. Zhao et al. \cite{ZHAO202018} proposed the InfGCN method and turned the influential node ranking problem into the binary classification problem according to a GCN algorithm. In the GCN model, neighbor networks and four classic structural features (degree centrality, closeness centrality, betweenness centrality, clustering coefficient \cite{ZhangPengandWang200709}) are used as node features, and classification results were compared with the ground truth derived from the SIR model. Yu et al. \cite{YuEnYuandWang202004} transformed the influential node ranking problem into a regression problem by a RCNN algorithm, in which the matrix forms of the nodes’ neighbor network are node features, and they are combined with the labels from all nodes’ infected scales by the SIR model to train the CNN model. Zhao et al. \cite{ZhaoGouhengandJia202003} introduced seven classic machine learning algorithms to the influential node ranking problem, including NB, DT, RF, SVM, KNN, LR and MLP.

\subsection{Summary}
The INRMs are firstly divided into four major-categories according to the computation process of feature influence, which are strcutral centrality method, iterative refinement method, MADM-based method and machine learning method. Then, each kind of method is deeply analyzed and meta-categories are divided. The conduct process of the INRMs based on epidemic model are discussed, which may be similar with the INRMs under other results validation models (cascade failure model, linear threshold model). However, we focus more on the ranking rationality of INMRs validated by epidemic model, which would be analyzed in next section.

\section{Evaluation Metrics} 
Among 81 INRMs we studied, 79 methods assign a value to each node in the network for influence ranking, and the rest directly assigns the nodes to certain ranks, for instance, the \textit{k-shell} decomposition method \cite{KitsakMaksim201001} decomposed the network and assigned a same rank to nodes within the same layer. GCN-based method \cite{ZHAO202018} transformed influential node ranking problem into a classification problem. Since each method yields a ranking list of node influence, a series of evaluation metrics were adopted for ranking lists to evaluate the efficiency of INRMs. 

To answer RQ3, We divided the evaluation metrics for ranking lists into the capability of ranking itself and the correctness of ranking from the ranking rationality perspective. The Capability metrics on the one hand evaluate the ranking granularity or distinguish the ranks with the same importance of nodes, and the ranking task of a coarse-grained ranking method is poorly accomplished. The capability metrics, on the other hand, verify the ranking credibility of INRMs by calculating similarity to classic or popular methods. The Correctness metrics evaluate the validity of a ranking list compared to the benchmark influence ranking list \textcolor{black}{and compare the consistency of the ranking list with the benchmark ranking list further generated by the robustness impact on a network after removing nodes.} \textcolor{black}{In addition, the comparison of INRMs based on the benchmark dissemination of nodes in the ranking list of each INRM generated by epidemic model can enhances the credibility of the comparison.} The classification of evaluation metrics are shown in Table \ref{tb_EvaluationMetrics}.

\begin{table}[htbp!]
\centering
\scriptsize
\caption{The Classification of Evaluation Metrics}
\label{tb_EvaluationMetrics}
\begin{tabular}{l|l|l} 
\hline
                             & \textbf{The Function}                                                                                                           & \textbf{Metrics}                                                                                 \\ 
\hline
\multirow{4}{*}{Capability}  & \multirow{3}{*}{\begin{tabular}[c]{@{}l@{}}The ability to differentiate  \\ equally important nodes \end{tabular}}                  & Distinct metric                                                                                  \\ 
\cline{3-3}
                             &                                                                                                                                 & Complementary  cumulative distribution function \\ 
\cline{3-3}
                             &                                                                                                                                 & Monotonicity function   \\ 
\cline{2-3}
                             & The ability to assess ranking credibility                                            & Ranking similarity                                                                               \\ 
\hline
\multirow{6}{*}{Correctness} & \multirow{5}{*}{\begin{tabular}[c]{@{}l@{}}The validity to approach \\the benchmarks \\(Epidemic model based)\end{tabular}} & Kendall’s tau correlation coefficient        \\ 
\cline{3-3}
                             &                                                                                                                                 & Spearman coefficient                                \\ 
\cline{3-3}
                             &                                                                                                                                 & Jaccard similarity coefficient             \\ 
\cline{3-3}
                             &                                                                                                                                 & Imprecision function                                  \\ 
\cline{3-3}
                             &                                                                                                                                 & Pearson coefficient                                                                              \\ 
\cline{2-3}
                             & \begin{tabular}[c]{@{}l@{}}The consistency with the \\ further benchmarks generated \\ 
by the robustness impact  \end{tabular}                                          & Robustness                                                                                       \\ 
\hline
\end{tabular}
\end{table}

\subsection{Capability}
\subsubsection{The capability to differentiate equally important nodes}
Assumed that the rank of a node is dependent on its centrality value, nodes with the same centrality value cannot be distinguished from each other. When a method is used to identify influential nodes and multiple nodes have the same rank, these nodes cannot be precisely distinguished despite the fact that they have different spreading influence. For this situation, a fine-granular method can assign fewer nodes to the same importance level, which means that it can more effectively differentiate the importance of different nodes. However, in view of the ranking results obtained by the coarse-granular methods, many nodes are assigned to the same importance level, and their importance is not distinguishable. Taking the Degree Centrality method as an example, a core node in a network generally has greater spreading influence than a periphery node with the same degree. Thus, the frequency of identical/different rank of nodes can be considered as a measure of the capability of methods, including \textbf{Distinct Metric (DM), Complementary Cumulative Distribution Function (CCDF)} and \textbf{Monotonicity Function (MF)} as follows.
\begin{equation}
    DM(R)=\frac{n}{\left | V \right |}, 
\end{equation}
where $R$ represents the ranking list, $n$ refers to the number of distinct ranks in ranking list $R$, and $\left | V \right |$ denotes the total number of nodes in the network. DM takes a value within the scope $\left [ 0,1 \right ]$. If a distinct rank is assigned to all nodes, the DM reaches a maximum value $1$. Minimum value $0$ means that all nodes are of equal rank. A higher DM value indicates better differentiation \cite{LIU20134154}. 

\begin{equation}
    CCDF(R)=\frac{{\left | V \right |}- {\textstyle \sum\limits_{r\in{R}}}{n_r}}{\left | V \right |},
\end{equation}
where $n_r$ is the number of nodes occupied at rank $r$ in a ranking list. The CCDF value decreases with the increase in $n_r$. The ranking distribution performance becomes worse when the function value decreases more rapidly \cite{Zhu_Jingcheng_and_Wang}. 

\begin{equation}
    MF(R)=\left(1-\frac{ {\textstyle \sum\limits_{r\in{R}}n_r\ast(n_r-1)} }{n\ast(n-1)} \right)^{2}.
\end{equation}
The ranking list is completely monotonic and each node is assigned a different rank if $MF(R)=1$. Otherwise, all nodes are assigned with the same rank when $MF(R)=0$. When the value of function $MF(R)$ is closer to $1$, the resolution/recognition of the ranking list will be better \cite{BAE2014549}. \textcolor{black}{Similarly, based on real-world networks like Karate \cite{WenTaoandJiang201902,DaiJinYingandWang201909}, Dolphins \cite{Ibnoulouafi201809,HuPingandMei201709}, Airport \cite{WenTaoandJiang201902,hindex201601} and Game Of Thrones \cite{Ibnoulouafi201809} in the Density centrality method \cite{Ibnoulouafi201809}, the MF metric demonstrates that their ranking results have better monotonicity ability.}

\subsubsection{The capability to assess ranking credibility}
In general, new proposed methods should be compared with classic methods to prove their credibility \cite{HU201673,ZhaoZhiyingandWang201504,BIAN2017422,FEI2017257}. On full consideration of information in the network, different INRMs could generate different ranking lists. Considering that nodes on the top rank play a vital role in information spreading, \textit{top-k} nodes are usually used to compute the similarity between different methods \cite{FeiLiguoandZhang201808,YangYuanzhi202005,BAE2014549}. Here, variable $k$ can be adjusted by the size of networks. Thus, the number of same nodes among 
the \textit{top-k} nodes in new proposed methods and classic methods can show the similarity of information considered and illustrate the credibility of the new proposed methods \cite{ZhangJunkai201910,YangYuanzhi201911}. \textcolor{black}{For example, Zhang et al. \cite{ZhangJunkai201910} applied Closeness Centrality\cite{Sabidussi196602}, PageRank\cite{PageL199901}, Eigenvector centrality \cite{Estrada200506}, \textit{H-index} and LeaderRank\cite{leaders201112} as ranking standards. The more same nodes between the \textit{top-k} node set or ranking list of the proposed INRM and the results obtained by these standards are, the better the ability of the INRM to assess ranking credibility is.}

\textbf{Ranking Similarity (RS)} metric is defined as follows.
\begin{equation}
    RS(k)=\frac{\left | R_1(k)\cap R_2(k) \right |}{\left | R_1(k)\cup  R_2(k) \right |},
\end{equation}
where $R_1$ and $R_2$ are the ranking lists of new proposed method and its comparison method, and $R_1(k)$ and $R_2(k)$ are the \textit{top-k} nodes in ranking lists $R_1$ and $R_2$. Once \textit{k} equals the total number of nodes in the network, a comprehensive comparison is conducted on these ranking lists. The closer the RS value is to 1, the higher Capability the proposed method is \cite{PuJunandChenXiaowu201407,DaiJinYingandWang201909,Ibnoulouafi201809}. \textcolor{black}{Identifying influential nodes in Powergrid \cite{DaiJinYingandWang201909,ZAREIE2017485}, Router \cite{hindex201601,CHEN20121777} and Blogs \cite{hindex201601,WEN2020549} networks under Community partitioning algorithms \cite{ZhaoZhiyingandWang201504}, the RS metric validated the similarity of proposed INRMs to industry-validated INRMs.}

\subsection{Correctness}
The actual propagation capability of each node in networks is usually difficult to measure. Therefore, various spreading models are used as benchmarks for comparison and evaluation of INRMs, of which popular models are Epidemic model, Independent Cascade Model, Linear Threshold Model, Cascading Failure model \cite{LU20161,2021ScChE..64..451L}. Among them, the Epidemic model is the most frequently used and the simplest \cite{LU20161,Efficient_analysis_of_node_influence,Effective_spreading_from_multiple}, including SI model, SIR model, and SIS model \cite{Contributions_to_the_mathematical}. 

To simulate the propagation ability of each node in the real-world network, all nodes are selected successively as the single initial node in an epidemic model to initiate infection \cite{Zhao_Xiang-Yu_and_Huang}. At the final stage of infection, the number of infected nodes is denoted as the propagation ability of a node. The ranking list generated by the epidemic model is used as the evaluation benchmark for the correctness of INRMs. \textcolor{black}{Further, the consistency with the ranking list generated by the robustness impact on a network after removed nodes are identified can also verify the correctness of INRMs. }

\subsubsection{The validity to approach the benchmarks (Epidemic model based)}
For a node ranking method, assumed that nodes with larger centrality value (higher rank) should have stronger propagation capability \cite{CHEN20121777}, the ranking list would be generated by different centrality values of each node. At this time, another benchmark ranking list can be obtained according to the node’s influence measured by SI, SIR or SIS model.  A strong positive correlation between the two ranking lists shows higher correctness of these methods \cite{WenTaoandJiang201902}, and validation for the following five metrics is based on these lists.

\textbf{Kendall’s tau Correlation Coefficient (KCC)}. Suppose $x_i,y_i$ are the ranking value from ranking lists $X$ and $Y$ respectively, then $\left ( x_1,y_1 \right ), \left ( x_2,y_2 \right ),\dots ,\left ( x_n,y_n \right )$ is a set of joint ranks. For any pair of ranks $(x_i,y_i)$ and $(x_j,y_j)$, if $(x_i>x_j\;\rm{and}\;y_i>y_j)\;or\;(x_i<x_j\;\rm{and}\;y_i<y_j)$, the pair is regarded as concordant $(c)$; otherwise, $(x_i>x_j\;\rm{and}\;y_i<y_j)\;or\;(x_i<x_j\;\rm{and}\;y_i>y_j)$ is taken as discordant $(d)$; $(x_i=x_j)\;or\;(y_i=y_j)$ refers to concordant. Kendall’s tau values \cite{A_New_Measure_of_Rank} for the two ranking lists $X$ and $Y$ are computed as follows.
\begin{equation}
    \tau=\frac{2(n_c-n_d)}{n(n-1)}, 
\end{equation}
where $n_c$ and $n_d$ are the numbers of concordant and discordant pairs in each list respectively, and $n$ is the size of the ranking list.

More specifically, KCC can be expressed by \cite{ZhongLinFengandLiu201807}.
\begin{equation}
    \tau=\frac{2}{n(n-1)}\sum_{i<j}sgn[(x_i-x_j)(y_i-y_j)],
\end{equation}
where $n$ is the size of a ranking list, $sgn(x)$ is the piecewise function, that is to say, when $x>0\;\rm{then}\;sgn(x)=+1;\;x<0\;\rm{then}\;sgn(x)=-1$, or when $x=0\;\rm{then}\;sgn(x)=0$. Kendall’s tau value $\tau$ is assigned between $\left [ -1,+1 \right ] $.

Then the improved $\tau$ ratio $\eta$ is calculated.
\begin{equation}
    \eta=\frac{\tau^{p}-\tau^{0}}{\tau^{0}}, 
\end{equation}
where $\tau^{p}$ is the Kendall’s $\tau$ of the proposed method, $\tau^{0}$ is the Kendall’s $\tau$ obtained by the benchmark methods. $\eta>0$ indicates that the proposed method has advantages. \cite{Chen_2013}. The KCC metric confirms the consistency of the ranking results of these INRMs with the simulation results of the epidemic model.

\textbf{Spearman Coefficient (SC)} \cite{Pearson_versus_Spearman} is a nonparametric measure of rank correlation between the two ranking lists, which is usually recorded as $\rho$.
\begin{equation}
    \rho=1-\frac{6 {\textstyle \sum_{i=1}^{n}}d^{2}_{i} }{n(n^{2}-1)}, 
\end{equation}
where $d_{i}$ represents the difference ($\left | x_i-y_i \right |$) between ranks for each $x_i$ and $y_i$, $n$ is the number of node set. The value of $\rho$ is close to $1$ when their ranking is consistent, and the value of $-1$ indicates that their ranking is completely opposite. The SC metric confirms the consistency between the ranking results and benchmarks. 

\textbf{Jaccard similarity coefficient (JSC)} \cite{Distance_Between_Sets} is used to compare the similarities and differences between the two ranking lists. JSC is similar to RS, but JSC compares the target ranking list with the one generated by the epidemic model rather than other ranking methods. It is defined as follows.
\begin{equation}
    J(k)=\frac{\left | X(k)\cap Y(k) \right | }{\left | X(k)\cup Y(k) \right | }, 
\end{equation}
where $X(k)$ and $Y(k)$ represents two sets of \textit{top-k} nodes in ranking lists \textit{X} and \textit{Y}. The closer the value of $J(k)$ is to $1$, the greater similarity the two ranking lists have. 

\textbf{Imprecision function (IF)} \cite{artiSehgal_Umesh_andcle} is applied to quantify how the propagation ability of nodes on the top rank in ranking methods is close to that generated by epidemic model simulations. Then the metric to evaluate the cumulative propagation ability of top-ranked nodes in different proportions is expressed as $\varepsilon(p)$. 
\begin{equation}
    \varepsilon_{\theta}(p)=1-\frac{M_{\theta}(p)}{M_{\rm{epi}}(p)}, 
\end{equation}
where $\theta$ represents a ranking method, $\rm{epi}$ represents simulations of epidemic models, $p$ represents a percentage of network size $N\;(p\in[0,1])$, and $M_{\theta}(p)$ represents the average propagation ability of the former $p*N$ nodes obtained by ranking algorithms. $M_{epi}(p)$ is the average propagation ability of $p*N$ nodes generated by comparing the actual propagation ability under epidemic models. The value of $\varepsilon(p)$ is assigned between $\rm0$ and $\rm1$. A smaller value of $\varepsilon(p)$ indicates higher Correctness of the method $\theta$ in identifying the most influential nodes. The IF metric shows the correctness of top-ranked nodes obtained under the simulation of the epidemic model.

\textbf{Pearson Coefficient (PC)} \cite{ZhaoZhiyingandWang201504,Distance_Between_Sets}, simply known as “the correlation coefficient”, is a measure of the linear correlation (dependence) between ranking list $X$ and ranking list $Y$.
\begin{equation}
    \psi=\frac{ {\textstyle \sum_{i=1}^{n}(x_{i}-\overline{X})(y_{i}-\overline{Y})} }{\sqrt{ {\textstyle \sum_{i=1}^{n}(x_{i}-\overline{X} )^2} }\sqrt{ {\textstyle \sum_{i=1}^{n}(y_{i}-\overline{Y} )^2} } },
\end{equation}
where $\overline{X}$ and $\overline{Y}$ are the average ranking value of ranking lists $X$ and $Y$ respectively. The value of $\psi$ is in $[-1,+1]$. The correlation between the two ranking lists is positive when $\psi>0$, and it is negative when $\psi<0$. The two ranking lists are uncorrelated when $\psi=0$. Hence, the closer the value of $|\psi|$ is to $\rm{1}$, the more similarity the two ranking lists has. In general, $|\psi|>\rm{0.8}$ indicates an extremely strong correlation, $\rm{0.6}<|\psi|<\rm{0.8}$ a strong correlation, and $\rm{0.4}<|\psi|<\rm{0.6}$ a moderate correlation. 

\subsubsection{The consistency with the further benchmarks generated by the robustness impact}
\textcolor{black}{Take the initial network as $N$ and the network after removing the node as $N^{'}$, the sum of the changes of a index of all nodes between $N$ and $N^{'}$ is the robustness impact on the network. The index can be represented by the ranking value changes under the epidemic model or centrality values, etc. Each node in the network would generated an impact value. According to these impact values, a further benchmark ranking list is obtained. Similarly, the higher consistency  between the two raking lists also shows the Correctness of method. Following indexes are the multi-form computation of the impact for a node in the network.}
 
\textbf{Robustness (R)}, 
% the tolerance of ranking against spurious and missing links, i.e., false positive and false negative connections, is crucial when network structure is subjected to noisy observations 
can be considered as the tolerance for ranking false and missing links, and it is critical when network structures are affected by noisy observations
\cite{YuHuiandCao201705,YangYuanzhi202005,Searching_for_superspreaders}. To study the tolerance of these noisy data, the impact from changes in centrality values and ranking values under the epidemic model can be measured when links and nodes are added or removed randomly \cite{Ibnoulouafi201809}. This kind of impact dependent on centrality values $I_c$ is described as follows.

\begin{equation}
    I_c= {\textstyle \sum\limits_{i=1}^{n}\left | C_{i}^{'}-C_i  \right | },    
\end{equation}
where $C_i$ and $C_{i}^{'}$ are the centrality values of a node in network $N$ and network $N^{'}$, and $n$ represents the total number of nodes in the network $N^{'}$. A smaller value of $I_c$ indicates smaller influence of the removed node and more tolerance against noisy data in the ranking method. Further, changes in centrality value do not directly correspond to changes in ranking values. A similar measure to examine the impact $I_r$ on ranking values under the epidemic model is proposed by Liu et al. \cite{LiuZhonghuaandJiangChengandWang201504}.

\begin{equation}
    I_r= {\textstyle \sum\limits_{i=1}^{N}\left | R_{i}^{'}-R_i  \right | },    
\end{equation}
where $R_{i}^{'}$ and $R_i$ are the ranking values of a node in network $N$ and  network $N^{'}$.

There also exists other calculation method \cite{ZhongLinFengandShang201806} for the robustness impact.
\begin{equation}
    \eta=\frac{1}{n(n-1)} {\textstyle \sum_{i\ne j\in{n}}e_{ij}},
\end{equation}
where $e_{ij}$ represents the efficiency between nodes $i$ and $j$, $e_{ij}=\frac{1}{d_{ij}}$, $d_{ij}$ represents the length of the shortest path between node $i$ and node $j$, and $n$ represents the number of nodes in the network. Then the decline rate of network efficiency $\mu$ is defined as $\mu$.
\begin{equation}
    \mu=1-\frac{\eta}{\eta_{o}}, 
\end{equation}
where $\eta$ represents the efficiency of  network $N^{'}$, and $\eta_{o}$ represents the initial efficiency of network $N$. A greater value of $\mu$ means worse network connectivity destroyed by the removed node and and more influence of the removed node.

Finally, based on the robustness impact values for all nodes in network $N$, a further benchmark ranking list is obtained. The calculation of consistency between the ranking list from an INRM and the benchmark ranking list is also based on equation (8)-(14).

\subsection{The strategies for the comparison of INRMs by using the epidemic model}

\textcolor{black}{The RS metric of Capability and the JSC of Correctness are all focused on the similarity between ranking lists directly based on simple operation of node sets in those ranking lists. To further increase the credibility of comparison, when comparing the efficiency of multiple INRMs, a part of nodes in the ranking list of each INRM would be selected to input into the epidemic model to get the propagation ability. This propagation ability is regard as benchmark propagation ability, which can be used to prove the advantages of the method over other methods. For an INRM, the greater the ability value is, the wider the propagation range in the epidemic model is, the better capability and correctness of the INRM are. However, as shown in Table \ref{tb_NetworkStatistics} , since the scale of real-world networks usually ranges from a single node to massive nodes, if all nodes in the large scale network are selected to be simulated in the epidemic model, the consumption would be unacceptable. Secondly, there are very few nodes affecting the communication of information over network, for example, internet celebrities in social networks have great influence on the direction of public opinion, a few genes in biological networks control the expression of the entire genome, and a few servers in DNS technology network affect the connectivity of the Internet. So selecting \textit{top-k} nodes for calculation has been proved feasible \cite{LI201447,ZhaoJieandSongYutong202003,spreadingdynamics201505,LIU20134154,WangJunyiandHou201702} considering that it can reduce the use of resources and improve efficiency. Thus, the \textit{top-k} nodes in the ranking list of each INRM can be simulated in the epidemic model to calculate their benchmark propagation ability. However, there are a number of ways to choose the \textit{top-k} nodes in networks. Different selection strategies may lead to different computation results. Then we summarized five strategies (I, II, III, IV, V) to select \textit{top-k} nodes in networks and calculate their benchmark propagation ability under epidemic model for the comparison of INRMs.}

For a more intuitive presentation, an example of two ranking lists $R_1$ and $R_2$ from two INRMs based on a same network is listed. There are 10 nodes ($A,B,C,D,E,F,G,H,I,J$) in the network. Descriptions of $R_1$ and $R_2$ are as follows.
\begin{align*}
R_1=\{B,C,D,F,J,E,A,G,H,I\}, 
\\
R_2=\{B,A,D,J,F,G,C,H,I,E\}.
\end{align*}

\begin{itemize}
    \item[\uppercase\expandafter{\romannumeral1}] 
   \textbf{Select all the \textit{top-k} nodes as initial nodes.} All the \textit{top-k} nodes are regard as as initial nodes together in the epidemic model. Then two propagation ability values, namely, the number of infected nodes and those of infected and recovered nodes in the epidemic model for two INRMs are calculated. Thus the INRM corresponding to the larger propagation ability value is better. In the above example, assuming that \textit{k=5}, then $R_1^{'}$ = \{$B,C,D,F,J$\} and $R_2^{'}$ = \{$B,A,D,J,F$\} based on the strategy I. Based on the five nodes in $R_1^{'}$ and $R_2^{'}$, the propagation ability value $p_1$ and $p_2$ are got. The comparison between them determines which method is better.
  
    \item[\uppercase\expandafter{\romannumeral2}]   
    \textbf{Select \textit{top-k} nodes as initial nodes separately.} Nodes existing in \textit{top-k} nodes of two INRMs would form $k$ node pairs including two nodes. In the above example, assuming that \textit{k=5}, the node pairs would be $(B,B),(C,A),(D,D),(F,J),(J,F)$. In each pair, the two nodes are input into epidemic model separately and two propagation ability values $p_1$ and $p_1^{'}$ are calculated. The INRM corresponding to the larger propagation ability value wins 1 score. After $k$ round of iteration, the INRM corresponding to greater score is better. 
    
   \item[\uppercase\expandafter{\romannumeral3}] 
\textbf{Select \textit{top-k} nodes as initial nodes but removing the overlapping nodes and filling the empty parts.}
     In most cases, there are the same nodes in the \textit{top-k} nodes. Some studies consider that it is not meaningful to compare the same nodes \cite{ZhangJunkai201910,8322554}. Therefore, the strategy is that the same nodes existing in the \textit{top-k} nodes will not be considered, but the empty parts of the \textit{top-k} nodes are filled by other nodes that follow sequentially in the ranking lists. 
   In the above example, assuming that $k=5$,  then $R_1^{'}$ = \{$C,E,A,G,H$\} and $R_2^{'}$ = \{$A,G,C,H,I$\}. Finally, the ways to calculate propagation ability values for $R_1^{'}$ and $R_2^{'}$ can be derived from strategy I or strategy II.
   
    \item[\uppercase\expandafter{\romannumeral4}] 
\textbf{Select \textit{top-k} nodes as initial nodes but removing the overlapping nodes and without filling the empty parts.}
    The difference from strategy II is that the overlapped nodes will be deleted directly, and the space left will not be replenished \cite{8322554,leaders201112,WEI20132564,FEI2017257,Chen_2013}. In the above example, assuming that \textit{k=5},  then $R_1^{'}$ = \{$C$\} and $R_2^{'}$ = \{$A$\}. Similarly, the ways to calculate propagation ability values for $R_1^{'}$ and $R_2^{'}$ are according to strategy I or strategy II.
   
    \item[\uppercase\expandafter{\romannumeral5}] 
   \textbf{Select \textit{top-k} nodes as initial nodes but removing the overlapping nodes on the same rank.}
    Some studies \cite{Mekonnen202003,PuJunandChenXiaowu201407} have found that the comparison of different nodes at the same level in those ranking lists obtained by different methods is meaningful. Therefore, in the selection of the top-k nodes, only different nodes at the same rank are considered. In the above example, assuming that \textit{k=5},  then $R_1^{'}$ = \{$C,F,J$\} and $R_2^{'}$ = \{$A,J,F$\}. The ways to calculate propagation ability values for $R_1^{'}$ and $R_2^{'}$ are also according to strategy I or strategy II.
\end{itemize}

\subsection{Summary}
Ranking capability and ranking correctness are two important metrics to measure the INRM method. The ranking capability tests whether the method can discriminate the importance of different nodes, and whether its results are close to the results of industry-certified methods. While the ranking correctness on the one hand crafts whether the ranking list produced by the method is consistent with the benchmark importance ranking list of the nodes. On the other hand, robustness crafts the changes of node features which can be the centrality value or ranks of importance under epidemic model, before and after a node is removed from the network. Then, the consistency of the ranking list with the one generated by the robustness verifies the correctness. In addition, in comparison of the efficiency of multiple INRMs, how to select \textit{top-k} nodes is particularly important. We summarized and analyzed the five selection strategies most used in INRMs.

\section{Results and findings}
\textcolor{black}{In this section, we analyzed the networks applied and the evaluation metrics used by each method, and a series of findings are given from whole to part. The results of the capability and correctness of 81 INRMs for various networks are listed in Table \ref{tb_ResultsSS}}. 

{ \tiny
\begin{longtable}{l|l|l|l|llll|llllll|l}
% \centering
% \tiny
\caption{The Capability and Correctness of INRMs on Different Types of Networks}
\label{tb_ResultsSS} \\
% \resizebox{\linewidth}{!}{
% \begin{tabular}{l|l|l|l|llll|llllll|l}
\hline
  &          &      &         & \multicolumn{4}{c|}{\textbf{Capability}}  & \multicolumn{6}{c|}{\textbf{Correctness}}                                                                                                          &                                                                                                                                                           \\ \hline
 \textbf{ID} & \textbf{Method}     & \textbf{Ref}                                                                                                                                                                                     & \textbf{Networks} & \multicolumn{1}{l|}{\textbf{DM}} & \multicolumn{1}{l|}{\textbf{MF}} & \multicolumn{1}{l|}{\textbf{CCDF}} & \textbf{RS} & \multicolumn{1}{l|}{\textbf{PC}} & \multicolumn{1}{l|}{\textbf{SC}} & \multicolumn{1}{l|}{\textbf{JSC}} & \multicolumn{1}{l|}{\textbf{IF}} & \multicolumn{1}{l|}{\textbf{KCC}} & \textbf{R} &  \textbf{Strategies} \\ \hline
\endhead

1 & \textcolor{black}{Local} & \cite{FreemanLinton197901}                          &        & \multicolumn{1}{l|}{}                & \multicolumn{1}{l|}{}                      & \multicolumn{1}{l|}{}                                               &                                    & \multicolumn{1}{l|}{}                    & \multicolumn{1}{l|}{}                     & \multicolumn{1}{l|}{}                               & \multicolumn{1}{l|}{}                     & \multicolumn{1}{l|}{}                               &                    &     \uppercase\expandafter{}                        \\ \hline
2 & Local      & \cite{Chen_2013}                                                                                                                                              & SN       & \multicolumn{1}{l|}{}                & \multicolumn{1}{l|}{}                      & \multicolumn{1}{l|}{}                                               &                                    & \multicolumn{1}{l|}{}                    & \multicolumn{1}{l|}{}                     & \multicolumn{1}{l|}{}                               & \multicolumn{1}{l|}{}                     & \multicolumn{1}{l|}{$\bullet$}                               &                    & \uppercase\expandafter{\romannumeral2}, \uppercase\expandafter{\romannumeral4}                \\
\end{longtable}
}

\subsection{Results}
\begin{figure*}[htbp!]
\centering
\includegraphics[width=1\textwidth]{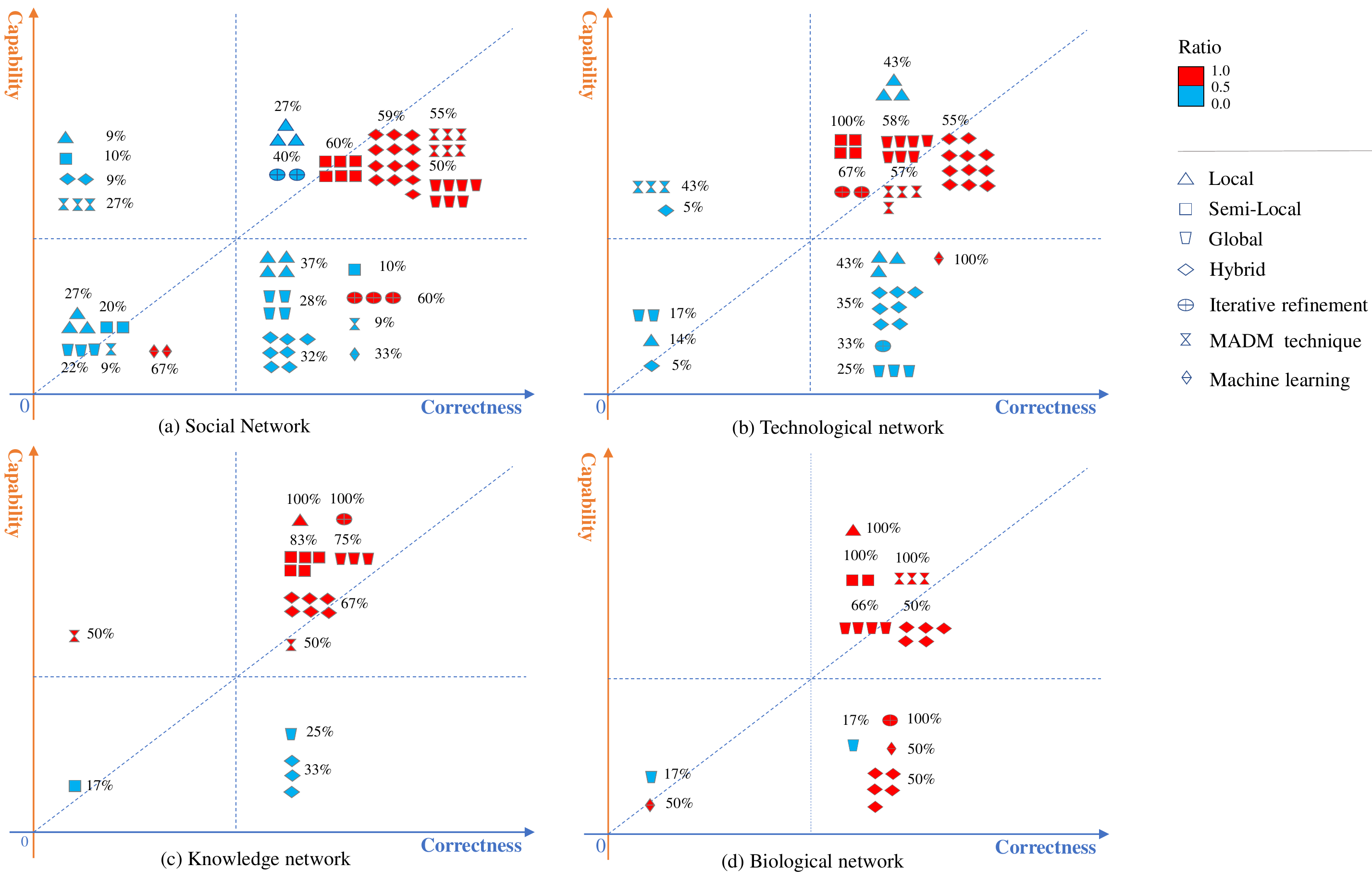}
\caption{\textsc{Results among Networks, Methods and Evaluation Metrics}}
\label{fig_Results}
\end{figure*}

In Table \ref{tb_ResultsSS}, because such classical methods as Degree centrality \cite{FreemanLinton197901}, betweenness centrality \cite{Brandes2001AFA}, closeness centrality \cite{Sabidussi196602} and eigenvector centrality \cite{Estrada200506} did not involve any real-world networks and evaluation metrics, the following analysis are focused on the remaining 77 INRMs. Almost all the seven kinds of INRMs are applicable to social networks, technological networks and biological networks. In social networks, there are 11 local methods, 10 simi-local methods, 14 global methods, 22 hybrid methods, 5 iterative refinement methods, 11 MADM methods, and 3 machine learning methods. In technological networks, there are 7 local methods, 4 semi-local methods, 12 global methods, 20 hybrid methods, 3 iterative refinement methods, 7 MADM methods and 1 machine learning methods. In biological networks, there are 1 local method, 2 semi-local methods, 6 global methods, 10 hybrid methods, 1 iterative refinement method, 3 MADM methods, and 2 machine learning methods. In knowledge networks, there are 1 local method, 6 semi-local methods, 4 global methods, 9 hybrid methods, 1 iterative refinement method and 2 MADM methods, but the machine learning method is not included. Referring to the data in Table \ref{tb_ResultsSS}, we further coarsely map the seven kinds of INRMs to the "Correctness-Capability" two-dimensional coordinate system by different shapes (Fig. \ref{fig_Results}), so as to better discover the common findings. 

\subsection{Overall findings}
\subsubsection{\textbf{The methods of Semi-local, Global, Hybrid, and MADM technique are well applied in the four types of networks (more than 50\%).}} In these networks, the topology of the network is an important quality to assess network function, and the function and characteristics of the network also depend on the network topology to a large extent \cite{Chen_2013,7978752,WenTaoandJiang201902,DaiJinYingandWang201909,Ibnoulouafi201809,8351889,BAO2017391,Garas_2012}. Therefore, the method based on structural centrality can better identify important nodes in complex networks \cite{FreemanLinton197901,HuPingandMei201709,KitsakMaksim201001,LIU2016289}. Local, semi-local, global and hybrid methods are meta-categories of structural centrality methods. Among them, the local method only applies the topological characteristics of nodes themselves in the network, so its limited use is to accurately identify the importance of nodes in large networks \cite{FreemanLinton197901,Chen_2013,7978752,Mekonnen202003,PuJunandChenXiaowu201407,WenTaoandJiang201902,DaiJinYingandWang201909,Ibnoulouafi201809,hindex201601,CHEN20121777,khop201502,MaQianandMaJun201608}. The semi-local method and the global method further consider the topological structure of the area near the node or even the entire network, so they can well identify important nodes in the network   \cite{HuPingandMei201709,HuiyuandZun201301,YuHuiandCao201705,theroleofclustering201310,ZAREIE2017485,BAO2017391,7925142,ShaoZengzhenandLiu201911,8351889,RePEc:eee:phsmap:v:403:y:2014:i:c:p:130-147}. Based on the local method, the semi-local method and the global method synthesized to conform to the topological characteristics of a certain type of network and identify important nodes, the hybrid method would accumulate their index values as the importance of nodes \cite{LIU2016289,e22080848,MaQianandMaJun201608,LIN20143279,spreadingdynamics201505,WEI2015277,LIU20134154,6477051,LiuDongandNie201810,YangYuanzhi202005,10.1155/2020/5903798,ZareieAhmad201811,8529839,QiuLiqing202107,ZhaoJieandWang202004,ShengJinfang201910,ZhangJunkai201910,ZhongLinFengandLiu201807,ZhongLinFengandShang201806,FuYuHsiangandHuang201504,WangJunyiandHou201702,QingChengHuanfYanShen201301,MA2016205,FeiLiguoandZhang201808}, while MADM technique would integrate the results by a decision maker hybrid method \cite{BIAN2017422,DU201457,HU201673,FEI2017257,YangYuanzhi201911,YangPingleandLiu201806,7979310,WEI20132564,ZhaoJieandSongYutong202003,HouBonanandYao201208,FeiLiguoandmo201708}. Both iterative refinement and machine learning methods are based on their own general rules, and do not fully consider the topological properties of the whole network, so their effect in identifying important nodes may only be good for a certain type of network   \cite{ZhaoJieandSong202008,leaders201112,LI201447,Ren_2014,ZHONG20152272,Estrada200506,ZHAO202018,YuEnYuandWang202004,ZhaoGouhengandJia202003}. Therefore, the semi-local, global, hybrid centrality methods and the MADM method can fully utilize the topological characteristics of the four types of networks, and they have wider applications than other methods.

\subsubsection{\textbf{The application of local centrality method and iterative refinement method in social networks and technological networks is higher than that in knowledge networks and biological networks.}} The local method only discusses the properties of the node itself, or it computes the sum of influence of neighbor nodes \cite{FreemanLinton197901,Chen_2013,7978752,Mekonnen202003,PuJunandChenXiaowu201407,WenTaoandJiang201902,DaiJinYingandWang201909,Ibnoulouafi201809}. In this case, the node is more important if the node has a higher property value or has more neighbor nodes with higher influence values \cite{hindex201601,CHEN20121777,khop201502,MaQianandMaJun201608}. The iterative refinement method gives each node an initial importance "score", and then goes through multiple rounds of iterations. In each round of iteration, each node will pass its own "score" to neighbor nodes and obtain "scores" from neighbor nodes at the same time \cite{ZhaoJieandSong202008,leaders201112}. At this point, if the node has more neighbors with higher scores, it will get more scores, so its importance will be higher \cite{LI201447,Ren_2014,ZHONG20152272,Estrada200506}. It can be concluded that the two methods are very similar in evaluating the importance of nodes determined by their own property values or neighbor nodes with higher importance. In social networks, the mode of spread is broadcast \cite{spreadingdynamics201505,ZareieAhmad201811,LiuDongandNie201810,LiuYingandTang201705,ZhaoZijuan201912,QiuLiqing202107,ZhangJunkai201910}. In this case, each node will transmit the information it owns to all neighbor nodes. If the node itself has more information and can get more information from neighbor nodes, its importance would be higher. In technological networks, this kind of trajectories is geodesics \cite{YangYuanzhi202005,LvZhiweiandZhao201902,DU201457,khop201502,Ibnoulouafi201809,WangShashaandDu201611,QiuLiqing202107,ZhaoJieandWang202004,FeiLiguoandZhang201808,ZhaoJieandSong202008,BIAN2017422,FEI2017257}. However, due to the cost control when building the technology network, the core in the technology network is usually on the geodesics with more nodes \cite{LiuDongandNie201810,ZhaoZhiyingandWang201504,QiuLiqing202107,ShengJinfang201910,FeiLiguoandZhang201808}. At this time, these core nodes have more neighbor nodes and can obtain more information on multiple paths. Through the above analyses, the spread mode or the kind of trajectories of social network and technological network fits well with the idea of local and iterative refinement methods, for which they are well applied to social networks and technological networks by practitioners and researchers. 

\subsubsection{\textbf{For machine learning methods, there is only correctness verification, but no capability verification in any network.}} INRMs based on machine learning usually transform the problem into a classification task, so by adjusting hyper-parameters, the granularity of ranking can be directly controlled \cite{ZHAO202018,YuEnYuandWang202004}. The ranking granularity, namely, the ability to differentiate equally important nodes, is one of capabilities of INRMs. Moreover, even if the ranking method based on machine learning converts the problem into a regression problem, the ranking granularity can also be optimized by increasing the complexity of the model or the number of training rounds \cite{YuEnYuandWang202004,ZhaoGouhengandJia202003}. Second, it can be applied to select arbitrary networks, generate the feature input for all nodes with no labels and get the output of them by the trained models. Then the ranking list  generated by output values can also be improved by the trained models that can be continuously optimized to get a higher ranking credibility. The ranking credibility is another capability of INRMs. Therefore, the methods based on machine learning can control the ranking granularity and the ranking list to achieve the optimal performance through some common operations in the field, so it is of little significance to evaluate the capability of such methods.

\subsubsection{\textbf{Further delicate findings}}
\textcolor{black}{The above analysis results are analyzed as a whole from Fig. \ref{fig_Results}. Based on these results, we further focus on the selection strategies of the methods and the fine-grained evaluation metrics. The delicate findings are analyzed as follows.}

\begin{figure}[htbp!]
\centering
\includegraphics[width=0.6\textwidth]{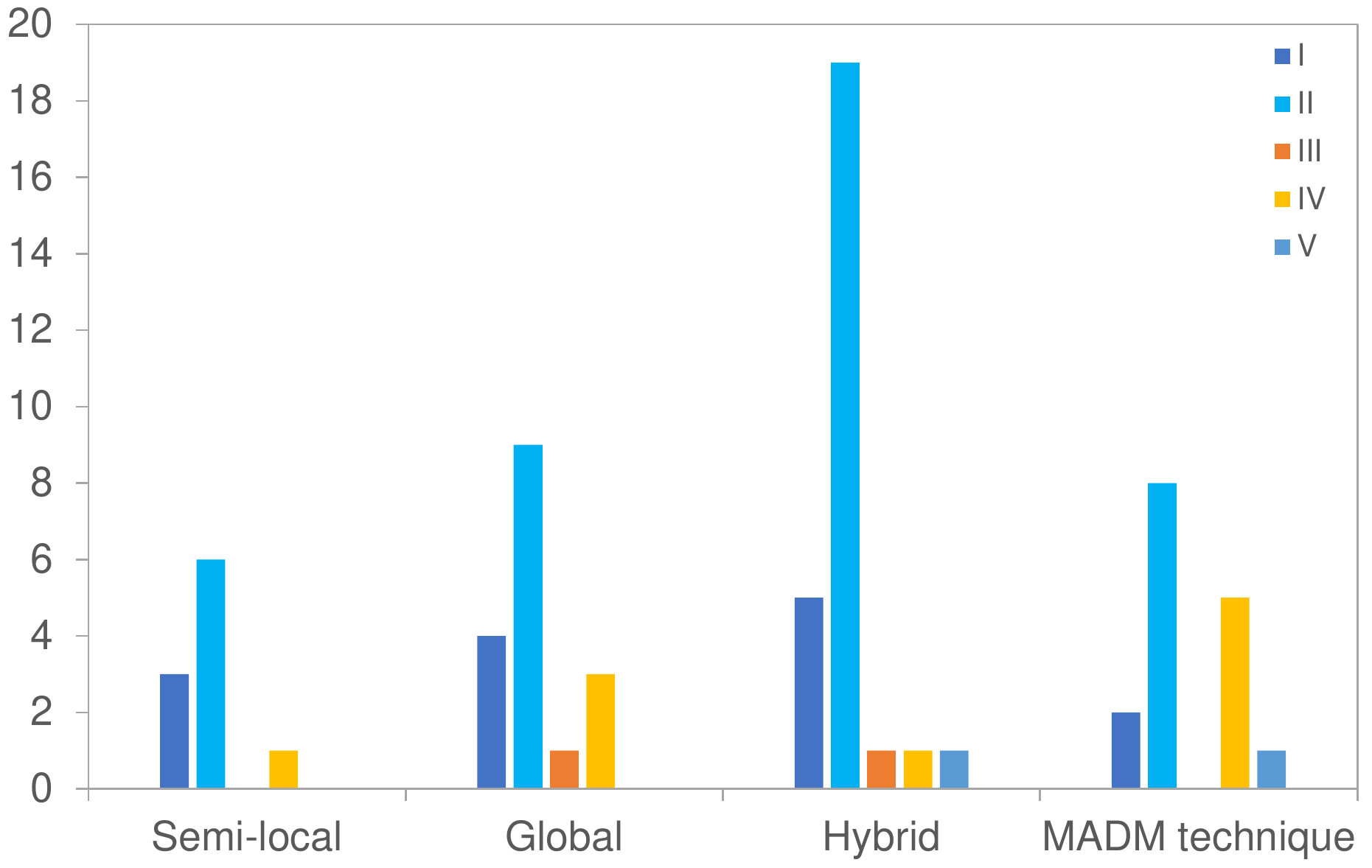}
\caption{\textsc{Selection Strategies Distribution of Semi-local, Global, Hybrid and MADM Methods}}
\label{fig_results2}
\end{figure}

(1) \textbf{As shown in Fig. \ref{fig_results2}, the Strategy II is significantly used in the four types of methods (Semi-local, global, hybrid, MADM methods) widely applied in the four types of networks}. A few nodes in the network are very influential to information transmission in the network, so in the verification of the ranking results, verifying the recognition accuracy of its \textit{top-k} nodes can help us assess the advantages and disadvantages of the method on the precondition of saving costs as much as possible. Strategy II applies an epidemic model to compare the importance of individual nodes in the \textit{top-k} node set rather than that of the entire \textit{top-k} node set (Strategy I) \cite{Chen_2013,7978752,WenTaoandJiang201902,DaiJinYingandWang201909,hindex201601,khop201502,MaQianandMaJun201608,7925142,HuPingandMei201709,RePEc:eee:phsmap:v:403:y:2014:i:c:p:130-147,ShaoZengzhenandLiu201911,ZAREIE2017485,BAO2017391,8322554}. Influenced by small-world characteristics of complex networks, the distance between nodes in the \textit{top-k} node set is often too close to each other \cite{ZhaoZijuan201912,8259501,KitsakMaksim201001,LiuYingandTang201705,Garas_2012,ZENG20131031,BAE2014549,ZhangRuishengandYang201608,6477051,8529839,YangYuanzhi202005,10.1155/2020/5903798,LiuDongandNie201810,FeiLiguoandZhang201808}. At this time, if the \textit{top-k} nodes are set as initial nodes at the same time in the SIR model, there will be overlaps between the influence coverage of the nodes \cite{DU201457,khop201502}. Thus the most important node will often make the comparison of other nodes meaningless. Strategy III,IV and V remove the same nodes in the \textit{top-k} node, which also leads to some information missing. Therefore, each time an individual node in the \textit{top-k} node set is defined as the initial node, the advantages and disadvantages of the method can be more rigorously and fully compared \cite{QingChengHuanfYanShen201301,WangJunyiandHou201702,FuYuHsiangandHuang201504,LIU2016289,spreadingdynamics201505}. 

\begin{figure}[htbp!]
\centering
\includegraphics[width=0.6\textwidth]{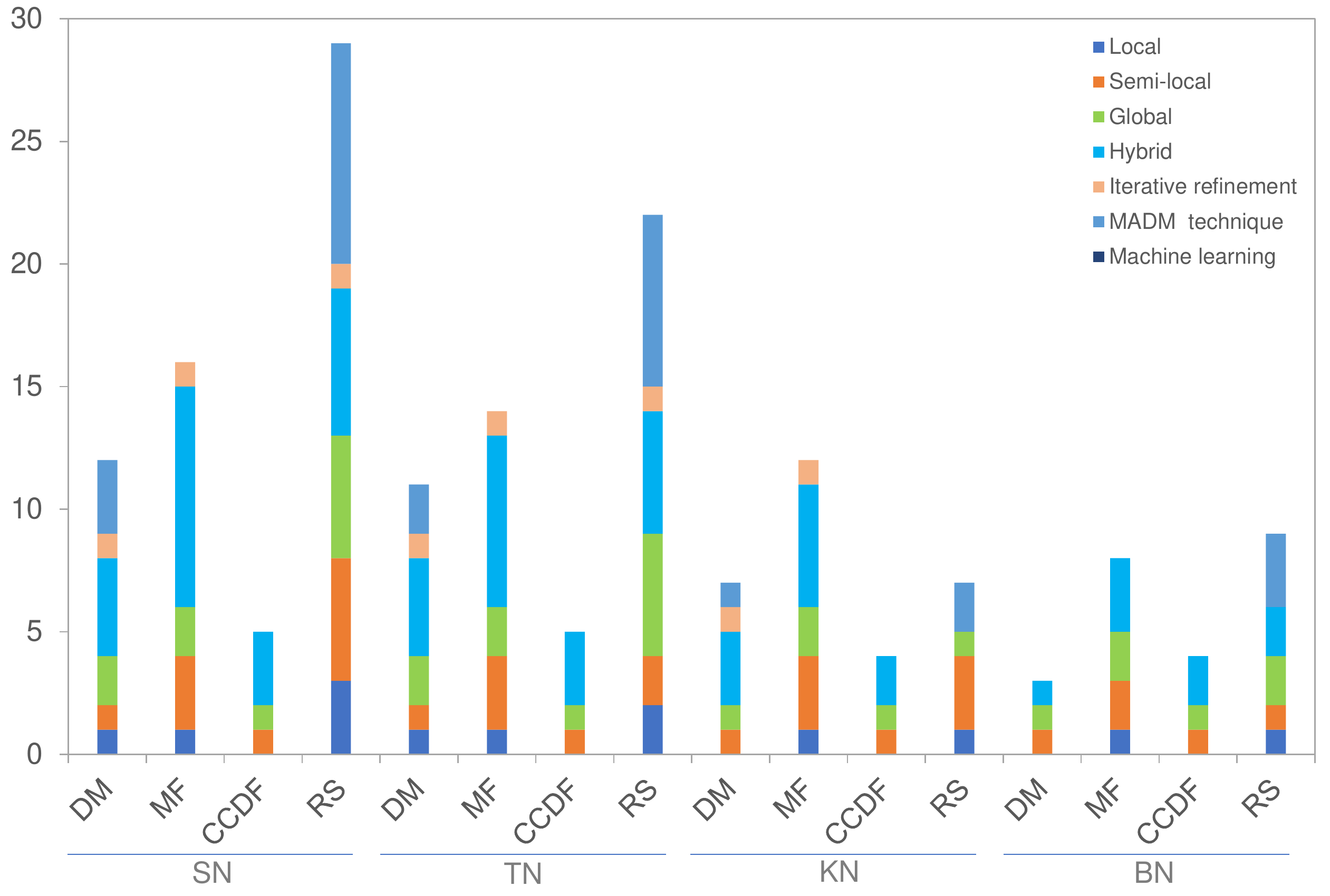}
\caption{\textsc{Distribution of Capability Evaluation Metrics}}
\label{fig_results3}
\end{figure}

\begin{figure}[htbp!]
\includegraphics[width=0.6\textwidth]{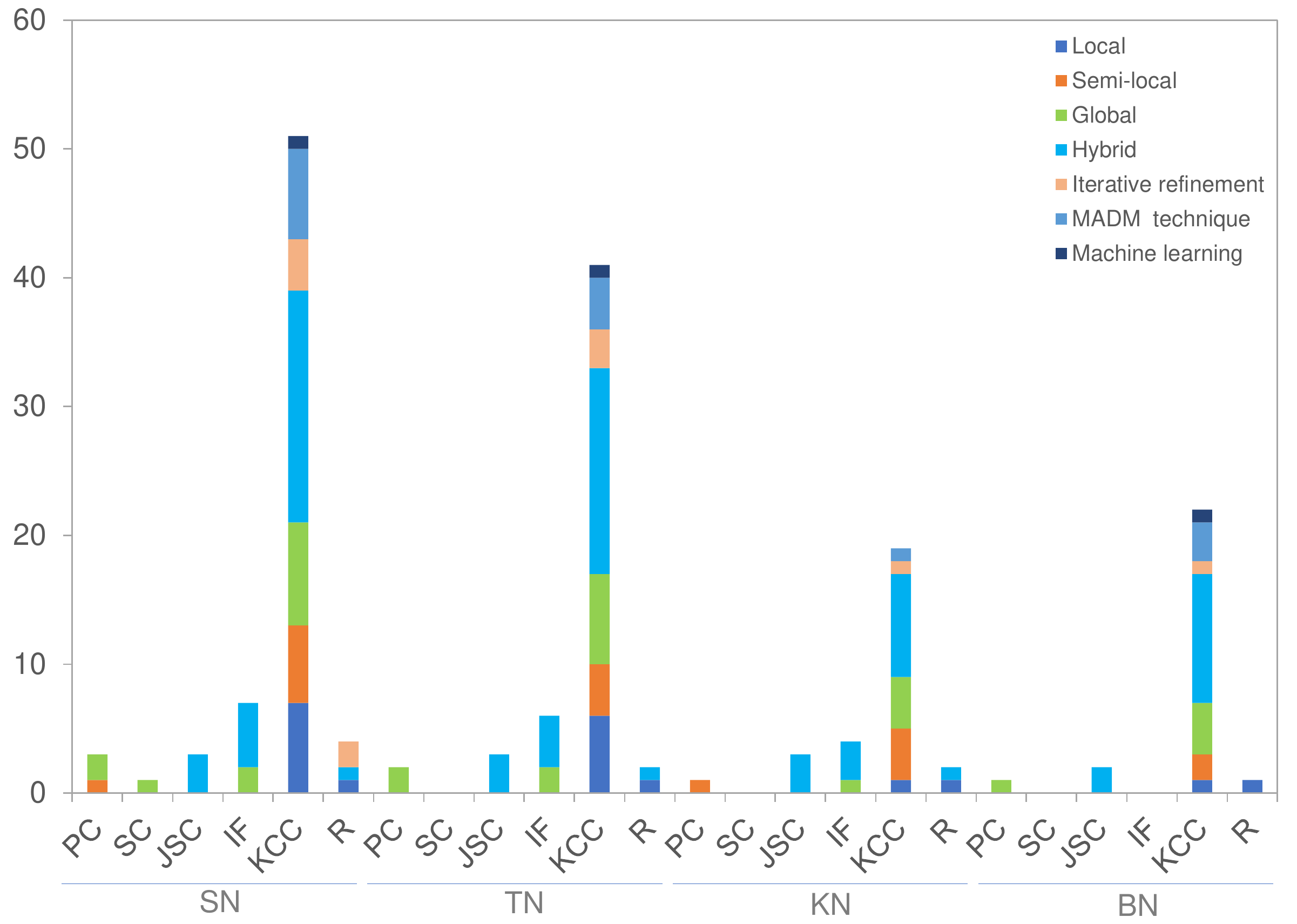}
\caption{\textsc{Distribution of Correctness Evaluation Metrics}}
\label{fig_results4}
\end{figure}

(2) \textbf{The analysis of all metrics of capability and correctness are conducted. In the instance of the metrics used
in social networks, as shown in Fig. \ref{fig_results3} and Fig. \ref{fig_results4}, RS and KCC metrics are the most frequently used in all methods.} \textcolor{black}{In the verification of the ranking results, it is particularly necessary to compare them with the latest methods or classic methods \cite{DaiJinYingandWang201909,HuPingandMei201709,RePEc:eee:phsmap:v:403:y:2014:i:c:p:130-147,ZAREIE2017485,8259501,BAE2014549}. First, the comparison of these methods have been verified by the industry, and the similarity with the results of these methods can roughly prove the  ranking credibility of the capability of the proposed methods \cite{YangYuanzhi202005,FeiLiguoandZhang201808,QingChengHuanfYanShen201301,FuYuHsiangandHuang201504,ZhangJunkai201910,ZhaoJieandSong202008,DU201457,7979310}. Compared with DM, CCDF and MF indicators that are used to differentiate the same importance of nodes, RS is applied more frequently in the capability evaluation metric.}
Similar metrics with KCC are PC and SC in correctness metrics. However, the use of PC and SC requires the data to meet certain conditions. PC requires data to conform to normal distribution, and the sample size should exceed a certain number. SC is suitable for monotonic relationship. While KCC does not have these restrictions, so it can be applied to various methods. In addition, KCC can measure the strength of the monotonous relationship between two ordered variables \cite{Chen_2013,WenTaoandJiang201902,DaiJinYingandWang201909,hindex201601,MaQianandMaJun201608,RePEc:eee:phsmap:v:403:y:2014:i:c:p:130-147,ShaoZengzhenandLiu201911,ZAREIE2017485,BAO2017391,ZhaoZijuan201912,LiuYingandTang201705,ZENG20131031,BAE2014549,ZhangRuishengandYang201608,8529839,YangYuanzhi202005,10.1155/2020/5903798,FeiLiguoandZhang201808,WangJunyiandHou201702,FuYuHsiangandHuang201504,LIU2016289,LIU20134154,WEI2015277,ZhongLinFengandLiu201807,ZhangJunkai201910,ZhongLinFengandShang201806,e22080848,LIN20143279,ZareieAhmad201811,MA2016205,ZhaoJieandSong202008,Ren_2014,ZHONG20152272,YangYuanzhi201911,FeiLiguoandmo201708,BIAN2017422,HU201673,YuEnYuandWang202004}. The concept of "pairing" used by KCC just meets the need to measure two ranking lists, which result in higher usability of KCC in different networks.

(3) \textbf{Similar to the SN network, as shown in Fig. \ref{fig_results3} and Fig. \ref{fig_results4}, KCC also has the same application results in the other three types of networks. RS is used most frequently in technological networks and biological networks, but MF is the most frequently used in knowledge networks.} Real-world knowledge networks prefer to ranking the knowledge nodes with fine grain \cite{RePEc:eee:phsmap:v:403:y:2014:i:c:p:130-147,ShaoZengzhenandLiu201911,ZAREIE2017485,BAO2017391,LiuYingandTang201705,ZENG20131031,BAE2014549}. For example, the literature database needs to rank and display the search results, and the citation network needs to be accurately ranked according to the importance of its nodes \cite{BAO2017391,ShaoZengzhenandLiu201911,ZENG20131031}. At the same time, the ranking in such networks usually sets some preconditions, such as time priority \cite{spreadingdynamics201505,LIU20134154,e22080848,LIN20143279}, correlation priority  
 \cite{ZareieAhmad201811}. Therefore, compared with the metrics DM and CCDF, to distinguish the same importance of nodes, it is more meaningful to examine the granularity of the ranking results evaluated by the metric MF. Because MF can get smaller ranking granularity and wider ranking range based on its calculation method in equation (6). For example, it is assumed that there are minimum 2 and maximum 9 identical node ranks in 10 nodes' ranking list. The Upper and lower bounds calculated by MF is 0.96 to 0.04. But for DM and CCDF, their Upper and lower bounds are 0.9 to 0.2 and  0.8 to 0.1 respectively. This advantages make MF be more frequently used.

\subsection{Findings from INRMs}

After summarizing the seven meta-categories of INRMs analyzed in Section \ref{SV}, we found that there are many techniques and crafts being used in attempts to enhance the performance of the methods. In this section, we first presented them and explained how they were used. Then, the findings observed from Fig. \ref{fig_Results} are also analyzed.

\subsubsection{Techniques and crafts used in INRMs}
\begin{enumerate}
    \item[\textit{(1)}] \textit{Techniques and crafts used by Structural centrality method}
\end{enumerate}

\textbf{Structural hole.} A structural hole can be explained as a bridge connecting two or more unconnected nodes. Lazega et al. \cite{LazegaEmmanuelandBurtRonald199510} introduced the network constraint index about structural hole. A smaller network constraint index indicates a bigger structure hole and higher importance of the node. In the structural hole theory, the important nodes obtained by the method have a larger propagation range in the network, and show higher correctness than the comparison methods under the indicators of KCC and PC \cite{HuPingandMei201709,HuiyuandZun201301,YuHuiandCao201705}.

\textbf{Neighbor contribution and Information entropy.} \textcolor{black}{A node must propagate information through neighbor nodes, so the propagation ability of neighbor nodes largely determines the importance of the node \cite{theroleofclustering201310,ZAREIE2017485,BAO2017391,7925142,ShaoZengzhenandLiu201911,8351889,RePEc:eee:phsmap:v:403:y:2014:i:c:p:130-147}. In a network, the neighbor contribution means that the more influential neighbor nodes a node has, the more influential it is \cite{10.1155/2020/5903798,ZhaoZhiyingandWang201504}.} However, some studies \cite{ZAREIE2017485,GuoChunguandYangLiangweiandChen202002,FEI2017257,e22080848,10.1155/2020/5903798} believe that a node's importance can be affected by the distribution of its neighbor nodes in the network. The more complex the distribution of neighbor nodes is, the wider the range of information dissemination is. Information entropy measures the uncertainty or the probability of discrete random events. In more chaos, the information entropy is higher, and vice versa. In the information world, higher entropy means that more information can be transmitted, and vice versa. The INRMs applying the neighbor theory have better effect than the comparison methods under the performance of the Pearson coefficient, and nodes in the network can be ranked more steadily and homogeneously \cite{theroleofclustering201310,ZAREIE2017485,BAO2017391,7925142,ShaoZengzhenandLiu201911,8351889,RePEc:eee:phsmap:v:403:y:2014:i:c:p:130-147}. After introducing information entropy in INRMs, it can not only improve the ranking Capability (under the DM, CCDF index, MF index and RS index), but also enhance the ranking Correctness (under KCC index) \cite{Chen_2013,Mekonnen202003,HuPingandMei201709,ZAREIE2017485,WEN2020549,e22080848,ZareieAhmad201811,8259501,FuYuHsiangandHuang201504,FEI2017257,WEI20132564}.

\textbf{Local Cluster coefficient.} The scale-free characteristics of complex networks cause the clustering phenomenon of node distribution in the network. The local Cluster coefficient of a node quantifies the extent to which its neighbors aggregate with each other to form clusters, causing a positive or negative impact on INRMs. For example, Qiu et al. \cite{QiuLiqing202107} considered that it is beneficial to information dissemination because closely connected neighbors have higher chance influencing other nodes, while the studies \cite{theroleofclustering201310,RePEc:eee:phsmap:v:403:y:2014:i:c:p:130-147,ZHAO202018} believed that it is not conducive to information transmission because it restricts the information dissemination in the network to a local area. By considering the local cluster coefficient, the INRMs can better differentiate the importance of nodes, thereby improving the capability (under the MF and DM index ) and correctness (KCC) of the method \cite{QiuLiqing202107,theroleofclustering201310,RePEc:eee:phsmap:v:403:y:2014:i:c:p:130-147,ZHAO202018}.

\begin{enumerate}
    \item[\textit{(2)}] \textit{Data Normalization used in INRMs}
\end{enumerate}

Data normalization is the adjustment of data by offset and scaling so that it falls into a small specific interval \cite{ZhaoZhiyingandWang201504,ZHAO202018,Chen_2013}. Normalization operations are often used in the processing of comparative and evaluation metrics where unit constraints are removed from the data and converted to dimensionless pure values, which allows comparisons and weights of different units or magnitudes. Data normalization can be roughly classified into three categories \cite{Centering_scaling_and_transformations}. 
\begin{itemize}
    \item Centering: subtracting all data from the mean so that the data are distributed around value 0 rather than around the mean. It focuses on the difference of the data.
    \item Scaling: multiplying or dividing the data by a factor uniformly so as to eliminate magnitude differences.
    \item Transformation: performing Log or Power transformations to eliminate heterogeneity.
\end{itemize}
Among node ranking methods, when a method considers multiple features, data normalization is used to scale the range of values of these features, so that the expression of features will not be affected by the order of magnitude difference between features  \cite{ZhaoZhiyingandWang201504,ZHAO202018,Chen_2013}.

\begin{enumerate}
    \item[\textit{(3)}] \textit{Intra-Parameter Setting in INRMs}
\end{enumerate}

Many of INRMs are not parameter free. The intra-parameters preset in INRMs attempt to improve the ranking effect. Those custom parameters are usually used for following purposes.

\begin{itemize}
    \item Some INRMs proposed multiple self-defined attributes or proposed fusion methods based on other centrality metrics. Since different attributes or methods have different magnitudes of influence on nodes' propagation ability, the role of intra-parameters in these methods is to adjust weights among attributes or methods \cite{DU201457,theroleofclustering201310,8351889,MaQianandMaJun201608,RePEc:eee:phsmap:v:403:y:2014:i:c:p:130-147,Garas_2012,10.1155/2020/5903798,ZareieAhmad201811,QiuLiqing202107,ZhongLinFengandLiu201807,ZhongLinFengandShang201806,QingChengHuanfYanShen201301},\cite{ZHONG20152272,LiuDongandNie201810}, or to scale the effect of individual attribute or methods \cite{ZENG20131031,YangYuanzhi202005,LiuYingandTang201705,ShengJinfang201910,Ren_2014}.
    \item Some INRMs replace network topology data difficult to quantify with parameters in the assessment of node importance. For example, Ibnoulouafi et al. \cite{Ibnoulouafi201809} discussed the farthest distance for information to be transmitted in the network by referring to the number of nodes in the network, so as to determine the maximum sphere of a node's influence. Moreover, the event of passing message between nodes is probabilistic, and this probability is usually called the transmission rate. Transmission rate is often difficult to quantify. 
   Therefore, the authors in \cite{khop201502,BAO2017391} described the average degree of nodes as the transmission rate.
    \item When considering the influence contribution of multi-hop neighbors, it is believed that the larger the distance between nodes is, the smaller the contribution is. Therefore, references \cite{ZAREIE2017485,LIU2016289} set the intra-parameters as attenuation coefficients to simulate this feature.
\end{itemize}

\subsubsection{Findings observed in INRMs}

Structural centrality methods has the largest proportion of all methods. 
% Structural centrality is usually a direct description of the network structure and characteristics, and the methods usually make assumptions, and then quantify the information propagation characteristics in the network \cite{Chen_2013,7978752,Mekonnen202003,PuJunandChenXiaowu201407,HuiyuandZun201301,theroleofclustering201310,7925142,8322554,ZhaoZijuan201912,8259501,CaiBiaoandTuoXianGuo2014,6477051,8529839,YangYuanzhi202005,10.1155/2020/5903798,LiuDongandNie201810,FeiLiguoandZhang201808,ZhaoJieandWang202004}. The ways such as information entropy  \cite{ZAREIE2017485,GuoChunguandYangLiangweiandChen202002,FEI2017257,e22080848,10.1155/2020/5903798}, gravity formula \cite{BAO2017391}, community discovery \cite{8322554,8259501,ZhaoZijuan201912,ZhaoZhiyingandWang201504,CaiBiaoandTuoXianGuo2014}, and structural holes \cite{HuPingandMei201709,HuiyuandZun201301,YuHuiandCao201705} are used in this process. Structural centrality methods are highly interpretable and make full use of the topology of the network and the flow preferences of information in the network. \textcolor{black}{They are more effective in identifying important nodes when the topological features of the network are obvious.} 
In this paper, structural centrality methods is divided into four subcategories, namely, local centrality, semi-local centrality, global centrality, and hybrid centrality. 
% In Fig. \ref{fig_Results}, structural centrality methods occupies an average of above 75\% of the number of methods in the four types of networks. Local centrality is an analysis of the characteristics of a node in the network itself or assumes that multi-step neighbors contribute directly to the propagation ability of the central node \cite{PuJunandChenXiaowu201407,WenTaoandJiang201902,DaiJinYingandWang201909,hindex201601,CHEN20121777,khop201502,MaQianandMaJun201608,Ibnoulouafi201809}. Unlike local centrality, semi-local centrality considers the influence between neighboring nodes \cite{7925142,HuPingandMei201709,YuHuiandCao201705,8351889,RePEc:eee:phsmap:v:403:y:2014:i:c:p:130-147,ShaoZengzhenandLiu201911,ZAREIE2017485,BAO2017391}. \textcolor{black}{Global methods evaluate the importance of nodes from the perspective of the entire network \cite{8259501,ZhaoZijuan201912,ZhaoZhiyingandWang201504,CaiBiaoandTuoXianGuo2014,LvZhiweiandZhao201902,WangShashaandDu201611,Sabidussi196602,FreemanLinton197901,WEN2020549}}. Hybrid centrality considers two or more of the three centrality methods simultaneously \cite{LiuDongandNie201810,FeiLiguoandZhang201808,ZhaoJieandWang202004,QingChengHuanfYanShen201301,WangJunyiandHou201702,FuYuHsiangandHuang201504,LIU2016289,spreadingdynamics201505,LIU20134154,QiuLiqing202107}. 
The local centrality method is simple and computationally efficient and has good applications in large networks and in scenarios where accuracy is not critical. Global centrality has high accuracy, but the obvious disadvantage is the same high computational complexity, which is often unacceptable in large networks. Local centrality combines the merits of the two centrality methods and ensures the computational efficiency of the algorithm at the expense of some accuracy. Hybrid centrality fully combines the advantages of the above three centrality methods, but its disadvantage also lies in the problem of computational complexity, but with the increasing arithmetic power and the demand for higher accuracy in recent years, hybrid centrality methods have gradually become the preferred choice  \cite{WEI2015277,ShengJinfang201910,ZhongLinFengandLiu201807,ZhangJunkai201910,ZhongLinFengandShang201806,e22080848,LIN20143279,ZareieAhmad201811,MA2016205}.
% The iterative refinement method first gives each node an initial score, and then distributes the scores of each node to its neighbors with some rules, and the nodes receive scores from other nodes while distributing scores, until the scores of nodes in the network reached stabilizing and the algorithm converges \cite{leaders201112,LI201447,ZhaoJieandSong202008}. 
The iterative refinement method \cite{leaders201112,LI201447,ZhaoJieandSong202008} is well used for node importance ranking in social networks because information in social networks is usually spread continuously by some central nodes, and those nodes being spread will also repeat this process continuously \cite{Ren_2014,ZHONG20152272}. Any two nodes can be connected by the broadcast transferring in social network. Secondly, iterative refinement methods are widely used in web page ranking systems and recommendation systems, and its algorithm has higher the complexity because of a large number of iterations in those systems, so iterative refinement-based algorithms are suitable for deployment in offline systems or for tasks that do not require high effectiveness. MADM method can combine the advantages of various types of methods and dynamically adjust the weight parameters according to the information propagation characteristics in different types of networks, so MADM method can well be applied to many types of networks and reach good performance. Machine learning methods have two solutions for the problem of influential node ranking, one is converting it into a regression problem \cite{YuEnYuandWang202004} and the other is converting it into a classification problem \cite{ZHAO202018,ZhaoGouhengandJia202003}. These two solutions can naturally deal with the characteristics of nodes and the connection between all nodes, and they well mine the coupling relationship between the importance of nodes and the topological features of the network. What's more, machine learning based approaches can automatically learn parameters to achieve good identification performance, which ensures its wide usage in social, technological and biological networks.

\subsection{Summary}
% We first summarize the quantitative relationships among networks, methods and evaluation metrics. Some phenomena are found and analyzed. Semi-local, global, hybrid centrality methods and MADM technique methods are widely applied in four kinds of networks. The local centrality and iterative refinement methods are more widely used in social networks and technical networks than in knowledge network and biological network. In machine learning methods, we illustrate why the capability verification is missing. Moreover, we mined the reasons behind that RS, KCC metris and strategy II are the most frequently used in different networks, and that the MF metrics appears more frequently in knowledge networks.  
% Second, we conduct summarized and methodical analysis from the perspectives of networks, methods and evaluation metrics, respectively, and some delicate findings are got, such as \textcolor{black}{Structural centrality methods usually have good performance and good interpretability for four kinds of networks, the iterative refinement methods are suitable for tasks that do not require high effectiveness such as offline computing for social networks, MADM methods and machine learning methods be well used to evaluate the importance of nodes from a global or multiperspective in biological networks. Moreover, the noteworthy techniques and crafts in INRMs are analyzed respectively. These techniques and crafts play a key role in method, which enhance the interpretability of INRMs and contributes to improving the performance of INRMs.} 

We first summarize the quantitative relationships among networks, methods and evaluation metrics. Some phenomena are found and analyzed. 
Semi-local, global, hybrid centrality methods and MADM technique methods are widely applied in four kinds of networks. The local centrality and iterative refinement methods are more widely used in social networks and technical networks than in knowledge networks and biological networks. In machine learning methods, we illustrate why the capability verification is missing. Moreover, we mined the reasons behind that RS, KCC metris and strategy II are the most frequently used in different networks, and that the MF metrics appears more frequently in knowledge networks.  
Second, we conduct summarized and methodical analysis from INRMs. The noteworthy techniques and crafts in INRMs are analyzed respectively. Such techniques and crafts matter in these methods, because they can enhance the interpretability of INRMs and contributes to improving the performance of INRMs.

\section{Comparison with Related work}
\textcolor{black}{Table \ref{tb_SurveyComparision} lists several surveys and literature reviews of  existing influential node ranking approaches. Then we compared our work with them definitely according to RQ1-RQ3. }

\begin{table*}[htbp]
\caption{\textsc{The Comparison Between The Related Surveys and Our Work}}
\label{tb_SurveyComparision}
\resizebox{\linewidth}{!}{
\begin{tabular}{l|llll|llll|ll}
\hline
\textbf{}    & \multicolumn{4}{c|}{\textbf{Networks}}                                                                                                                                                 & \multicolumn{4}{c|}{\textbf{Methods}}     & \multicolumn{2}{c}{\textbf{Evaluation metrics involved}}                \\ \hline
\textbf{Research articles} & \multicolumn{1}{l|}{\textbf{SN}} & \multicolumn{1}{l|}{\textbf{TN}} & \multicolumn{1}{l|}{\textbf{KN}} & \textbf{BN} & \multicolumn{1}{l|}{\textbf{Structural centrality}} & \multicolumn{1}{l|}{\textbf{Iterative refinement}} & \multicolumn{1}{l|}{\textbf{MADM technique}} & \textbf{Machine learning} & \multicolumn{1}{l|}{\textbf{Capability}} & \textbf{Correctness} \\ \hline
\multicolumn{11}{c}{\textbf{Studies focusing on single kind of networks}} \\ \hline

{Bian et al. \cite{10.1145/3301286}}    & \multicolumn{1}{l|}{$\surd$}                  & \multicolumn{1}{l|}{}                                & \multicolumn{1}{l|}{}                            &                              & \multicolumn{1}{l|}{$\surd$}                                & \multicolumn{1}{l|}{$\surd$}                               & \multicolumn{1}{l|}{}                                &                                   & \multicolumn{1}{l|}{}                    & $\surd$              \\ \hline
{Hafiene et al. \cite{HAFIENE2020113642}}    & \multicolumn{1}{l|}{$\surd$}                  & \multicolumn{1}{l|}{}                                & \multicolumn{1}{l|}{}                            &                              & \multicolumn{1}{l|}{$\surd$}                                & \multicolumn{1}{l|}{$\surd$}                               & \multicolumn{1}{l|}{}                                &                                   & \multicolumn{1}{l|}{}                    &                      \\ \hline
{Tulu et al. \cite{TULU2020102768}}    & \multicolumn{1}{l|}{$\surd$}                  & \multicolumn{1}{l|}{}                                & \multicolumn{1}{l|}{}                            &                              & \multicolumn{1}{l|}{$\surd$}                                & \multicolumn{1}{l|}{$\surd$}                               & \multicolumn{1}{l|}{}                                &                                   & \multicolumn{1}{l|}{}                    &                      \\ \hline

{Xiong et al. \cite{7854364}}    & \multicolumn{1}{l|}{}                         & \multicolumn{1}{l|}{$\surd$}                         & \multicolumn{1}{l|}{}                            &                              & \multicolumn{1}{l|}{$\surd$}                                & \multicolumn{1}{l|}{}                                      & \multicolumn{1}{l|}{$\surd$}                         &                                   & \multicolumn{1}{l|}{}                    & $\surd$              \\ \hline

\multicolumn{11}{c}{\textbf{Studies focusing on comprehensive kinds of networks}}  \\ \hline

{Maji et al. \cite{MAJI2020113681}}    & \multicolumn{1}{l|}{$\surd$}                  & \multicolumn{1}{l|}{$\surd$}                         & \multicolumn{1}{l|}{$\surd$}                     & $\surd$                      & \multicolumn{1}{l|}{$\surd$}                                & \multicolumn{1}{l|}{}                                      & \multicolumn{1}{l|}{}                                &                                   & \multicolumn{1}{l|}{$\surd$}             & $\surd$              \\ \hline
{Lü et al. \cite{LU20161}}    & \multicolumn{1}{l|}{$\surd$}                  & \multicolumn{1}{l|}{}                                & \multicolumn{1}{l|}{$\surd$}                            & $\surd$                             & \multicolumn{1}{l|}{$\surd$}                                & \multicolumn{1}{l|}{$\surd$}                               & \multicolumn{1}{l|}{$\surd$}                         &                                   & \multicolumn{1}{l|}{}                    & $\surd$              \\ \hline
{Liu et al. \cite{2021ScChE..64..451L}}    & \multicolumn{1}{l|}{$\surd$}                  & \multicolumn{1}{l|}{$\surd$}                         & \multicolumn{1}{l|}{$\surd$}                     &                              & \multicolumn{1}{l|}{$\surd$}                                & \multicolumn{1}{l|}{$\surd$}                               & \multicolumn{1}{l|}{}                                &                                   & \multicolumn{1}{l|}{}                    &                      \\ \hline
{Lalou et al. \cite{LALOU201892}}    & \multicolumn{1}{l|}{$\surd$}                  & \multicolumn{1}{l|}{}                                & \multicolumn{1}{l|}{}                            & $\surd$                      & \multicolumn{1}{l|}{$\surd$}                                & \multicolumn{1}{l|}{$\surd$}                               & \multicolumn{1}{l|}{}                                &                                   & \multicolumn{1}{l|}{}                    &                      \\ \hline
Our work     & \multicolumn{1}{l|}{$\surd$}                  & \multicolumn{1}{l|}{$\surd$}                         & \multicolumn{1}{l|}{$\surd$}                     & $\surd$                      & \multicolumn{1}{l|}{$\surd$}                                & \multicolumn{1}{l|}{$\surd$}                               & \multicolumn{1}{l|}{$\surd$}                         & $\surd$                           & \multicolumn{1}{l|}{$\surd$}             & $\surd$              \\ \hline
\end{tabular}
}
\end{table*}
% Please add the following required packages to your document preamble:
% \usepackage{graphicx}

\begin{itemize}
\item[$\bullet$]\textbf{Studies focus on surveying and classifying INRMs on single kind of networks.} Bian et al. \cite{10.1145/3301286} found that researchers and practitioners drew increasing attention to the use of \textit{top-k} nodes in the effective identification of key players in social networks. Thus they reviewed and classified the existing literature on the identification of  \textit{top-k} nodes, and summarized the application of \textit{top-k} nodes in Twitter, Facebook, Blogosphere, Misinformation Control, Community Question Answering, Networks with Complex Topologies, and Miscellaneous Applications networks.
Hafiene et al. \cite{HAFIENE2020113642} gave a general review of the state-of-the-art methods in dynamic social networks, including heuristic, greedy and hybrid approaches focused on static networks, snapshots networks or dynamic networks. Tulu et al. \cite{TULU2020102768} summarized influential node identification methods for enhancing information dissemination and the characteristics of the mobile social networks (MSNs), and discussed the influential node identification methods in MSNs from different characteristics, which are social relationship, mobility pattern, overlay networking, heterogeneity of mobile devices. Xiong et al. \cite{7854364} outlined the characteristics of world city networks and assessed the centrality and power of control correspondingly. Then, from the perspectives of node relevance (Degree), shortest path (betweenness, closeness centrality, eigenevector centrality, Bonacich centrality, Alter-based centrality), and others (node deletion, node contraction), node importance evaluation methods based on network topologies were analyzed. \textcolor{black}{All these studies preferred to surveying the approaches classifying into structural centrality method or Iterative refinement method by our classification specification, on one kind of network. What's more, they haven't considered the capability metrics.}

 \item[$\bullet$]\textbf{Studies focus on surveying and classifying INRMs on comprehensive kinds of networks.} Maji et al.\cite{MAJI2020113681} studied the main variations of the \textit{k-shell} method together with hybrid techniques focused on representative network topology, which belongs to Structural centrality method by our classification specification. Then, various performance metrics were discussed, and simulation models like the SIR epidemic model were performed with a comparison analysis of state-of-the-art methods for several standard real-world networks. Lü et al.\cite{LU20161} suggested a comprehensive approach to discover critical nodes on different networks on an individual basis. Then, they carried out a wide range of empirical analyses of representative methods, and examined the merits, demerits, and applicability of these methods to various networks and target functions. Liu et al. \cite{2021ScChE..64..451L} classified INRMs into three types. Node ranking methods based on centrality were devoted to measuring the importance of a node by evaluating that of its neighbors. The PageRank algorithm and the HITS algorithm were designed to solve the web page ranking problem. Then, the most recent extensions and improvement of the representative methods provided with some major applications were discussed. Lalou et al. \cite{LALOU201892} presented an algorithm for critical node identification in search of a set of optimal nodes. The absence of such nodes would cause a dramatic decrease in network connectivity. They focused on greedy methods to solve The Critical Node Detection Problem, and summarized the framework of greedy approach and discusses some variant methods. \textcolor{black}{The above studies pay more attention to surveying the application of important node identification method. Lü et al.\cite{LU20161} considered the applicability to different networks and objective functions, which is one of our targets. Besides, we further analyzed the convince of the results ranked by the methods, namely, the analysis of the capability and correctness metrics.}
\end{itemize}

\textcolor{black}{In summary, references \cite{LU20161,2021ScChE..64..451L,LALOU201892} statistically analyzed influential node identification methods for some kinds of networks, only reference \cite{MAJI2020113681} analyzed Structural centrality methods targeting for all kinds of networks. Our work focused on all these four types of networks classified in Section 3. For INRMs, none of these surveys or literature reviews covered the full range of methods. Our classifications are more comprehensive and finer-grained (7 categories) than those of other approaches, as illustrated in Section 4 and Section 5. As for evaluation metrics, references \cite{7854364,10.1145/3301286,LU20161} only analyzed the correctness metrics, and references \cite{HAFIENE2020113642,TULU2020102768} didn't address the evaluation metrics. Only reference \cite{MAJI2020113681} involved part of evaluation metrics in capability and correctness set we defined. In Section 5, our work conducted a systematic analysis of the capability and correctness of each INRM.}

\section{DISCUSSION}
\textcolor{black}{When service engineers conduct a series of operations in specific networks, e.g., identifying influential spreaders in social networks to introduce a new product to customers, or predicting essential proteins in the biomedical network, it is sometimes necessary to perform a generic search to identify the most significant nodes (customers or proteins). Engineers need to know what techniques are available and how well the techniques can function. Our findings from RQ1, RQ2 and RQ3 summed up to which networks INRMs were applicable and presented the effectiveness of a method in identifying influential nodes in a specific network. Knowing what networks INRMs focused can help engineers and researchers narrow the choosing range of INRMs. For example, in the results analysis, we reveal that semi-local, global, hybrid and MADM technique methods are widely applied in four kinds of networks. The local centrality and iterative refinement method are more widely used in social networks and technical networks. Knowing the capability and correctness effects of the INRM can help engineers and researchers prioritize and select appropriate tools, for instance, in our results, we found that RS and KCC metric in Capability and Correctness are used frequently and strategy II is the most frequently used in all methods. Then we can select the INRM according to higher values on these metrics. }

\textcolor{black}{In addition, our results from RQ1 identified real-world networks that were used to evaluate the primary studies. INRM-developing companies can use them as benchmarks for different tools. Results of RQ1 also indicate that high quality networks with complete information are insufficient and that the majority of INRMs are concentrated on undirected and unweighted networks. It is necessary for industry professionals and researchers to jointly deliver a more successful test suite and network for the benefit of the complex network community.}

For INRMs, we extract notable techniques and crafts from all kinds of methods. Some of these techniques or crafts are the characterization of network topological features, such as structural holes, neighbor contributions, local Cluster coefficient. Some introduced information theory, such as information entropy. Some introduced computational methods or models from other fields like machine learning and deep learning. There are also crafts from data perspectives, such as regularization and hyper-parameter setting. The use of these techniques and crafts enhances the interpretability of the methods and contributes to the performance enhancement of these methods. In the future studies, from these perspectives, INRMs will be further explored to promote the development of research and multi-domain integration.

\section{Conclusion and future work}
Because of the increased application of critical nodes in real-world networks, many INRMs have been developed to detect influential nodes over the past 20 years, especially influential nodes detecting under epidemic model. In some surveys and literature reviews, commonly used approaches have been summarized and classified. As far as we know, however, there is no systematic analysis and comparison of the capability and correctness of INRMs. In this paper, we reviewed the INRMs proposed based on epidemic model and their capability and correctness. Instead of classifying existing literature reviews, we selected four types of real-world networks for analyses of INRMs and identified seven categories of INRM meta-methods based on how the feature influence for nodes in network are computed. Then, we compared the capability and correctness of each category of the INRM to detect influential nodes by a series of metrics such as MF, RS, KCC. Finally, we identified quantitative relationships among networks, methods and evaluation metrics. The result analyses show that a few methods can be well applied in any kinds of network and the capability and correctness of INRMs is influenced by the scale, regularity, information flow etc. of networks and even related to the INRM itself, such as the machine learning methods can adjust parameter to control the ranking granularity. The results also show that a few metrics are applied most frequently to verify INRMs in any network. We also posed several meaningful research questions for further investigation. High-quality networks with full information and more directed and weighed networks need to be developed by industry practitioners and researcher to further verify INRMs. The techniques and crafts in INRMs can be deeply reviewed to improve the performance of INRMs. More studies verified by other models such as cascade failure also need to be analyzed.

\bibliographystyle{ACM-Reference-Format}
\bibliography{CSUR}

%%% -*-BibTeX-*-
%%% Do NOT edit. File created by BibTeX with style
%%% ACM-Reference-Format-Journals [18-Jan-2012].

\begin{thebibliography}{153}

%%% ====================================================================
%%% NOTE TO THE USER: you can override these defaults by providing
%%% customized versions of any of these macros before the \bibliography
%%% command.  Each of them MUST provide its own final punctuation,
%%% except for \shownote{}, \showDOI{}, and \showURL{}.  The latter two
%%% do not use final punctuation, in order to avoid confusing it with
%%% the Web address.
%%%
%%% To suppress output of a particular field, define its macro to expand
%%% to an empty string, or better, \unskip, like this:
%%%
%%% \newcommand{\showDOI}[1]{\unskip}   % LaTeX syntax
%%%
%%% \def \showDOI #1{\unskip}           % plain TeX syntax
%%%
%%% ====================================================================

\ifx \showCODEN    \undefined \def \showCODEN     #1{\unskip}     \fi
\ifx \showDOI      \undefined \def \showDOI       #1{#1}\fi
\ifx \showISBNx    \undefined \def \showISBNx     #1{\unskip}     \fi
\ifx \showISBNxiii \undefined \def \showISBNxiii  #1{\unskip}     \fi
\ifx \showISSN     \undefined \def \showISSN      #1{\unskip}     \fi
\ifx \showLCCN     \undefined \def \showLCCN      #1{\unskip}     \fi
\ifx \shownote     \undefined \def \shownote      #1{#1}          \fi
\ifx \showarticletitle \undefined \def \showarticletitle #1{#1}   \fi
\ifx \showURL      \undefined \def \showURL       {\relax}        \fi
% The following commands are used for tagged output and should be
% invisible to TeX
\providecommand\bibfield[2]{#2}
\providecommand\bibinfo[2]{#2}
\providecommand\natexlab[1]{#1}
\providecommand\showeprint[2][]{arXiv:#2}

\bibitem[Ahmad et~al\mbox{.}(2019)]%
        {AhmadAmreenandAhmad201911}
\bibfield{author}{\bibinfo{person}{Amreen Ahmad}, \bibinfo{person}{Tanvir
  Ahmad}, {and} \bibinfo{person}{Abhishek Bhatt}.}
  \bibinfo{year}{2019}\natexlab{}.
\newblock \showarticletitle{HWMSCB: A community-based hybrid approach for
  identifying influential nodes in the social network}.
\newblock \bibinfo{journal}{\emph{Physica A: Statistical Mechanics and its
  Applications}}  \bibinfo{volume}{545} (\bibinfo{date}{11}
  \bibinfo{year}{2019}).
\newblock
\urldef\tempurl%
\url{https://doi.org/10.1016/j.physa.2019.123590}
\showDOI{\tempurl}


\bibitem[Bae and Kim(2014)]%
        {BAE2014549}
\bibfield{author}{\bibinfo{person}{Joonhyun Bae} {and}
  \bibinfo{person}{Sangwook Kim}.} \bibinfo{year}{2014}\natexlab{}.
\newblock \showarticletitle{Identifying and ranking influential spreaders in
  complex networks by neighborhood coreness}.
\newblock \bibinfo{journal}{\emph{Physica A: Statistical Mechanics and its
  Applications}}  \bibinfo{volume}{395} (\bibinfo{year}{2014}),
  \bibinfo{pages}{549--559}.
\newblock
\showISSN{0378-4371}
\urldef\tempurl%
\url{https://doi.org/10.1016/j.physa.2013.10.047}
\showDOI{\tempurl}


\bibitem[Bao et~al\mbox{.}(2017)]%
        {BAO2017391}
\bibfield{author}{\bibinfo{person}{Zhong-Kui Bao}, \bibinfo{person}{Chuang Ma},
  \bibinfo{person}{Bing-Bing Xiang}, {and} \bibinfo{person}{Hai-Feng Zhang}.}
  \bibinfo{year}{2017}\natexlab{}.
\newblock \showarticletitle{Identification of influential nodes in complex
  networks: Method from spreading probability viewpoint}.
\newblock \bibinfo{journal}{\emph{Physica A: Statistical Mechanics and its
  Applications}}  \bibinfo{volume}{468} (\bibinfo{year}{2017}),
  \bibinfo{pages}{391--397}.
\newblock
\showISSN{0378-4371}
\urldef\tempurl%
\url{https://doi.org/10.1016/j.physa.2016.10.086}
\showDOI{\tempurl}


\bibitem[Basaras et~al\mbox{.}(2013)]%
        {6477051}
\bibfield{author}{\bibinfo{person}{Pavlos Basaras}, \bibinfo{person}{Dimitrios
  Katsaros}, {and} \bibinfo{person}{Leandros Tassiulas}.}
  \bibinfo{year}{2013}\natexlab{}.
\newblock \showarticletitle{Detecting Influential Spreaders in Complex, Dynamic
  Networks}.
\newblock \bibinfo{journal}{\emph{Computer}} \bibinfo{volume}{46},
  \bibinfo{number}{4} (\bibinfo{date}{April} \bibinfo{year}{2013}),
  \bibinfo{pages}{24--29}.
\newblock
\showISSN{1558-0814}
\urldef\tempurl%
\url{https://doi.org/10.1109/MC.2013.75}
\showDOI{\tempurl}


\bibitem[Berahmand et~al\mbox{.}(2018)]%
        {BerahmandKamalandBouyer201803}
\bibfield{author}{\bibinfo{person}{Kamal Berahmand}, \bibinfo{person}{Asgarali
  Bouyer}, {and} \bibinfo{person}{Negin Samadi}.}
  \bibinfo{year}{2018}\natexlab{}.
\newblock \showarticletitle{A new centrality measure based on the negative and
  positive effects of clustering coefficient for identifying influential
  spreaders in complex networks}.
\newblock \bibinfo{journal}{\emph{Chaos, Solitons and Fractals}}
  \bibinfo{volume}{110} (\bibinfo{date}{03} \bibinfo{year}{2018}).
\newblock
\urldef\tempurl%
\url{https://doi.org/10.1016/j.chaos.2018.03.014}
\showDOI{\tempurl}


\bibitem[Bian et~al\mbox{.}(2019)]%
        {10.1145/3301286}
\bibfield{author}{\bibinfo{person}{Ranran Bian}, \bibinfo{person}{Yun~Sing
  Koh}, \bibinfo{person}{Gillian Dobbie}, {and} \bibinfo{person}{Anna Divoli}.}
  \bibinfo{year}{2019}\natexlab{}.
\newblock \showarticletitle{Identifying Top-<i>k</i> Nodes in Social Networks:
  A Survey}.
\newblock \bibinfo{journal}{\emph{ACM Comput. Surv.}} \bibinfo{volume}{52},
  \bibinfo{number}{1}, Article \bibinfo{articleno}{22} (\bibinfo{date}{feb}
  \bibinfo{year}{2019}), \bibinfo{numpages}{33}~pages.
\newblock
\showISSN{0360-0300}
\urldef\tempurl%
\url{https://doi.org/10.1145/3301286}
\showDOI{\tempurl}


\bibitem[Bian et~al\mbox{.}(2017)]%
        {BIAN2017422}
\bibfield{author}{\bibinfo{person}{Tian Bian}, \bibinfo{person}{Jiantao Hu},
  {and} \bibinfo{person}{Yong Deng}.} \bibinfo{year}{2017}\natexlab{}.
\newblock \showarticletitle{Identifying influential nodes in complex networks
  based on AHP}.
\newblock \bibinfo{journal}{\emph{Physica A: Statistical Mechanics and its
  Applications}}  \bibinfo{volume}{479} (\bibinfo{year}{2017}),
  \bibinfo{pages}{422--436}.
\newblock
\showISSN{0378-4371}
\urldef\tempurl%
\url{https://doi.org/10.1016/j.physa.2017.02.085}
\showDOI{\tempurl}


\bibitem[Bjornstad et~al\mbox{.}(2020)]%
        {BjornstadOttarandShea2020}
\bibfield{author}{\bibinfo{person}{Ottar Bjornstad}, \bibinfo{person}{Katriona
  Shea}, \bibinfo{person}{Martin Krzywinski}, {and} \bibinfo{person}{Naomi
  Altman}.} \bibinfo{year}{2020}\natexlab{}.
\newblock \showarticletitle{Modeling infectious epidemics}.
\newblock \bibinfo{journal}{\emph{Nature Methods}}  \bibinfo{volume}{17}
  (\bibinfo{date}{04} \bibinfo{year}{2020}), \bibinfo{pages}{1--2}.
\newblock
\urldef\tempurl%
\url{https://doi.org/10.1038/s41592-020-0822-z}
\showDOI{\tempurl}


\bibitem[Blondel et~al\mbox{.}(2008)]%
        {londel200804}
\bibfield{author}{\bibinfo{person}{Vincent Blondel}, \bibinfo{person}{Jean-Loup
  Guillaume}, \bibinfo{person}{Renaud Lambiotte}, {and}
  \bibinfo{person}{Etienne Lefebvre}.} \bibinfo{year}{2008}\natexlab{}.
\newblock \showarticletitle{Fast Unfolding of Communities in Large Networks}.
\newblock \bibinfo{journal}{\emph{Journal of Statistical Mechanics Theory and
  Experiment}}  \bibinfo{volume}{2008} (\bibinfo{date}{04}
  \bibinfo{year}{2008}).
\newblock
\urldef\tempurl%
\url{https://doi.org/10.1088/1742-5468/2008/10/P10008}
\showDOI{\tempurl}


\bibitem[Boccaletti et~al\mbox{.}(2006)]%
        {BoccalettiStefano200602}
\bibfield{author}{\bibinfo{person}{Stefano Boccaletti}, \bibinfo{person}{Vito
  Latora}, \bibinfo{person}{Yamir Moreno}, \bibinfo{person}{Maisa Chávez},
  {and} \bibinfo{person}{Dong-Uk Hwang}.} \bibinfo{year}{2006}\natexlab{}.
\newblock \showarticletitle{Complex networks: Structure and dynamics}.
\newblock \bibinfo{journal}{\emph{Physics Reports}}  \bibinfo{volume}{424}
  (\bibinfo{date}{02} \bibinfo{year}{2006}), \bibinfo{pages}{175--308}.
\newblock
\urldef\tempurl%
\url{https://doi.org/10.1016/j.physrep.2005.10.009}
\showDOI{\tempurl}


\bibitem[Bolboaca and Jäntschi(2006)]%
        {Pearson_versus_Spearman}
\bibfield{author}{\bibinfo{person}{Sorana Bolboaca} {and}
  \bibinfo{person}{Lorentz Jäntschi}.} \bibinfo{year}{2006}\natexlab{}.
\newblock \showarticletitle{Pearson versus Spearman, Kendall's Tau Correlation
  Analysis on Structure-Activity Relationships of Biologic Active Compounds}.
\newblock \bibinfo{journal}{\emph{Leonardo Journal of Sciences}}
  \bibinfo{volume}{9} (\bibinfo{date}{07} \bibinfo{year}{2006}).
\newblock


\bibitem[Borgatti(2005)]%
        {Borgatti200501}
\bibfield{author}{\bibinfo{person}{Stephen Borgatti}.}
  \bibinfo{year}{2005}\natexlab{}.
\newblock \showarticletitle{Centrality and Network Flow}.
\newblock \bibinfo{journal}{\emph{Social Networks}}  \bibinfo{volume}{27}
  (\bibinfo{date}{01} \bibinfo{year}{2005}), \bibinfo{pages}{55--71}.
\newblock
\urldef\tempurl%
\url{https://doi.org/10.1016/j.socnet.2004.11.008}
\showDOI{\tempurl}


\bibitem[Brandes(2001)]%
        {Brandes2001AFA}
\bibfield{author}{\bibinfo{person}{Ulrik Brandes}.}
  \bibinfo{year}{2001}\natexlab{}.
\newblock \showarticletitle{A faster algorithm for betweenness centrality}.
\newblock \bibinfo{journal}{\emph{The Journal of Mathematical Sociology}}
  \bibinfo{volume}{25} (\bibinfo{year}{2001}), \bibinfo{pages}{163 -- 177}.
\newblock


\bibitem[Cai et~al\mbox{.}(2014)]%
        {CaiBiaoandTuoXianGuo2014}
\bibfield{author}{\bibinfo{person}{Biao Cai}, \bibinfo{person}{Xian-Guo Tuo},
  \bibinfo{person}{Kai-Xue Yang}, {and} \bibinfo{person}{Ming-Zhe Liu}.}
  \bibinfo{year}{2014}\natexlab{}.
\newblock \showarticletitle{Community centrality for node's influential ranking
  in complex network}.
\newblock \bibinfo{journal}{\emph{International Journal of Modern Physics C}}
  \bibinfo{volume}{25} (\bibinfo{date}{02} \bibinfo{year}{2014}).
\newblock
\urldef\tempurl%
\url{https://doi.org/10.1142/S0129183113500964}
\showDOI{\tempurl}


\bibitem[Cai et~al\mbox{.}(2017)]%
        {7979310}
\bibfield{author}{\bibinfo{person}{Die Cai}, \bibinfo{person}{Zhixuan Wang},
  \bibinfo{person}{Ningkui Wang}, {and} \bibinfo{person}{Daijun Wei}.}
  \bibinfo{year}{2017}\natexlab{}.
\newblock \showarticletitle{A new method for identifying influential nodes
  based on D-S evidence theory}. In \bibinfo{booktitle}{\emph{2017 29th Chinese
  Control And Decision Conference (CCDC)}}. \bibinfo{pages}{4603--4609}.
\newblock
\showISSN{1948-9447}
\urldef\tempurl%
\url{https://doi.org/10.1109/CCDC.2017.7979310}
\showDOI{\tempurl}


\bibitem[Chen et~al\mbox{.}(2013a)]%
        {theroleofclustering201310}
\bibfield{author}{\bibinfo{person}{Duanbing Chen}, \bibinfo{person}{Hui Gao},
  \bibinfo{person}{Linyuan Lü}, {and} \bibinfo{person}{Tao Zhou}.}
  \bibinfo{year}{2013}\natexlab{a}.
\newblock \showarticletitle{Identifying Influential Nodes in Large-Scale
  Directed Networks: The Role of Clustering}.
\newblock \bibinfo{journal}{\emph{PloS one}}  \bibinfo{volume}{8}
  (\bibinfo{date}{10} \bibinfo{year}{2013}), \bibinfo{pages}{e77455}.
\newblock
\urldef\tempurl%
\url{https://doi.org/10.1371/journal.pone.0077455}
\showDOI{\tempurl}


\bibitem[Chen et~al\mbox{.}(2012)]%
        {CHEN20121777}
\bibfield{author}{\bibinfo{person}{Duanbing Chen}, \bibinfo{person}{Linyuan
  Lü}, \bibinfo{person}{Ming-Sheng Shang}, \bibinfo{person}{Yi-Cheng Zhang},
  {and} \bibinfo{person}{Tao Zhou}.} \bibinfo{year}{2012}\natexlab{}.
\newblock \showarticletitle{Identifying influential nodes in complex networks}.
\newblock \bibinfo{journal}{\emph{Physica A: Statistical Mechanics and its
  Applications}} \bibinfo{volume}{391}, \bibinfo{number}{4}
  (\bibinfo{year}{2012}), \bibinfo{pages}{1777--1787}.
\newblock
\showISSN{0378-4371}
\urldef\tempurl%
\url{https://doi.org/10.1016/j.physa.2011.09.017}
\showDOI{\tempurl}


\bibitem[Chen et~al\mbox{.}(2020a)]%
        {ChenDongmingandPanpanDuandFang2020}
\bibfield{author}{\bibinfo{person}{Dongming Chen}, \bibinfo{person}{Du Panpan},
  \bibinfo{person}{Bo Fang}, \bibinfo{person}{Dongqi Wang}, {and}
  \bibinfo{person}{Xinyu Huang}.} \bibinfo{year}{2020}\natexlab{a}.
\newblock \showarticletitle{A Node Embedding-Based Influential Spreaders
  Identification Approach}.
\newblock  (\bibinfo{date}{09} \bibinfo{year}{2020}).
\newblock
\urldef\tempurl%
\url{https://doi.org/10.3390/math8091554}
\showDOI{\tempurl}


\bibitem[Chen et~al\mbox{.}(2013b)]%
        {Chen_2013}
\bibfield{author}{\bibinfo{person}{Duan-Bing Chen}, \bibinfo{person}{Rui Xiao},
  \bibinfo{person}{An Zeng}, {and} \bibinfo{person}{Yi-Cheng Zhang}.}
  \bibinfo{year}{2013}\natexlab{b}.
\newblock \showarticletitle{Path diversity improves the identification of
  influential spreaders}.
\newblock \bibinfo{journal}{\emph{{EPL} (Europhysics Letters)}}
  \bibinfo{volume}{104}, \bibinfo{number}{6} (\bibinfo{date}{dec}
  \bibinfo{year}{2013}), \bibinfo{pages}{68006}.
\newblock
\urldef\tempurl%
\url{https://doi.org/10.1209/0295-5075/104/68006}
\showDOI{\tempurl}


\bibitem[Chen et~al\mbox{.}(2020b)]%
        {e22080848}
\bibfield{author}{\bibinfo{person}{Xuegong Chen}, \bibinfo{person}{Jie Zhou},
  \bibinfo{person}{Zhifang Liao}, \bibinfo{person}{Shengzong Liu}, {and}
  \bibinfo{person}{Yan Zhang}.} \bibinfo{year}{2020}\natexlab{b}.
\newblock \showarticletitle{A Novel Method to Rank Influential Nodes in Complex
  Networks Based on Tsallis Entropy}.
\newblock \bibinfo{journal}{\emph{Entropy}} \bibinfo{volume}{22},
  \bibinfo{number}{8} (\bibinfo{year}{2020}).
\newblock
\showISSN{1099-4300}
\urldef\tempurl%
\url{https://doi.org/10.3390/e22080848}
\showDOI{\tempurl}


\bibitem[Cheng et~al\mbox{.}(2014)]%
        {IMRankInfluenceMaximization}
\bibfield{author}{\bibinfo{person}{Suqi Cheng}, \bibinfo{person}{Hua-Wei Shen},
  \bibinfo{person}{Junming Huang}, {and} \bibinfo{person}{Xue-Qi Cheng}.}
  \bibinfo{year}{2014}\natexlab{}.
\newblock \showarticletitle{IMRank: Influence Maximization via Finding
  Self-Consistent Ranking}.
\newblock  (\bibinfo{date}{02} \bibinfo{year}{2014}).
\newblock
\urldef\tempurl%
\url{https://doi.org/10.1145/2600428.2609592}
\showDOI{\tempurl}


\bibitem[Dai et~al\mbox{.}(2019)]%
        {DaiJinYingandWang201909}
\bibfield{author}{\bibinfo{person}{JinYing Dai}, \bibinfo{person}{Bin Wang},
  \bibinfo{person}{JinFang Sheng}, \bibinfo{person}{Zejun Sun},
  \bibinfo{person}{Faiza Khawaja}, \bibinfo{person}{Aman Ullah},
  \bibinfo{person}{Dawit Aklilu}, {and} \bibinfo{person}{GuiHua Duan}.}
  \bibinfo{year}{2019}\natexlab{}.
\newblock \showarticletitle{Identifying Influential Nodes in Complex Networks
  Based on Local Neighbor Contribution}.
\newblock \bibinfo{journal}{\emph{IEEE Access}}  \bibinfo{volume}{PP}
  (\bibinfo{date}{09} \bibinfo{year}{2019}), \bibinfo{pages}{1--1}.
\newblock
\urldef\tempurl%
\url{https://doi.org/10.1109/ACCESS.2019.2939804}
\showDOI{\tempurl}


\bibitem[Deng et~al\mbox{.}(2012)]%
        {FuzzyDijkstraalgorithm}
\bibfield{author}{\bibinfo{person}{Yong Deng}, \bibinfo{person}{Yuxin Chen},
  \bibinfo{person}{Yajuan Zhang}, {and} \bibinfo{person}{Sankaran Mahadevan}.}
  \bibinfo{year}{2012}\natexlab{}.
\newblock \showarticletitle{Fuzzy Dijkstra algorithm for shortest path problem
  under uncertain environment}.
\newblock \bibinfo{journal}{\emph{Appl. Soft Comput.}}  \bibinfo{volume}{12}
  (\bibinfo{date}{03} \bibinfo{year}{2012}), \bibinfo{pages}{1231--1237}.
\newblock
\urldef\tempurl%
\url{https://doi.org/10.1016/j.asoc.2011.11.011}
\showDOI{\tempurl}


\bibitem[Dong et~al\mbox{.}(2018)]%
        {8351889}
\bibfield{author}{\bibinfo{person}{Jiali Dong}, \bibinfo{person}{Fanghua Ye},
  \bibinfo{person}{Wuhui Chen}, {and} \bibinfo{person}{Jiajing Wu}.}
  \bibinfo{year}{2018}\natexlab{}.
\newblock \showarticletitle{Identifying Influential Nodes in Complex Networks
  via Semi-Local Centrality}. In \bibinfo{booktitle}{\emph{2018 IEEE
  International Symposium on Circuits and Systems (ISCAS)}}.
  \bibinfo{pages}{1--5}.
\newblock
\urldef\tempurl%
\url{https://doi.org/10.1109/ISCAS.2018.8351889}
\showDOI{\tempurl}


\bibitem[Du et~al\mbox{.}(2014)]%
        {DU201457}
\bibfield{author}{\bibinfo{person}{Yuxian Du}, \bibinfo{person}{Cai Gao},
  \bibinfo{person}{Yong Hu}, \bibinfo{person}{Sankaran Mahadevan}, {and}
  \bibinfo{person}{Yong Deng}.} \bibinfo{year}{2014}\natexlab{}.
\newblock \showarticletitle{A new method of identifying influential nodes in
  complex networks based on TOPSIS}.
\newblock \bibinfo{journal}{\emph{Physica A: Statistical Mechanics and its
  Applications}}  \bibinfo{volume}{399} (\bibinfo{year}{2014}),
  \bibinfo{pages}{57--69}.
\newblock
\showISSN{0378-4371}
\urldef\tempurl%
\url{https://doi.org/10.1016/j.physa.2013.12.031}
\showDOI{\tempurl}


\bibitem[Duch and Arenas(2005)]%
        {Duch200509}
\bibfield{author}{\bibinfo{person}{Jordi Duch} {and} \bibinfo{person}{Alex
  Arenas}.} \bibinfo{year}{2005}\natexlab{}.
\newblock \showarticletitle{Community Detection in Complex Networks Using
  Extremal Optimization}.
\newblock \bibinfo{journal}{\emph{Physical review. E, Statistical, nonlinear,
  and soft matter physics}}  \bibinfo{volume}{72} (\bibinfo{date}{09}
  \bibinfo{year}{2005}), \bibinfo{pages}{027104}.
\newblock
\urldef\tempurl%
\url{https://doi.org/10.1103/PhysRevE.72.027104}
\showDOI{\tempurl}


\bibitem[Estrada and Rodriguez-Velazquez(2005)]%
        {Estrada200506}
\bibfield{author}{\bibinfo{person}{Ernesto Estrada} {and}
  \bibinfo{person}{Juan~Alberto Rodriguez-Velazquez}.}
  \bibinfo{year}{2005}\natexlab{}.
\newblock \showarticletitle{Subgraph Centrality in Complex Networks}.
\newblock \bibinfo{journal}{\emph{Physical review. E, Statistical, nonlinear,
  and soft matter physics}}  \bibinfo{volume}{71} (\bibinfo{date}{06}
  \bibinfo{year}{2005}), \bibinfo{pages}{056103}.
\newblock
\urldef\tempurl%
\url{https://doi.org/10.1103/PhysRevE.71.056103}
\showDOI{\tempurl}


\bibitem[Fei and Deng(2017)]%
        {FEI2017257}
\bibfield{author}{\bibinfo{person}{Liguo Fei} {and} \bibinfo{person}{Yong
  Deng}.} \bibinfo{year}{2017}\natexlab{}.
\newblock \showarticletitle{A new method to identify influential nodes based on
  relative entropy}.
\newblock \bibinfo{journal}{\emph{Chaos, Solitons \& Fractals}}
  \bibinfo{volume}{104} (\bibinfo{year}{2017}), \bibinfo{pages}{257--267}.
\newblock
\showISSN{0960-0779}
\urldef\tempurl%
\url{https://doi.org/10.1016/j.chaos.2017.08.010}
\showDOI{\tempurl}


\bibitem[Fei et~al\mbox{.}(2017)]%
        {FeiLiguoandmo201708}
\bibfield{author}{\bibinfo{person}{Liguo Fei}, \bibinfo{person}{Hongming mo},
  {and} \bibinfo{person}{Yong Deng}.} \bibinfo{year}{2017}\natexlab{}.
\newblock \showarticletitle{A new method to identify influential nodes based on
  combining of existing centrality measures}.
\newblock \bibinfo{journal}{\emph{Modern Physics Letters B}}
  \bibinfo{volume}{31} (\bibinfo{date}{08} \bibinfo{year}{2017}),
  \bibinfo{pages}{1750243}.
\newblock
\urldef\tempurl%
\url{https://doi.org/10.1142/S0217984917502438}
\showDOI{\tempurl}


\bibitem[Fei et~al\mbox{.}(2018)]%
        {FeiLiguoandZhang201808}
\bibfield{author}{\bibinfo{person}{Liguo Fei}, \bibinfo{person}{Qi Zhang},
  {and} \bibinfo{person}{Yong Deng}.} \bibinfo{year}{2018}\natexlab{}.
\newblock \showarticletitle{Identifying influential nodes in complex networks
  based on the inverse-square law}.
\newblock \bibinfo{journal}{\emph{Physica A: Statistical Mechanics and its
  Applications}}  \bibinfo{volume}{512} (\bibinfo{date}{08}
  \bibinfo{year}{2018}).
\newblock
\urldef\tempurl%
\url{https://doi.org/10.1016/j.physa.2018.08.135}
\showDOI{\tempurl}


\bibitem[Freeman(1979)]%
        {FreemanLinton197901}
\bibfield{author}{\bibinfo{person}{Linton Freeman}.}
  \bibinfo{year}{1979}\natexlab{}.
\newblock \showarticletitle{Centrality in Social Networks' Conceptual
  Clarification}.
\newblock \bibinfo{journal}{\emph{Social Networks}}  \bibinfo{volume}{1}
  (\bibinfo{date}{01} \bibinfo{year}{1979}), \bibinfo{pages}{215--239}.
\newblock
\urldef\tempurl%
\url{https://doi.org/10.1016/0378-8733(78)90021-7}
\showDOI{\tempurl}


\bibitem[Fu et~al\mbox{.}(2015)]%
        {FuYuHsiangandHuang201504}
\bibfield{author}{\bibinfo{person}{Yu-Hsiang Fu}, \bibinfo{person}{Chung-Yuan
  Huang}, {and} \bibinfo{person}{Chuen-Tsai Sun}.}
  \bibinfo{year}{2015}\natexlab{}.
\newblock \showarticletitle{Using global diversity and local features to
  identify influential social network spreaders}.
\newblock \bibinfo{journal}{\emph{Physica A: Statistical Mechanics and its
  Applications}}  \bibinfo{volume}{104} (\bibinfo{date}{04}
  \bibinfo{year}{2015}).
\newblock
\urldef\tempurl%
\url{https://doi.org/10.1016/j.physa.2015.03.042}
\showDOI{\tempurl}


\bibitem[Gao et~al\mbox{.}(2015)]%
        {GaoChaoandZhongLuandLi201510}
\bibfield{author}{\bibinfo{person}{Chao Gao}, \bibinfo{person}{Lu Zhong},
  \bibinfo{person}{Xianghua Li}, \bibinfo{person}{Zili Zhang}, {and}
  \bibinfo{person}{Ning Shi}.} \bibinfo{year}{2015}\natexlab{}.
\newblock \showarticletitle{Combination methods for identifying influential
  nodes in networks}.
\newblock \bibinfo{journal}{\emph{International Journal of Modern Physics C}}
  \bibinfo{volume}{26} (\bibinfo{date}{10} \bibinfo{year}{2015}).
\newblock
\urldef\tempurl%
\url{https://doi.org/10.1142/S0129183115500679}
\showDOI{\tempurl}


\bibitem[Gao et~al\mbox{.}(2014)]%
        {RePEc:eee:phsmap:v:403:y:2014:i:c:p:130-147}
\bibfield{author}{\bibinfo{person}{Shuai Gao}, \bibinfo{person}{Jun Ma},
  \bibinfo{person}{Zhumin Chen}, \bibinfo{person}{Guanghui Wang}, {and}
  \bibinfo{person}{Changming Xing}.} \bibinfo{year}{2014}\natexlab{}.
\newblock \showarticletitle{{Ranking the spreading ability of nodes in complex
  networks based on local structure}}.
\newblock \bibinfo{journal}{\emph{Physica A: Statistical Mechanics and its
  Applications}} \bibinfo{volume}{403}, \bibinfo{number}{C}
  (\bibinfo{year}{2014}), \bibinfo{pages}{130--147}.
\newblock
\urldef\tempurl%
\url{https://doi.org/10.1016/j.physa.2014.02.0}
\showDOI{\tempurl}


\bibitem[Garas et~al\mbox{.}(2012)]%
        {Garas_2012}
\bibfield{author}{\bibinfo{person}{Antonios Garas}, \bibinfo{person}{Frank
  Schweitzer}, {and} \bibinfo{person}{Shlomo Havlin}.}
  \bibinfo{year}{2012}\natexlab{}.
\newblock \showarticletitle{A k-shell decomposition method for weighted
  networks}.
\newblock \bibinfo{journal}{\emph{New Journal of Physics}}
  \bibinfo{volume}{14}, \bibinfo{number}{8} (\bibinfo{date}{aug}
  \bibinfo{year}{2012}), \bibinfo{pages}{083030}.
\newblock
\urldef\tempurl%
\url{https://doi.org/10.1088/1367-2630/14/8/083030}
\showDOI{\tempurl}


\bibitem[Goldenberg et~al\mbox{.}(2001)]%
        {GoldenbergJacobandLibaiBarakandMuller200108}
\bibfield{author}{\bibinfo{person}{Jacob Goldenberg}, \bibinfo{person}{Barak
  Libai}, {and} \bibinfo{person}{Eitan Muller}.}
  \bibinfo{year}{2001}\natexlab{}.
\newblock \showarticletitle{Talk of the Network: A Complex Systems Look at the
  Underlying Process of Word-of-Mouth}.
\newblock \bibinfo{journal}{\emph{Marketing Letters}}  \bibinfo{volume}{12}
  (\bibinfo{date}{08} \bibinfo{year}{2001}), \bibinfo{pages}{211--223}.
\newblock
\urldef\tempurl%
\url{https://doi.org/10.1023/A:1011122126881}
\showDOI{\tempurl}


\bibitem[Guo et~al\mbox{.}(2020)]%
        {GuoChunguandYangLiangweiandChen202002}
\bibfield{author}{\bibinfo{person}{Chungu Guo}, \bibinfo{person}{Liangwei
  Yang}, \bibinfo{person}{Xiao Chen}, \bibinfo{person}{Duanbing Chen},
  \bibinfo{person}{Hui Gao}, {and} \bibinfo{person}{Jing Ma}.}
  \bibinfo{year}{2020}\natexlab{}.
\newblock \showarticletitle{Influential Nodes Identification in Complex
  Networks via Information Entropy}.
\newblock \bibinfo{journal}{\emph{Entropy}}  \bibinfo{volume}{22}
  (\bibinfo{date}{02} \bibinfo{year}{2020}), \bibinfo{pages}{242}.
\newblock
\urldef\tempurl%
\url{https://doi.org/10.3390/e22020242}
\showDOI{\tempurl}


\bibitem[Guzzi and Roy(2020)]%
        {inbookKim201905}
\bibfield{author}{\bibinfo{person}{Pietro~Hiram Guzzi} {and}
  \bibinfo{person}{Swarup Roy}.} \bibinfo{year}{2020}\natexlab{}.
\newblock \showarticletitle{4 - Complex network models}.
\newblock In \bibinfo{booktitle}{\emph{Biological Network Analysis}},
  \bibfield{editor}{\bibinfo{person}{Pietro~Hiram Guzzi} {and}
  \bibinfo{person}{Swarup Roy}} (Eds.). \bibinfo{publisher}{Academic Press},
  \bibinfo{pages}{53--75}.
\newblock
\showISBNx{978-0-12-819350-1}
\urldef\tempurl%
\url{https://doi.org/10.1016/B978-0-12-819350-1.00010-4}
\showDOI{\tempurl}


\bibitem[Hafiene et~al\mbox{.}(2020)]%
        {HAFIENE2020113642}
\bibfield{author}{\bibinfo{person}{Nesrine Hafiene}, \bibinfo{person}{Wafa
  Karoui}, {and} \bibinfo{person}{Lotfi {Ben Romdhane}}.}
  \bibinfo{year}{2020}\natexlab{}.
\newblock \showarticletitle{Influential nodes detection in dynamic social
  networks: A survey}.
\newblock \bibinfo{journal}{\emph{Expert Systems with Applications}}
  \bibinfo{volume}{159} (\bibinfo{year}{2020}), \bibinfo{pages}{113642}.
\newblock
\showISSN{0957-4174}
\urldef\tempurl%
\url{https://doi.org/10.1016/j.eswa.2020.113642}
\showDOI{\tempurl}


\bibitem[He et~al\mbox{.}(2015)]%
        {HeJialinandFuYanandChen2015}
\bibfield{author}{\bibinfo{person}{Jialin He}, \bibinfo{person}{Yan Fu}, {and}
  \bibinfo{person}{Duanbing Chen}.} \bibinfo{year}{2015}\natexlab{}.
\newblock \showarticletitle{A Novel Top-k Strategy for Influence Maximization
  in Complex Networks with Community Structure}.
\newblock \bibinfo{journal}{\emph{PloS one}}  \bibinfo{volume}{10}
  (\bibinfo{date}{12} \bibinfo{year}{2015}), \bibinfo{pages}{e0145283}.
\newblock
\urldef\tempurl%
\url{https://doi.org/10.1371/journal.pone.0145283}
\showDOI{\tempurl}


\bibitem[Hou et~al\mbox{.}(2012)]%
        {HouBonanandYao201208}
\bibfield{author}{\bibinfo{person}{Bonan Hou}, \bibinfo{person}{Yiping Yao},
  {and} \bibinfo{person}{Dongsheng Liao}.} \bibinfo{year}{2012}\natexlab{}.
\newblock \showarticletitle{Identifying all-around nodes for spreading dynamics
  in complex networks}.
\newblock \bibinfo{journal}{\emph{Physica A: Statistical Mechanics and its
  Applications}}  \bibinfo{volume}{391} (\bibinfo{date}{08}
  \bibinfo{year}{2012}), \bibinfo{pages}{4012–4017}.
\newblock
\urldef\tempurl%
\url{https://doi.org/10.1016/j.physa.2012.02.033}
\showDOI{\tempurl}


\bibitem[Hu et~al\mbox{.}(2016)]%
        {HU201673}
\bibfield{author}{\bibinfo{person}{Jiantao Hu}, \bibinfo{person}{Yuxian Du},
  \bibinfo{person}{Hongming Mo}, \bibinfo{person}{Daijun Wei}, {and}
  \bibinfo{person}{Yong Deng}.} \bibinfo{year}{2016}\natexlab{}.
\newblock \showarticletitle{A modified weighted TOPSIS to identify influential
  nodes in complex networks}.
\newblock \bibinfo{journal}{\emph{Physica A: Statistical Mechanics and its
  Applications}}  \bibinfo{volume}{444} (\bibinfo{year}{2016}),
  \bibinfo{pages}{73--85}.
\newblock
\showISSN{0378-4371}
\urldef\tempurl%
\url{https://doi.org/10.1016/j.physa.2015.09.028}
\showDOI{\tempurl}


\bibitem[Hu and Mei(2017)]%
        {HuPingandMei201709}
\bibfield{author}{\bibinfo{person}{Ping Hu} {and} \bibinfo{person}{Ting Mei}.}
  \bibinfo{year}{2017}\natexlab{}.
\newblock \showarticletitle{Ranking influential nodes in complex networks with
  structural holes}.
\newblock \bibinfo{journal}{\emph{Physica A: Statistical Mechanics and its
  Applications}}  \bibinfo{volume}{490} (\bibinfo{date}{09}
  \bibinfo{year}{2017}).
\newblock
\urldef\tempurl%
\url{https://doi.org/10.1016/j.physa.2017.08.049}
\showDOI{\tempurl}


\bibitem[Huang and Yu(2017)]%
        {HuangDawenYu2017}
\bibfield{author}{\bibinfo{person}{Da-wen Huang} {and} \bibinfo{person}{Zuguo
  Yu}.} \bibinfo{year}{2017}\natexlab{}.
\newblock \showarticletitle{Dynamic-Sensitive centrality of nodes in temporal
  networks}.
\newblock \bibinfo{journal}{\emph{Scientific Reports}}  \bibinfo{volume}{7}
  (\bibinfo{date}{02} \bibinfo{year}{2017}), \bibinfo{pages}{41454}.
\newblock
\urldef\tempurl%
\url{https://doi.org/10.1038/srep41454}
\showDOI{\tempurl}


\bibitem[Hui et~al\mbox{.}(2013)]%
        {HuiyuandZun201301}
\bibfield{author}{\bibinfo{person}{Yu Hui}, \bibinfo{person}{Liu Zun}, {and}
  \bibinfo{person}{Li Yongjun}.} \bibinfo{year}{2013}\natexlab{}.
\newblock \showarticletitle{Using Local Improved Structural Holes Method to
  Identify Key Nodes in Complex Networks}. In \bibinfo{booktitle}{\emph{2013
  Fifth International Conference on Measuring Technology and Mechatronics
  Automation}}. \bibinfo{pages}{1292--1295}.
\newblock
\urldef\tempurl%
\url{https://doi.org/10.1109/ICMTMA.2013.317}
\showDOI{\tempurl}


\bibitem[Hwang and Yoon(1981)]%
        {Hwang1981MultipleAD}
\bibfield{author}{\bibinfo{person}{Ching-Lai Hwang} {and}
  \bibinfo{person}{Kwangsun Yoon}.} \bibinfo{year}{1981}\natexlab{}.
\newblock \showarticletitle{Multiple Attribute Decision Making: Methods and
  Applications - A State-of-the-Art Survey}. In
  \bibinfo{booktitle}{\emph{Lecture Notes in Economics and Mathematical
  Systems}}.
\newblock


\bibitem[Ibnoulouafi and Haziti(2018)]%
        {Ibnoulouafi201809}
\bibfield{author}{\bibinfo{person}{Ahmed Ibnoulouafi} {and}
  \bibinfo{person}{Mohamed Haziti}.} \bibinfo{year}{2018}\natexlab{}.
\newblock \showarticletitle{Density centrality: identifying influential nodes
  based on area density formula}.
\newblock \bibinfo{journal}{\emph{Chaos Solitons \& Fractals}}
  \bibinfo{volume}{114} (\bibinfo{date}{09} \bibinfo{year}{2018}),
  \bibinfo{pages}{69 -- 80}.
\newblock
\urldef\tempurl%
\url{https://doi.org/10.1016/j.chaos.2018.06.022}
\showDOI{\tempurl}


\bibitem[Iyer et~al\mbox{.}(2013)]%
        {IyerSwamiandKillingback}
\bibfield{author}{\bibinfo{person}{Swami Iyer}, \bibinfo{person}{Timothy
  Killingback}, \bibinfo{person}{Bala Sundaram}, {and} \bibinfo{person}{Zhen
  Wang}.} \bibinfo{year}{2013}\natexlab{}.
\newblock \showarticletitle{Attack Robustness and Centrality of Complex
  Networks}.
\newblock \bibinfo{journal}{\emph{PloS one}}  \bibinfo{volume}{8}
  (\bibinfo{date}{04} \bibinfo{year}{2013}), \bibinfo{pages}{e59613}.
\newblock
\urldef\tempurl%
\url{https://doi.org/10.1371/journal.pone.0059613}
\showDOI{\tempurl}


\bibitem[Ji et~al\mbox{.}(2017a)]%
        {JiShenggongand2017}
\bibfield{author}{\bibinfo{person}{Shenggong Ji}, \bibinfo{person}{Linyuan
  Lü}, \bibinfo{person}{Chi~Ho Yeung}, {and} \bibinfo{person}{Yanqing Hu}.}
  \bibinfo{year}{2017}\natexlab{a}.
\newblock \showarticletitle{Effective spreading from multiple leaders
  identified by percolation in the susceptible-infected-recovered (SIR) model}.
\newblock \bibinfo{journal}{\emph{New Journal of Physics}}
  \bibinfo{volume}{19} (\bibinfo{date}{07} \bibinfo{year}{2017}),
  \bibinfo{pages}{073020}.
\newblock
\urldef\tempurl%
\url{https://doi.org/10.1088/1367-2630/aa76b0}
\showDOI{\tempurl}


\bibitem[Ji et~al\mbox{.}(2017b)]%
        {Effective_spreading_from_multiple}
\bibfield{author}{\bibinfo{person}{Shenggong Ji}, \bibinfo{person}{Linyuan
  Lü}, \bibinfo{person}{Chi~Ho Yeung}, {and} \bibinfo{person}{Yanqing Hu}.}
  \bibinfo{year}{2017}\natexlab{b}.
\newblock \showarticletitle{Effective spreading from multiple leaders
  identified by percolation in the susceptible-infected-recovered (SIR) model}.
\newblock \bibinfo{journal}{\emph{New Journal of Physics}}
  \bibinfo{volume}{19} (\bibinfo{date}{07} \bibinfo{year}{2017}),
  \bibinfo{pages}{073020}.
\newblock
\urldef\tempurl%
\url{https://doi.org/10.1088/1367-2630/aa76b0}
\showDOI{\tempurl}


\bibitem[Jung et~al\mbox{.}(2011)]%
        {IRIEScalableandRobustInfluence}
\bibfield{author}{\bibinfo{person}{Kyomin Jung}, \bibinfo{person}{Wooram Heo},
  {and} \bibinfo{person}{Wei Chen}.} \bibinfo{year}{2011}\natexlab{}.
\newblock \showarticletitle{IRIE: Scalable and Robust Influence Maximization in
  Social Networks}.
\newblock \bibinfo{journal}{\emph{Proceedings - IEEE International Conference
  on Data Mining, ICDM}} (\bibinfo{date}{11} \bibinfo{year}{2011}).
\newblock
\urldef\tempurl%
\url{https://doi.org/10.1109/ICDM.2012.79}
\showDOI{\tempurl}


\bibitem[Kang et~al\mbox{.}(2016)]%
        {7925142}
\bibfield{author}{\bibinfo{person}{Wenfeng Kang}, \bibinfo{person}{Guangming
  Tang}, \bibinfo{person}{Yifeng Sun}, {and} \bibinfo{person}{Shuo Wang}.}
  \bibinfo{year}{2016}\natexlab{}.
\newblock \showarticletitle{Identifying influential nodes in complex network
  based on weighted semi-local centrality}. In \bibinfo{booktitle}{\emph{2016
  2nd IEEE International Conference on Computer and Communications (ICCC)}}.
  \bibinfo{pages}{2467--2471}.
\newblock
\urldef\tempurl%
\url{https://doi.org/10.1109/CompComm.2016.7925142}
\showDOI{\tempurl}


\bibitem[Karrer et~al\mbox{.}(2014)]%
        {KarrerBrian201405}
\bibfield{author}{\bibinfo{person}{Brian Karrer}, \bibinfo{person}{M. Newman},
  {and} \bibinfo{person}{Lenka Zdeborova}.} \bibinfo{year}{2014}\natexlab{}.
\newblock \showarticletitle{Percolation on Sparse Networks}.
\newblock \bibinfo{journal}{\emph{Physical review letters}}
  \bibinfo{volume}{113} (\bibinfo{date}{05} \bibinfo{year}{2014}).
\newblock
\urldef\tempurl%
\url{https://doi.org/10.1103/PhysRevLett.113.208702}
\showDOI{\tempurl}


\bibitem[Keeling and Eames(2005)]%
        {KeelingMattandEames}
\bibfield{author}{\bibinfo{person}{Matt Keeling} {and} \bibinfo{person}{Ken
  Eames}.} \bibinfo{year}{2005}\natexlab{}.
\newblock \showarticletitle{Networks and Epidemic Models}.
\newblock \bibinfo{journal}{\emph{Journal of the Royal Society, Interface / the
  Royal Society}}  \bibinfo{volume}{2} (\bibinfo{date}{10}
  \bibinfo{year}{2005}), \bibinfo{pages}{295--307}.
\newblock
\urldef\tempurl%
\url{https://doi.org/10.1098/rsif.2005.0051}
\showDOI{\tempurl}


\bibitem[Kendall(1938)]%
        {A_New_Measure_of_Rank}
\bibfield{author}{\bibinfo{person}{Maurice Kendall}.}
  \bibinfo{year}{1938}\natexlab{}.
\newblock \showarticletitle{A New Measure of Rank Correlation}.
\newblock \bibinfo{journal}{\emph{Biometrika}}  \bibinfo{volume}{30}
  (\bibinfo{date}{01} \bibinfo{year}{1938}).
\newblock
\urldef\tempurl%
\url{https://doi.org/10.1093/biomet/30.1-2.81}
\showDOI{\tempurl}


\bibitem[Kermack and McKendrick(1991)]%
        {Contributions_to_the_mathematical}
\bibfield{author}{\bibinfo{person}{W.O. Kermack} {and} \bibinfo{person}{A.G.
  McKendrick}.} \bibinfo{year}{1991}\natexlab{}.
\newblock \showarticletitle{Contributions to the mathematical theory of
  epidemics—I}.
\newblock \bibinfo{journal}{\emph{Bulletin of mathematical biology}}
  \bibinfo{volume}{53} (\bibinfo{date}{02} \bibinfo{year}{1991}),
  \bibinfo{pages}{33--55}.
\newblock
\urldef\tempurl%
\url{https://doi.org/10.1016/S0092-8240(05)80040-0}
\showDOI{\tempurl}


\bibitem[Kimura et~al\mbox{.}(2015)]%
        {Efficient_analysis_of_node_influence}
\bibfield{author}{\bibinfo{person}{Masahiro Kimura}, \bibinfo{person}{Kazumi
  Saito}, \bibinfo{person}{Kouzou Ohara}, {and} \bibinfo{person}{Hiroshi
  Motoda}.} \bibinfo{year}{2015}\natexlab{}.
\newblock \showarticletitle{Efficient analysis of node influence based on SIR
  model over huge complex networks}.
\newblock \bibinfo{journal}{\emph{DSAA 2014 - Proceedings of the 2014 IEEE
  International Conference on Data Science and Advanced Analytics}}
  (\bibinfo{date}{03} \bibinfo{year}{2015}), \bibinfo{pages}{216--222}.
\newblock
\urldef\tempurl%
\url{https://doi.org/10.1109/DSAA.2014.7058076}
\showDOI{\tempurl}


\bibitem[Kitsak et~al\mbox{.}(2010)]%
        {KitsakMaksim201001}
\bibfield{author}{\bibinfo{person}{Maksim Kitsak}, \bibinfo{person}{Lazaros
  Gallos}, \bibinfo{person}{Shlomo Havlin}, \bibinfo{person}{Fredrik Liljeros},
  \bibinfo{person}{Lev Muchnik}, \bibinfo{person}{H. Stanley}, {and}
  \bibinfo{person}{Hernan Makse}.} \bibinfo{year}{2010}\natexlab{}.
\newblock \showarticletitle{Identification of influential spreaders in complex
  networks}.
\newblock \bibinfo{journal}{\emph{Nature Physics}}  \bibinfo{volume}{6}
  (\bibinfo{date}{01} \bibinfo{year}{2010}).
\newblock
\urldef\tempurl%
\url{https://doi.org/10.1038/nphys1746}
\showDOI{\tempurl}


\bibitem[Kumar and Panda(2022)]%
        {KumarSanjayandPanda}
\bibfield{author}{\bibinfo{person}{Sanjay Kumar} {and} \bibinfo{person}{Ankit
  Panda}.} \bibinfo{year}{2022}\natexlab{}.
\newblock \showarticletitle{Identifying influential nodes in weighted complex
  networks using an improved WVoteRank approach}.
\newblock \bibinfo{journal}{\emph{Applied Intelligence}}  \bibinfo{volume}{52}
  (\bibinfo{date}{01} \bibinfo{year}{2022}).
\newblock
\urldef\tempurl%
\url{https://doi.org/10.1007/s10489-021-02403-5}
\showDOI{\tempurl}


\bibitem[Kumar and Panda(2020)]%
        {KumarSanjayandPanda2020}
\bibfield{author}{\bibinfo{person}{Sanjay Kumar} {and}
  \bibinfo{person}{Bishwajit Panda}.} \bibinfo{year}{2020}\natexlab{}.
\newblock \showarticletitle{Identifying influential nodes in Social Networks:
  Neighborhood Coreness based voting approach}.
\newblock \bibinfo{journal}{\emph{Physica A: Statistical Mechanics and its
  Applications}}  \bibinfo{volume}{553} (\bibinfo{date}{01}
  \bibinfo{year}{2020}), \bibinfo{pages}{124215}.
\newblock
\urldef\tempurl%
\url{https://doi.org/10.1016/j.physa.2020.124215}
\showDOI{\tempurl}


\bibitem[Lalou et~al\mbox{.}(2018)]%
        {LALOU201892}
\bibfield{author}{\bibinfo{person}{Mohammed Lalou},
  \bibinfo{person}{Mohammed~Amin Tahraoui}, {and} \bibinfo{person}{Hamamache
  Kheddouci}.} \bibinfo{year}{2018}\natexlab{}.
\newblock \showarticletitle{The Critical Node Detection Problem in networks: A
  survey}.
\newblock \bibinfo{journal}{\emph{Computer Science Review}}
  \bibinfo{volume}{28} (\bibinfo{year}{2018}), \bibinfo{pages}{92--117}.
\newblock
\showISSN{1574-0137}
\urldef\tempurl%
\url{https://doi.org/10.1016/j.cosrev.2018.02.002}
\showDOI{\tempurl}


\bibitem[Lazega and Burt(1995)]%
        {LazegaEmmanuelandBurtRonald199510}
\bibfield{author}{\bibinfo{person}{Emmanuel Lazega} {and}
  \bibinfo{person}{Ronald Burt}.} \bibinfo{year}{1995}\natexlab{}.
\newblock \showarticletitle{Structural Holes: The Social Structure of
  Competition}.
\newblock \bibinfo{journal}{\emph{Revue Française de Sociologie}}
  \bibinfo{volume}{36} (\bibinfo{date}{10} \bibinfo{year}{1995}),
  \bibinfo{pages}{779}.
\newblock
\urldef\tempurl%
\url{https://doi.org/10.2307/3322456}
\showDOI{\tempurl}


\bibitem[LEVANDOWSKY and WINTER(1971)]%
        {Distance_Between_Sets}
\bibfield{author}{\bibinfo{person}{MICHAEL LEVANDOWSKY} {and}
  \bibinfo{person}{DAVID WINTER}.} \bibinfo{year}{1971}\natexlab{}.
\newblock \showarticletitle{Distance Between Sets}.
\newblock \bibinfo{journal}{\emph{Nature}}  \bibinfo{volume}{234}
  (\bibinfo{date}{11} \bibinfo{year}{1971}).
\newblock
\urldef\tempurl%
\url{https://doi.org/10.1038/234034a0}
\showDOI{\tempurl}


\bibitem[Li et~al\mbox{.}(2014)]%
        {LI201447}
\bibfield{author}{\bibinfo{person}{Qian Li}, \bibinfo{person}{Tao Zhou},
  \bibinfo{person}{Linyuan Lü}, {and} \bibinfo{person}{Duanbing Chen}.}
  \bibinfo{year}{2014}\natexlab{}.
\newblock \showarticletitle{Identifying influential spreaders by weighted
  LeaderRank}.
\newblock \bibinfo{journal}{\emph{Physica A: Statistical Mechanics and its
  Applications}}  \bibinfo{volume}{404} (\bibinfo{year}{2014}),
  \bibinfo{pages}{47--55}.
\newblock
\showISSN{0378-4371}
\urldef\tempurl%
\url{https://doi.org/10.1016/j.physa.2014.02.041}
\showDOI{\tempurl}


\bibitem[Li et~al\mbox{.}(2018a)]%
        {LiYuchenandFan2018}
\bibfield{author}{\bibinfo{person}{Yuchen Li}, \bibinfo{person}{Ju Fan},
  \bibinfo{person}{Yanhao Wang}, {and} \bibinfo{person}{Kian-Lee Tan}.}
  \bibinfo{year}{2018}\natexlab{a}.
\newblock \showarticletitle{Influence Maximization on Social Graphs: A Survey}.
\newblock \bibinfo{journal}{\emph{IEEE Transactions on Knowledge and Data
  Engineering}} \bibinfo{volume}{30}, \bibinfo{number}{10}
  (\bibinfo{year}{2018}), \bibinfo{pages}{1852--1872}.
\newblock
\urldef\tempurl%
\url{https://doi.org/10.1109/TKDE.2018.2807843}
\showDOI{\tempurl}


\bibitem[Li et~al\mbox{.}(2018b)]%
        {LiYuchenandFan}
\bibfield{author}{\bibinfo{person}{Yuchen Li}, \bibinfo{person}{Ju Fan},
  \bibinfo{person}{Yanhao Wang}, {and} \bibinfo{person}{Kian-Lee Tan}.}
  \bibinfo{year}{2018}\natexlab{b}.
\newblock \showarticletitle{Influence Maximization on Social Graphs: A Survey}.
\newblock \bibinfo{journal}{\emph{IEEE Transactions on Knowledge and Data
  Engineering}}  \bibinfo{volume}{30} (\bibinfo{date}{02}
  \bibinfo{year}{2018}), \bibinfo{pages}{1852--1872}.
\newblock
\urldef\tempurl%
\url{https://doi.org/10.1109/TKDE.2018.2807843}
\showDOI{\tempurl}


\bibitem[Lin et~al\mbox{.}(2014)]%
        {LIN20143279}
\bibfield{author}{\bibinfo{person}{Jian-Hong Lin}, \bibinfo{person}{Qiang Guo},
  \bibinfo{person}{Wen-Zhao Dong}, \bibinfo{person}{Li-Ying Tang}, {and}
  \bibinfo{person}{Jian-Guo Liu}.} \bibinfo{year}{2014}\natexlab{}.
\newblock \showarticletitle{Identifying the node spreading influence with
  largest k-core values}.
\newblock \bibinfo{journal}{\emph{Physics Letters A}} \bibinfo{volume}{378},
  \bibinfo{number}{45} (\bibinfo{year}{2014}), \bibinfo{pages}{3279--3284}.
\newblock
\showISSN{0375-9601}
\urldef\tempurl%
\url{https://doi.org/10.1016/j.physleta.2014.09.054}
\showDOI{\tempurl}


\bibitem[ling Ma et~al\mbox{.}(2016)]%
        {MA2016205}
\bibfield{author}{\bibinfo{person}{Ling ling Ma}, \bibinfo{person}{Chuang Ma},
  \bibinfo{person}{Hai-Feng Zhang}, {and} \bibinfo{person}{Bing-Hong Wang}.}
  \bibinfo{year}{2016}\natexlab{}.
\newblock \showarticletitle{Identifying influential spreaders in complex
  networks based on gravity formula}.
\newblock \bibinfo{journal}{\emph{Physica A: Statistical Mechanics and its
  Applications}}  \bibinfo{volume}{451} (\bibinfo{year}{2016}),
  \bibinfo{pages}{205--212}.
\newblock
\showISSN{0378-4371}
\urldef\tempurl%
\url{https://doi.org/10.1016/j.physa.2015.12.162}
\showDOI{\tempurl}


\bibitem[Liu et~al\mbox{.}(2019)]%
        {Computational_network_biology}
\bibfield{author}{\bibinfo{person}{Chuang Liu}, \bibinfo{person}{Yifang Ma},
  \bibinfo{person}{Jing Zhao}, \bibinfo{person}{Ruth Nussinov},
  \bibinfo{person}{Yi-Cheng Zhang}, \bibinfo{person}{Feixiong Cheng}, {and}
  \bibinfo{person}{Zi-Ke Zhang}.} \bibinfo{year}{2019}\natexlab{}.
\newblock \showarticletitle{Computational network biology: Data, model, and
  applications}.
\newblock \bibinfo{journal}{\emph{Physics Reports}}  \bibinfo{volume}{846}
  (\bibinfo{date}{12} \bibinfo{year}{2019}).
\newblock
\urldef\tempurl%
\url{https://doi.org/10.1016/j.physrep.2019.12.004}
\showDOI{\tempurl}


\bibitem[Liu et~al\mbox{.}(2018a)]%
        {LiuDongandNie201810}
\bibfield{author}{\bibinfo{person}{Dong Liu}, \bibinfo{person}{Hao Nie}, {and}
  \bibinfo{person}{Baowen Zhang}.} \bibinfo{year}{2018}\natexlab{a}.
\newblock \showarticletitle{A novel method for identifying influential nodes in
  complex networks based on multiple attributes}.
\newblock \bibinfo{journal}{\emph{International Journal of Modern Physics B}}
  \bibinfo{volume}{32} (\bibinfo{date}{10} \bibinfo{year}{2018}),
  \bibinfo{pages}{1850307}.
\newblock
\urldef\tempurl%
\url{https://doi.org/10.1142/S0217979218503071}
\showDOI{\tempurl}


\bibitem[Liu et~al\mbox{.}(2018b)]%
        {8529839}
\bibfield{author}{\bibinfo{person}{Fengzeng Liu}, \bibinfo{person}{Bing Xiao},
  \bibinfo{person}{Hao Li}, {and} \bibinfo{person}{Junjie Xue}.}
  \bibinfo{year}{2018}\natexlab{b}.
\newblock \showarticletitle{Complex Network Node Centrality Measurement Based
  on Multiple Attributes}. In \bibinfo{booktitle}{\emph{2018 10th International
  Conference on Modelling, Identification and Control (ICMIC)}}.
  \bibinfo{pages}{1--5}.
\newblock
\urldef\tempurl%
\url{https://doi.org/10.1109/ICMIC.2018.8529839}
\showDOI{\tempurl}


\bibitem[{Liu} et~al\mbox{.}(2021)]%
        {2021ScChE..64..451L}
\bibfield{author}{\bibinfo{person}{JiaQi {Liu}}, \bibinfo{person}{XueRong
  {Li}}, {and} \bibinfo{person}{JiChang {Dong}}.}
  \bibinfo{year}{2021}\natexlab{}.
\newblock \showarticletitle{{A survey on network node ranking algorithms:
  Representative methods, extensions, and applications}}.
\newblock \bibinfo{journal}{\emph{Science in China E: Technological Sciences}}
  \bibinfo{volume}{64}, \bibinfo{number}{3} (\bibinfo{date}{March}
  \bibinfo{year}{2021}), \bibinfo{pages}{451--461}.
\newblock
\urldef\tempurl%
\url{https://doi.org/10.1007/s11431-020-1683-2}
\showDOI{\tempurl}


\bibitem[Liu et~al\mbox{.}(2013)]%
        {LIU20134154}
\bibfield{author}{\bibinfo{person}{Jian-Guo Liu}, \bibinfo{person}{Zhuo-Ming
  Ren}, {and} \bibinfo{person}{Qiang Guo}.} \bibinfo{year}{2013}\natexlab{}.
\newblock \showarticletitle{Ranking the spreading influence in complex
  networks}.
\newblock \bibinfo{journal}{\emph{Physica A: Statistical Mechanics and its
  Applications}} \bibinfo{volume}{392}, \bibinfo{number}{18}
  (\bibinfo{year}{2013}), \bibinfo{pages}{4154--4159}.
\newblock
\showISSN{0378-4371}
\urldef\tempurl%
\url{https://doi.org/10.1016/j.physa.2013.04.037}
\showDOI{\tempurl}


\bibitem[Liu et~al\mbox{.}(2017a)]%
        {LiuYingandTangMingandDo2017}
\bibfield{author}{\bibinfo{person}{Ying Liu}, \bibinfo{person}{Ming Tang},
  \bibinfo{person}{Younghae Do}, {and} \bibinfo{person}{Pak~Ming Hui}.}
  \bibinfo{year}{2017}\natexlab{a}.
\newblock \showarticletitle{Accurate ranking of influential spreaders in
  networks based on dynamically asymmetric link-impact}.
\newblock \bibinfo{journal}{\emph{Physical Review E}}  \bibinfo{volume}{96}
  (\bibinfo{date}{05} \bibinfo{year}{2017}).
\newblock
\urldef\tempurl%
\url{https://doi.org/10.1103/PhysRevE.96.022323}
\showDOI{\tempurl}


\bibitem[Liu et~al\mbox{.}(2017b)]%
        {LiuYingandTang201705}
\bibfield{author}{\bibinfo{person}{Ying Liu}, \bibinfo{person}{Ming Tang},
  \bibinfo{person}{Younghae Do}, {and} \bibinfo{person}{Pak~Ming Hui}.}
  \bibinfo{year}{2017}\natexlab{b}.
\newblock \showarticletitle{Accurate ranking of influential spreaders in
  networks based on dynamically asymmetric link-impact}.
\newblock \bibinfo{journal}{\emph{Physical Review E}}  \bibinfo{volume}{96}
  (\bibinfo{date}{05} \bibinfo{year}{2017}).
\newblock
\urldef\tempurl%
\url{https://doi.org/10.1103/PhysRevE.96.022323}
\showDOI{\tempurl}


\bibitem[Liu et~al\mbox{.}(2015b)]%
        {spreadingdynamics201505}
\bibfield{author}{\bibinfo{person}{Ying Liu}, \bibinfo{person}{Ming Tang},
  \bibinfo{person}{Tao Zhou}, {and} \bibinfo{person}{Younghae Do}.}
  \bibinfo{year}{2015}\natexlab{b}.
\newblock \showarticletitle{Improving the accuracy of the k-shell method by
  removing redundant links: From a perspective of spreading dynamics}.
\newblock \bibinfo{journal}{\emph{Scientific reports}}  \bibinfo{volume}{5}
  (\bibinfo{date}{05} \bibinfo{year}{2015}).
\newblock
\urldef\tempurl%
\url{https://doi.org/10.1038/srep13172}
\showDOI{\tempurl}


\bibitem[Liu et~al\mbox{.}(2016)]%
        {LIU2016289}
\bibfield{author}{\bibinfo{person}{Ying Liu}, \bibinfo{person}{Ming Tang},
  \bibinfo{person}{Tao Zhou}, {and} \bibinfo{person}{Younghae Do}.}
  \bibinfo{year}{2016}\natexlab{}.
\newblock \showarticletitle{Identify influential spreaders in complex networks,
  the role of neighborhood}.
\newblock \bibinfo{journal}{\emph{Physica A: Statistical Mechanics and its
  Applications}}  \bibinfo{volume}{452} (\bibinfo{year}{2016}),
  \bibinfo{pages}{289--298}.
\newblock
\showISSN{0378-4371}
\urldef\tempurl%
\url{https://doi.org/10.1016/j.physa.2016.02.028}
\showDOI{\tempurl}


\bibitem[Liu et~al\mbox{.}(2015a)]%
        {LiuZhonghuaandJiangChengandWang201504}
\bibfield{author}{\bibinfo{person}{Zhonghua Liu}, \bibinfo{person}{Cheng
  Jiang}, \bibinfo{person}{Juyun Wang}, \bibinfo{person}{Huihui Zhang}, {and}
  \bibinfo{person}{Hua Yu}.} \bibinfo{year}{2015}\natexlab{a}.
\newblock \showarticletitle{The node importance in actual complex networks
  based on a multi-attribute ranking method}.
\newblock \bibinfo{journal}{\emph{Knowledge-Based Systems}}
  \bibinfo{volume}{84} (\bibinfo{date}{04} \bibinfo{year}{2015}).
\newblock
\urldef\tempurl%
\url{https://doi.org/10.1016/j.knosys.2015.03.026}
\showDOI{\tempurl}


\bibitem[Liu and Tian(2020)]%
        {LiuZuhanandTian2020}
\bibfield{author}{\bibinfo{person}{Zuhan Liu} {and} \bibinfo{person}{Canrong
  Tian}.} \bibinfo{year}{2020}\natexlab{}.
\newblock \showarticletitle{A weighted networked SIRS epidemic model}.
\newblock \bibinfo{journal}{\emph{Journal of Differential Equations}}
  \bibinfo{volume}{269} (\bibinfo{date}{12} \bibinfo{year}{2020}),
  \bibinfo{pages}{10995--11019}.
\newblock
\urldef\tempurl%
\url{https://doi.org/10.1016/j.jde.2020.07.038}
\showDOI{\tempurl}


\bibitem[Lv et~al\mbox{.}(2019)]%
        {LvZhiweiandZhao201902}
\bibfield{author}{\bibinfo{person}{Zhiwei Lv}, \bibinfo{person}{Nan Zhao},
  \bibinfo{person}{Fei Xiong}, {and} \bibinfo{person}{Nan Chen}.}
  \bibinfo{year}{2019}\natexlab{}.
\newblock \showarticletitle{A novel measure of identifying influential nodes in
  complex networks}.
\newblock \bibinfo{journal}{\emph{Physica A: Statistical Mechanics and its
  Applications}}  \bibinfo{volume}{523} (\bibinfo{date}{02}
  \bibinfo{year}{2019}).
\newblock
\urldef\tempurl%
\url{https://doi.org/10.1016/j.physa.2019.01.136}
\showDOI{\tempurl}


\bibitem[Lü et~al\mbox{.}(2016a)]%
        {LU20161}
\bibfield{author}{\bibinfo{person}{Linyuan Lü}, \bibinfo{person}{Duanbing
  Chen}, \bibinfo{person}{Xiao-Long Ren}, \bibinfo{person}{Qian-Ming Zhang},
  \bibinfo{person}{Yi-Cheng Zhang}, {and} \bibinfo{person}{Tao Zhou}.}
  \bibinfo{year}{2016}\natexlab{a}.
\newblock \showarticletitle{Vital nodes identification in complex networks}.
\newblock \bibinfo{journal}{\emph{Physics Reports}}  \bibinfo{volume}{650}
  (\bibinfo{year}{2016}), \bibinfo{pages}{1--63}.
\newblock
\showISSN{0370-1573}
\urldef\tempurl%
\url{https://doi.org/10.1016/j.physrep.2016.06.007}
\showDOI{\tempurl}
\newblock
\shownote{Vital nodes identification in complex networks}.


\bibitem[Lü et~al\mbox{.}(2011)]%
        {leaders201112}
\bibfield{author}{\bibinfo{person}{Linyuan Lü}, \bibinfo{person}{Yi-Cheng
  Zhang}, \bibinfo{person}{Chi~Ho Yeung}, {and} \bibinfo{person}{Tao Zhou}.}
  \bibinfo{year}{2011}\natexlab{}.
\newblock \showarticletitle{Leaders in Social Networks, the Delicious Case}.
\newblock \bibinfo{journal}{\emph{PloS one}}  \bibinfo{volume}{6}
  (\bibinfo{date}{12} \bibinfo{year}{2011}), \bibinfo{pages}{e21202}.
\newblock
\urldef\tempurl%
\url{https://doi.org/10.1371/journal.pone.0021202}
\showDOI{\tempurl}


\bibitem[Lü et~al\mbox{.}(2016b)]%
        {hindex201601}
\bibfield{author}{\bibinfo{person}{Linyuan Lü}, \bibinfo{person}{Tao Zhou},
  \bibinfo{person}{Qian-Ming Zhang}, {and} \bibinfo{person}{H. Stanley}.}
  \bibinfo{year}{2016}\natexlab{b}.
\newblock \showarticletitle{The H-index of a network node and its relation to
  degree and coreness}.
\newblock \bibinfo{journal}{\emph{Nature Communications}}  \bibinfo{volume}{7}
  (\bibinfo{date}{01} \bibinfo{year}{2016}), \bibinfo{pages}{10168}.
\newblock
\urldef\tempurl%
\url{https://doi.org/10.1038/ncomms10168}
\showDOI{\tempurl}


\bibitem[Ma and Ma(2016)]%
        {MaQianandMaJun201608}
\bibfield{author}{\bibinfo{person}{Qian Ma} {and} \bibinfo{person}{Jun Ma}.}
  \bibinfo{year}{2016}\natexlab{}.
\newblock \showarticletitle{Identifying and ranking influential spreaders in
  complex networks with consideration of spreading probability}.
\newblock \bibinfo{journal}{\emph{Physica A: Statistical Mechanics and its
  Applications}}  \bibinfo{volume}{465} (\bibinfo{date}{08}
  \bibinfo{year}{2016}).
\newblock
\urldef\tempurl%
\url{https://doi.org/10.1016/j.physa.2016.08.041}
\showDOI{\tempurl}


\bibitem[Maji et~al\mbox{.}(2020)]%
        {MAJI2020113681}
\bibfield{author}{\bibinfo{person}{Giridhar Maji}, \bibinfo{person}{Sharmistha
  Mandal}, {and} \bibinfo{person}{Soumya Sen}.}
  \bibinfo{year}{2020}\natexlab{}.
\newblock \showarticletitle{A systematic survey on influential spreaders
  identification in complex networks with a focus on K-shell based techniques}.
\newblock \bibinfo{journal}{\emph{Expert Systems with Applications}}
  \bibinfo{volume}{161} (\bibinfo{year}{2020}), \bibinfo{pages}{113681}.
\newblock
\showISSN{0957-4174}
\urldef\tempurl%
\url{https://doi.org/10.1016/j.eswa.2020.113681}
\showDOI{\tempurl}


\bibitem[Mekonnen et~al\mbox{.}(2020)]%
        {Mekonnen202003}
\bibfield{author}{\bibinfo{person}{Muluneh Mekonnen}, \bibinfo{person}{Sultan
  Feisso}, \bibinfo{person}{Hou Ronghui}, {and} \bibinfo{person}{Talha
  Younas}.} \bibinfo{year}{2020}\natexlab{}.
\newblock \showarticletitle{CSE: A Content Spreading Efficiency Based
  Influential Nodes Selection Method in 5G Mobile Social Networks}.
  \bibinfo{pages}{475--479}.
\newblock
\urldef\tempurl%
\url{https://doi.org/10.1109/ICICT50521.2020.00082}
\showDOI{\tempurl}


\bibitem[Mohapatra et~al\mbox{.}(2018)]%
        {MohapatraDebasisandPradhan201812}
\bibfield{author}{\bibinfo{person}{Debasis Mohapatra},
  \bibinfo{person}{Soubhagya Pradhan}, \bibinfo{person}{Hahnemann Lenka},
  \bibinfo{person}{Rojalini Tripathy}, \bibinfo{person}{Anjana Panda}, {and}
  \bibinfo{person}{Monalisa Sethy}.} \bibinfo{year}{2018}\natexlab{}.
\newblock \showarticletitle{Establishing Correlation Between Structural and
  Spectral Property in K-Shell Structure}. \bibinfo{pages}{96--100}.
\newblock
\urldef\tempurl%
\url{https://doi.org/10.1109/ICIT.2018.00030}
\showDOI{\tempurl}


\bibitem[Moler(1967)]%
        {Moler_Cleve}
\bibfield{author}{\bibinfo{person}{Cleve Moler}.}
  \bibinfo{year}{1967}\natexlab{}.
\newblock \showarticletitle{Iterative Refinement in Floating Point}.
\newblock \bibinfo{journal}{\emph{J. ACM}}  \bibinfo{volume}{14}
  (\bibinfo{date}{04} \bibinfo{year}{1967}), \bibinfo{pages}{316--321}.
\newblock
\urldef\tempurl%
\url{https://doi.org/10.1145/321386.321394}
\showDOI{\tempurl}


\bibitem[Motter and Lai(2003)]%
        {MotterAdilsonandLai200301}
\bibfield{author}{\bibinfo{person}{Adilson Motter} {and}
  \bibinfo{person}{Ying-Cheng Lai}.} \bibinfo{year}{2003}\natexlab{}.
\newblock \showarticletitle{Cascade-based Attacks on Complex Networks}.
\newblock \bibinfo{journal}{\emph{Physical review. E, Statistical, nonlinear,
  and soft matter physics}}  \bibinfo{volume}{66} (\bibinfo{date}{01}
  \bibinfo{year}{2003}), \bibinfo{pages}{065102}.
\newblock
\urldef\tempurl%
\url{https://doi.org/10.1103/PhysRevE.66.065102}
\showDOI{\tempurl}


\bibitem[Nath et~al\mbox{.}(2017)]%
        {NathDilipandDas201704}
\bibfield{author}{\bibinfo{person}{Dilip Nath}, \bibinfo{person}{Kishore Das},
  {and} \bibinfo{person}{Tandrima Chakraborty}.}
  \bibinfo{year}{2017}\natexlab{}.
\newblock \showarticletitle{A Modified Epidemic Chain Binomial Model (MECBM)
  and Its 2,3-Introductory Probabilities}.
\newblock \bibinfo{journal}{\emph{Open Journal of Statistics}}
  \bibinfo{volume}{7} (\bibinfo{date}{04} \bibinfo{year}{2017}),
  \bibinfo{pages}{225--239}.
\newblock
\urldef\tempurl%
\url{https://doi.org/10.4236/ojs.2017.72018}
\showDOI{\tempurl}


\bibitem[Newman(2003)]%
        {doi10.1137S003614450342480}
\bibfield{author}{\bibinfo{person}{M.~E.~J. Newman}.}
  \bibinfo{year}{2003}\natexlab{}.
\newblock \showarticletitle{The Structure and Function of Complex Networks}.
\newblock \bibinfo{journal}{\emph{SIAM Rev.}} \bibinfo{volume}{45},
  \bibinfo{number}{2} (\bibinfo{year}{2003}), \bibinfo{pages}{167--256}.
\newblock
\urldef\tempurl%
\url{https://doi.org/10.1137/S003614450342480}
\showDOI{\tempurl}
\showeprint{https://doi.org/10.1137/S003614450342480}


\bibitem[Niu et~al\mbox{.}(2015)]%
        {khop201502}
\bibfield{author}{\bibinfo{person}{Jianwei Niu}, \bibinfo{person}{Jinyang Fan},
  \bibinfo{person}{Lei Wang}, {and} \bibinfo{person}{Milica Stojinenovic}.}
  \bibinfo{year}{2015}\natexlab{}.
\newblock \showarticletitle{K-hop centrality metric for identifying influential
  spreaders in dynamic large-scale social networks}.
\newblock \bibinfo{journal}{\emph{2014 IEEE Global Communications Conference,
  GLOBECOM 2014}} (\bibinfo{date}{02} \bibinfo{year}{2015}),
  \bibinfo{pages}{2954--2959}.
\newblock
\urldef\tempurl%
\url{https://doi.org/10.1109/GLOCOM.2014.7037257}
\showDOI{\tempurl}


\bibitem[Page(1999)]%
        {PageL199901}
\bibfield{author}{\bibinfo{person}{L. Page}.} \bibinfo{year}{1999}\natexlab{}.
\newblock \showarticletitle{The pagerank citation ranking: Bringing order to
  the web}.
\newblock \bibinfo{journal}{\emph{Stanford InfoLab}} (\bibinfo{date}{01}
  \bibinfo{year}{1999}), \bibinfo{pages}{1--14}.
\newblock


\bibitem[Pei et~al\mbox{.}(2014)]%
        {Searching_for_superspreaders}
\bibfield{author}{\bibinfo{person}{Sen Pei}, \bibinfo{person}{Lev Muchnik},
  \bibinfo{person}{Jose Jr}, \bibinfo{person}{Zhiming Zheng}, {and}
  \bibinfo{person}{Hernan Makse}.} \bibinfo{year}{2014}\natexlab{}.
\newblock \showarticletitle{Searching for superspreaders of information in
  real-world social media}.
\newblock \bibinfo{journal}{\emph{Scientific Reports}}  \bibinfo{volume}{4}
  (\bibinfo{date}{05} \bibinfo{year}{2014}), \bibinfo{pages}{5547}.
\newblock
\urldef\tempurl%
\url{https://doi.org/10.1038/srep05547}
\showDOI{\tempurl}


\bibitem[Piccardi(2011)]%
        {Piccardi201111}
\bibfield{author}{\bibinfo{person}{Carlo Piccardi}.}
  \bibinfo{year}{2011}\natexlab{}.
\newblock \showarticletitle{Finding and Testing Network Communities by Lumped
  Markov Chains}.
\newblock \bibinfo{journal}{\emph{PloS one}}  \bibinfo{volume}{6}
  (\bibinfo{date}{11} \bibinfo{year}{2011}), \bibinfo{pages}{e27028}.
\newblock
\urldef\tempurl%
\url{https://doi.org/10.1371/journal.pone.0027028}
\showDOI{\tempurl}


\bibitem[Pu et~al\mbox{.}(2014)]%
        {PuJunandChenXiaowu201407}
\bibfield{author}{\bibinfo{person}{Jun Pu}, \bibinfo{person}{Xiaowu Chen},
  \bibinfo{person}{Daijun Wei}, \bibinfo{person}{qi Liu}, {and}
  \bibinfo{person}{Yong Deng}.} \bibinfo{year}{2014}\natexlab{}.
\newblock \showarticletitle{Identifying influential nodes based on local
  dimension}.
\newblock \bibinfo{journal}{\emph{EPL (Europhysics Letters)}}
  \bibinfo{volume}{107} (\bibinfo{date}{07} \bibinfo{year}{2014}),
  \bibinfo{pages}{10010}.
\newblock
\urldef\tempurl%
\url{https://doi.org/10.1209/0295-5075/107/10010}
\showDOI{\tempurl}


\bibitem[Qing-Cheng et~al\mbox{.}(2013)]%
        {QingChengHuanfYanShen201301}
\bibfield{author}{\bibinfo{person}{Hu Qing-Cheng}, \bibinfo{person}{Yin
  Yan-Shen}, \bibinfo{person}{Ma Peng-Fei}, \bibinfo{person}{Gao Yang},
  \bibinfo{person}{Zhang Yong}, {and} \bibinfo{person}{Xing Chun-Xiao}.}
  \bibinfo{year}{2013}\natexlab{}.
\newblock \showarticletitle{A new approach to identify influential spreaders in
  complex networks}.
\newblock \bibinfo{journal}{\emph{Acta Physica Sinica}}  \bibinfo{volume}{62}
  (\bibinfo{date}{01} \bibinfo{year}{2013}), \bibinfo{pages}{140101}.
\newblock
\urldef\tempurl%
\url{https://doi.org/10.7498/aps.62.140101}
\showDOI{\tempurl}


\bibitem[Qiu et~al\mbox{.}(2021)]%
        {QiuLiqing202107}
\bibfield{author}{\bibinfo{person}{Liqing Qiu}, \bibinfo{person}{Jianyi Zhang},
  {and} \bibinfo{person}{Xiangbo Tian}.} \bibinfo{year}{2021}\natexlab{}.
\newblock \showarticletitle{Ranking influential nodes in complex networks based
  on local and global structures}.
\newblock \bibinfo{journal}{\emph{Applied Intelligence}}  \bibinfo{volume}{51}
  (\bibinfo{date}{07} \bibinfo{year}{2021}), \bibinfo{pages}{1--14}.
\newblock
\urldef\tempurl%
\url{https://doi.org/10.1007/s10489-020-02132-1}
\showDOI{\tempurl}


\bibitem[Rak and Rak(2020)]%
        {RakRafalandRak202004}
\bibfield{author}{\bibinfo{person}{Rafal Rak} {and} \bibinfo{person}{Ewa Rak}.}
  \bibinfo{year}{2020}\natexlab{}.
\newblock \showarticletitle{The Fractional Preferential Attachment Scale-Free
  Network Model}.
\newblock \bibinfo{journal}{\emph{Entropy}}  \bibinfo{volume}{22}
  (\bibinfo{date}{04} \bibinfo{year}{2020}), \bibinfo{pages}{509}.
\newblock
\urldef\tempurl%
\url{https://doi.org/10.3390/e22050509}
\showDOI{\tempurl}


\bibitem[Ren et~al\mbox{.}(2014)]%
        {Ren_2014}
\bibfield{author}{\bibinfo{person}{Zhuo-Ming Ren}, \bibinfo{person}{An Zeng},
  \bibinfo{person}{Duan-Bing Chen}, \bibinfo{person}{Hao Liao}, {and}
  \bibinfo{person}{Jian-Guo Liu}.} \bibinfo{year}{2014}\natexlab{}.
\newblock \showarticletitle{Iterative resource allocation for ranking spreaders
  in complex networks}.
\newblock \bibinfo{journal}{\emph{{EPL} (Europhysics Letters)}}
  \bibinfo{volume}{106}, \bibinfo{number}{4} (\bibinfo{date}{may}
  \bibinfo{year}{2014}), \bibinfo{pages}{48005}.
\newblock
\urldef\tempurl%
\url{https://doi.org/10.1209/0295-5075/106/48005}
\showDOI{\tempurl}


\bibitem[Sabidussi(1966)]%
        {Sabidussi196602}
\bibfield{author}{\bibinfo{person}{Gert Sabidussi}.}
  \bibinfo{year}{1966}\natexlab{}.
\newblock \showarticletitle{The Centrality Index of a Graph}.
\newblock \bibinfo{journal}{\emph{Psychometrika}}  \bibinfo{volume}{31}
  (\bibinfo{date}{02} \bibinfo{year}{1966}), \bibinfo{pages}{581--603}.
\newblock
\urldef\tempurl%
\url{https://doi.org/10.1007/BF02289527}
\showDOI{\tempurl}


\bibitem[Sehgal et~al\mbox{.}(2009)]%
        {artiSehgal_Umesh_andcle}
\bibfield{author}{\bibinfo{person}{Umesh Sehgal}, \bibinfo{person}{Kuljeet
  Kaur}, {and} \bibinfo{person}{Pawan Kumar}.} \bibinfo{year}{2009}\natexlab{}.
\newblock \showarticletitle{The Anatomy of a Large-Scale Hyper Textual Web
  Search Engine}.
\newblock \bibinfo{journal}{\emph{Computer and Electrical Engineering,
  International Conference on}}  \bibinfo{volume}{2} (\bibinfo{date}{12}
  \bibinfo{year}{2009}), \bibinfo{pages}{491--495}.
\newblock
\showISBNx{978-0-7695-3925-6}
\urldef\tempurl%
\url{https://doi.org/10.1109/ICCEE.2009.59}
\showDOI{\tempurl}


\bibitem[Shao et~al\mbox{.}(2019)]%
        {ShaoZengzhenandLiu201911}
\bibfield{author}{\bibinfo{person}{Zengzhen Shao}, \bibinfo{person}{Shulei
  Liu}, \bibinfo{person}{Yanyu Zhao}, {and} \bibinfo{person}{Yanxiu Liu}.}
  \bibinfo{year}{2019}\natexlab{}.
\newblock \showarticletitle{Identifying influential nodes in complex networks
  based on Neighbours and edges}.
\newblock \bibinfo{journal}{\emph{Peer-to-Peer Networking and Applications}}
  \bibinfo{volume}{12} (\bibinfo{date}{11} \bibinfo{year}{2019}).
\newblock
\urldef\tempurl%
\url{https://doi.org/10.1007/s12083-018-0681-x}
\showDOI{\tempurl}


\bibitem[Sheikhahmadi et~al\mbox{.}(2017)]%
        {SheikhahmadiAmir2017}
\bibfield{author}{\bibinfo{person}{Amir Sheikhahmadi},
  \bibinfo{person}{Mohammad Nematbakhsh}, {and} \bibinfo{person}{Ahmad
  Zareie}.} \bibinfo{year}{2017}\natexlab{}.
\newblock \showarticletitle{Identification of influential users by neighbors in
  online social networks}.
\newblock \bibinfo{journal}{\emph{Physica A: Statistical Mechanics and its
  Applications}}  \bibinfo{volume}{486} (\bibinfo{date}{06}
  \bibinfo{year}{2017}).
\newblock
\urldef\tempurl%
\url{https://doi.org/10.1016/j.physa.2017.05.098}
\showDOI{\tempurl}


\bibitem[Sheng et~al\mbox{.}(2019)]%
        {ShengJinfang201910}
\bibfield{author}{\bibinfo{person}{Jinfang Sheng}, \bibinfo{person}{Jinying
  Dai}, \bibinfo{person}{Bin Wang}, \bibinfo{person}{Guihua Duan},
  \bibinfo{person}{Jun Long}, \bibinfo{person}{Junkai Zhang},
  \bibinfo{person}{Kerong Guan}, \bibinfo{person}{Sheng Hu},
  \bibinfo{person}{Long Chen}, {and} \bibinfo{person}{Wanghao Guan}.}
  \bibinfo{year}{2019}\natexlab{}.
\newblock \showarticletitle{Identifying influential nodes in complex networks
  based on global and local structure}.
\newblock \bibinfo{journal}{\emph{Physica A: Statistical Mechanics and its
  Applications}}  \bibinfo{volume}{541} (\bibinfo{date}{10}
  \bibinfo{year}{2019}), \bibinfo{pages}{123262}.
\newblock
\urldef\tempurl%
\url{https://doi.org/10.1016/j.physa.2019.123262}
\showDOI{\tempurl}


\bibitem[Sun et~al\mbox{.}(2021)]%
        {SunHangandShengYuhong2021}
\bibfield{author}{\bibinfo{person}{Hang Sun}, \bibinfo{person}{Yuhong Sheng},
  {and} \bibinfo{person}{Qing Cui}.} \bibinfo{year}{2021}\natexlab{}.
\newblock \showarticletitle{An uncertain SIR rumor spreading model}.
\newblock \bibinfo{journal}{\emph{Advances in Difference Equations}}
  \bibinfo{volume}{2021} (\bibinfo{date}{06} \bibinfo{year}{2021}).
\newblock
\urldef\tempurl%
\url{https://doi.org/10.1186/s13662-021-03386-w}
\showDOI{\tempurl}


\bibitem[Tulu et~al\mbox{.}(2017)]%
        {8322554}
\bibfield{author}{\bibinfo{person}{Muluneh~Mekonnen Tulu},
  \bibinfo{person}{Ronghui Hou}, {and} \bibinfo{person}{Talha Younas}.}
  \bibinfo{year}{2017}\natexlab{}.
\newblock \showarticletitle{Finding important nodes based on community
  structure and degree of neighbor nodes to disseminate information in complex
  networks}. In \bibinfo{booktitle}{\emph{2017 3rd IEEE International
  Conference on Computer and Communications (ICCC)}}.
  \bibinfo{pages}{269--273}.
\newblock
\urldef\tempurl%
\url{https://doi.org/10.1109/CompComm.2017.8322554}
\showDOI{\tempurl}


\bibitem[Tulu et~al\mbox{.}(2018)]%
        {8259501}
\bibfield{author}{\bibinfo{person}{Muluneh~Mekonnen Tulu},
  \bibinfo{person}{Ronghui Hou}, {and} \bibinfo{person}{Talha Younas}.}
  \bibinfo{year}{2018}\natexlab{}.
\newblock \showarticletitle{Identifying Influential Nodes Based on Community
  Structure to Speed up the Dissemination of Information in Complex Network}.
\newblock \bibinfo{journal}{\emph{IEEE Access}}  \bibinfo{volume}{6}
  (\bibinfo{year}{2018}), \bibinfo{pages}{7390--7401}.
\newblock
\showISSN{2169-3536}
\urldef\tempurl%
\url{https://doi.org/10.1109/ACCESS.2018.2794324}
\showDOI{\tempurl}


\bibitem[Tulu et~al\mbox{.}(2020)]%
        {TULU2020102768}
\bibfield{author}{\bibinfo{person}{Muluneh~Mekonnen Tulu},
  \bibinfo{person}{Mbazingwa~E. Mkiramweni}, \bibinfo{person}{Ronghui Hou},
  \bibinfo{person}{Sultan Feisso}, {and} \bibinfo{person}{Talha Younas}.}
  \bibinfo{year}{2020}\natexlab{}.
\newblock \showarticletitle{Influential nodes selection to enhance data
  dissemination in mobile social networks: A survey}.
\newblock \bibinfo{journal}{\emph{Journal of Network and Computer
  Applications}}  \bibinfo{volume}{169} (\bibinfo{year}{2020}),
  \bibinfo{pages}{102768}.
\newblock
\showISSN{1084-8045}
\urldef\tempurl%
\url{https://doi.org/10.1016/j.jnca.2020.102768}
\showDOI{\tempurl}


\bibitem[van~den Berg et~al\mbox{.}(2006)]%
        {Centering_scaling_and_transformations}
\bibfield{author}{\bibinfo{person}{Robert van~den Berg}, \bibinfo{person}{Huub
  Hoefsloot}, \bibinfo{person}{Johan Westerhuis}, \bibinfo{person}{Age Smilde},
  {and} \bibinfo{person}{Mariet van~der Werf}.}
  \bibinfo{year}{2006}\natexlab{}.
\newblock \showarticletitle{Centering, scaling, and transformations: improving
  the biological information content of metabolomics data. BMC Genomics 7:
  142-157}.
\newblock \bibinfo{journal}{\emph{BMC genomics}}  \bibinfo{volume}{7}
  (\bibinfo{date}{02} \bibinfo{year}{2006}), \bibinfo{pages}{142}.
\newblock
\urldef\tempurl%
\url{https://doi.org/10.1186/1471-2164-7-142}
\showDOI{\tempurl}


\bibitem[Wang et~al\mbox{.}(2017a)]%
        {WangJunyiandHou201702}
\bibfield{author}{\bibinfo{person}{Junyi Wang}, \bibinfo{person}{Xiaoni Hou},
  \bibinfo{person}{Kezan Li}, {and} \bibinfo{person}{Yong Ding}.}
  \bibinfo{year}{2017}\natexlab{a}.
\newblock \showarticletitle{A novel weight neighborhood centrality algorithm
  for identifying influential spreaders in complex networks}.
\newblock \bibinfo{journal}{\emph{Physica A: Statistical Mechanics and its
  Applications}}  \bibinfo{volume}{475} (\bibinfo{date}{02}
  \bibinfo{year}{2017}).
\newblock
\urldef\tempurl%
\url{https://doi.org/10.1016/j.physa.2017.02.007}
\showDOI{\tempurl}


\bibitem[Wang et~al\mbox{.}(2016a)]%
        {WangShashaandDu201611}
\bibfield{author}{\bibinfo{person}{Shasha Wang}, \bibinfo{person}{Yuxian Du},
  {and} \bibinfo{person}{Yong Deng}.} \bibinfo{year}{2016}\natexlab{a}.
\newblock \showarticletitle{A new measure of identifying influential nodes:
  Efficiency centrality}.
\newblock \bibinfo{journal}{\emph{Communications in Nonlinear Science and
  Numerical Simulation}}  \bibinfo{volume}{47} (\bibinfo{date}{11}
  \bibinfo{year}{2016}).
\newblock
\urldef\tempurl%
\url{https://doi.org/10.1016/j.cnsns.2016.11.008}
\showDOI{\tempurl}


\bibitem[Wang et~al\mbox{.}(2016b)]%
        {WangXiaojieandSu2016}
\bibfield{author}{\bibinfo{person}{Xiaojie Wang}, \bibinfo{person}{Yanyuan Su},
  \bibinfo{person}{Chengli Zhao}, {and} \bibinfo{person}{Dongyun Yi}.}
  \bibinfo{year}{2016}\natexlab{b}.
\newblock \showarticletitle{Effective identification of multiple influential
  spreaders by DegreePunishment}.
\newblock \bibinfo{journal}{\emph{Physica A: Statistical Mechanics and its
  Applications}}  \bibinfo{volume}{461} (\bibinfo{date}{05}
  \bibinfo{year}{2016}).
\newblock
\urldef\tempurl%
\url{https://doi.org/10.1016/j.physa.2016.05.020}
\showDOI{\tempurl}


\bibitem[Wang et~al\mbox{.}(2017b)]%
        {7978752}
\bibfield{author}{\bibinfo{person}{Zhiqiang Wang}, \bibinfo{person}{Xubin Pei},
  \bibinfo{person}{Yanbo Wang}, {and} \bibinfo{person}{Yiyang Yao}.}
  \bibinfo{year}{2017}\natexlab{b}.
\newblock \showarticletitle{Ranking the key nodes with temporal degree
  deviation centrality on complex networks}. In \bibinfo{booktitle}{\emph{2017
  29th Chinese Control And Decision Conference (CCDC)}}.
  \bibinfo{pages}{1484--1489}.
\newblock
\showISSN{1948-9447}
\urldef\tempurl%
\url{https://doi.org/10.1109/CCDC.2017.7978752}
\showDOI{\tempurl}


\bibitem[Watts and Strogatz(1998)]%
        {WattsDuncan201112}
\bibfield{author}{\bibinfo{person}{Duncan~J. Watts} {and}
  \bibinfo{person}{Steven~H. Strogatz}.} \bibinfo{year}{1998}\natexlab{}.
\newblock \showarticletitle{Collective dynamics of `small-world' networks}.
\newblock \bibinfo{journal}{\emph{Nature}} \bibinfo{volume}{393},
  \bibinfo{number}{6684} (\bibinfo{date}{01 Jun} \bibinfo{year}{1998}),
  \bibinfo{pages}{440--442}.
\newblock
\showISSN{1476-4687}
\urldef\tempurl%
\url{https://doi.org/10.1038/30918}
\showDOI{\tempurl}


\bibitem[Wei et~al\mbox{.}(2015)]%
        {WEI2015277}
\bibfield{author}{\bibinfo{person}{Bo Wei}, \bibinfo{person}{Jie Liu},
  \bibinfo{person}{Daijun Wei}, \bibinfo{person}{Cai Gao}, {and}
  \bibinfo{person}{Yong Deng}.} \bibinfo{year}{2015}\natexlab{}.
\newblock \showarticletitle{Weighted k-shell decomposition for complex networks
  based on potential edge weights}.
\newblock \bibinfo{journal}{\emph{Physica A: Statistical Mechanics and its
  Applications}}  \bibinfo{volume}{420} (\bibinfo{year}{2015}),
  \bibinfo{pages}{277--283}.
\newblock
\showISSN{0378-4371}
\urldef\tempurl%
\url{https://doi.org/10.1016/j.physa.2014.11.012}
\showDOI{\tempurl}


\bibitem[Wei et~al\mbox{.}(2013)]%
        {WEI20132564}
\bibfield{author}{\bibinfo{person}{Daijun Wei}, \bibinfo{person}{Xinyang Deng},
  \bibinfo{person}{Xiaoge Zhang}, \bibinfo{person}{Yong Deng}, {and}
  \bibinfo{person}{Sankaran Mahadevan}.} \bibinfo{year}{2013}\natexlab{}.
\newblock \showarticletitle{Identifying influential nodes in weighted networks
  based on evidence theory}.
\newblock \bibinfo{journal}{\emph{Physica A: Statistical Mechanics and its
  Applications}} \bibinfo{volume}{392}, \bibinfo{number}{10}
  (\bibinfo{year}{2013}), \bibinfo{pages}{2564--2575}.
\newblock
\showISSN{0378-4371}
\urldef\tempurl%
\url{https://doi.org/10.1016/j.physa.2013.01.054}
\showDOI{\tempurl}


\bibitem[Wen and Deng(2020)]%
        {WEN2020549}
\bibfield{author}{\bibinfo{person}{Tao Wen} {and} \bibinfo{person}{Yong Deng}.}
  \bibinfo{year}{2020}\natexlab{}.
\newblock \showarticletitle{Identification of influencers in complex networks
  by local information dimensionality}.
\newblock \bibinfo{journal}{\emph{Information Sciences}}  \bibinfo{volume}{512}
  (\bibinfo{year}{2020}), \bibinfo{pages}{549--562}.
\newblock
\showISSN{0020-0255}
\urldef\tempurl%
\url{https://doi.org/10.1016/j.ins.2019.10.003}
\showDOI{\tempurl}


\bibitem[Wen and Jiang(2019)]%
        {WenTaoandJiang201902}
\bibfield{author}{\bibinfo{person}{Tao Wen} {and} \bibinfo{person}{Wen Jiang}.}
  \bibinfo{year}{2019}\natexlab{}.
\newblock \showarticletitle{Identifying influential nodes based on fuzzy local
  dimension in complex networks}.
\newblock \bibinfo{journal}{\emph{Chaos, Solitons \& Fractals}}
  \bibinfo{volume}{119} (\bibinfo{date}{02} \bibinfo{year}{2019}),
  \bibinfo{pages}{332--342}.
\newblock
\urldef\tempurl%
\url{https://doi.org/10.1016/j.chaos.2019.01.011}
\showDOI{\tempurl}


\bibitem[Wu and Zhang(2016)]%
        {Epidemicthresholdofnodeweighted}
\bibfield{author}{\bibinfo{person}{Qingchu Wu} {and} \bibinfo{person}{Hai-Feng
  Zhang}.} \bibinfo{year}{2016}\natexlab{}.
\newblock \showarticletitle{Epidemic threshold of node-weighted
  susceptible-infected-susceptible models on networks}.
\newblock \bibinfo{journal}{\emph{Journal of Physics A: Mathematical and
  Theoretical}}  \bibinfo{volume}{49} (\bibinfo{date}{08}
  \bibinfo{year}{2016}), \bibinfo{pages}{345601}.
\newblock
\urldef\tempurl%
\url{https://doi.org/10.1088/1751-8113/49/34/345601}
\showDOI{\tempurl}


\bibitem[Xiong et~al\mbox{.}(2016)]%
        {7854364}
\bibfield{author}{\bibinfo{person}{Li Xiong}, \bibinfo{person}{Lu Zhao}, {and}
  \bibinfo{person}{Shan Xue}.} \bibinfo{year}{2016}\natexlab{}.
\newblock \showarticletitle{Node importance evaluation of world city networks:
  A survey}. In \bibinfo{booktitle}{\emph{2016 International Conference on
  Logistics, Informatics and Service Sciences (LISS)}}. \bibinfo{pages}{1--6}.
\newblock
\urldef\tempurl%
\url{https://doi.org/10.1109/LISS.2016.7854364}
\showDOI{\tempurl}


\bibitem[Yang et~al\mbox{.}(2018)]%
        {YangPingleandLiu201806}
\bibfield{author}{\bibinfo{person}{Pingle Yang}, \bibinfo{person}{Xin Liu},
  {and} \bibinfo{person}{Guiqiong Xu}.} \bibinfo{year}{2018}\natexlab{}.
\newblock \showarticletitle{A dynamic weighted TOPSIS method for identifying
  influential nodes in complex networks}.
\newblock \bibinfo{journal}{\emph{Modern Physics Letters B}}
  \bibinfo{volume}{32} (\bibinfo{date}{06} \bibinfo{year}{2018}),
  \bibinfo{pages}{1850216}.
\newblock
\urldef\tempurl%
\url{https://doi.org/10.1142/S0217984918502160}
\showDOI{\tempurl}


\bibitem[Yang et~al\mbox{.}(2017)]%
        {YangXandHuangDC}
\bibfield{author}{\bibinfo{person}{X. Yang}, \bibinfo{person}{D.-C Huang},
  {and} \bibinfo{person}{Z.-K Zhang}.} \bibinfo{year}{2017}\natexlab{}.
\newblock \showarticletitle{Neighborhood coreness algorithm for identifying a
  set of influential spreaders in complex networks}.
\newblock \bibinfo{journal}{\emph{KSII Transactions on Internet and Information
  Systems}}  \bibinfo{volume}{11} (\bibinfo{date}{06} \bibinfo{year}{2017}),
  \bibinfo{pages}{2979--2995}.
\newblock
\urldef\tempurl%
\url{https://doi.org/10.3837/tiis.2017.06.010}
\showDOI{\tempurl}


\bibitem[Yang et~al\mbox{.}(2019)]%
        {YangYuanzhi201911}
\bibfield{author}{\bibinfo{person}{Yuanzhi Yang}, \bibinfo{person}{Lei Yu},
  \bibinfo{person}{Xing Wang}, \bibinfo{person}{Siyi Chen},
  \bibinfo{person}{You Chen}, {and} \bibinfo{person}{Yipeng Zhou}.}
  \bibinfo{year}{2019}\natexlab{}.
\newblock \showarticletitle{A novel method to identify influential nodes in
  complex networks}.
\newblock \bibinfo{journal}{\emph{International Journal of Modern Physics C}}
  \bibinfo{volume}{31} (\bibinfo{date}{11} \bibinfo{year}{2019}).
\newblock
\urldef\tempurl%
\url{https://doi.org/10.1142/S0129183120500229}
\showDOI{\tempurl}


\bibitem[Yang et~al\mbox{.}(2020)]%
        {YangYuanzhi202005}
\bibfield{author}{\bibinfo{person}{Yuan-zhi Yang}, \bibinfo{person}{Min Hu},
  {and} \bibinfo{person}{Tai-yu Huang}.} \bibinfo{year}{2020}\natexlab{}.
\newblock \showarticletitle{Influential nodes identification in complex
  networks based on global and local information}.
\newblock \bibinfo{journal}{\emph{Chinese Physics B}}  \bibinfo{volume}{29}
  (\bibinfo{date}{05} \bibinfo{year}{2020}).
\newblock
\urldef\tempurl%
\url{https://doi.org/10.1088/1674-1056/ab969f}
\showDOI{\tempurl}


\bibitem[Yu et~al\mbox{.}(2020a)]%
        {YuEnyuandFuYanandTang202001}
\bibfield{author}{\bibinfo{person}{Enyu Yu}, \bibinfo{person}{Yan Fu},
  \bibinfo{person}{Qing Tang}, \bibinfo{person}{Jun-Yan Zhao}, {and}
  \bibinfo{person}{Duanbing Chen}.} \bibinfo{year}{2020}\natexlab{a}.
\newblock \showarticletitle{A Re-Ranking Algorithm for Identifying Influential
  Nodes in Complex Networks}.
\newblock \bibinfo{journal}{\emph{IEEE Access}}  \bibinfo{volume}{8}
  (\bibinfo{date}{01} \bibinfo{year}{2020}), \bibinfo{pages}{211281--211290}.
\newblock
\urldef\tempurl%
\url{https://doi.org/10.1109/ACCESS.2020.3038791}
\showDOI{\tempurl}


\bibitem[Yu et~al\mbox{.}(2020b)]%
        {YuEnYuandWang202004}
\bibfield{author}{\bibinfo{person}{En-Yu Yu}, \bibinfo{person}{Yue-Ping Wang},
  \bibinfo{person}{Yan Fu}, \bibinfo{person}{Duanbing Chen}, {and}
  \bibinfo{person}{Mei Xie}.} \bibinfo{year}{2020}\natexlab{b}.
\newblock \showarticletitle{Identifying critical nodes in complex networks via
  graph convolutional networks}.
\newblock \bibinfo{journal}{\emph{Knowledge-Based Systems}}
  \bibinfo{volume}{198} (\bibinfo{date}{04} \bibinfo{year}{2020}),
  \bibinfo{pages}{105893}.
\newblock
\urldef\tempurl%
\url{https://doi.org/10.1016/j.knosys.2020.105893}
\showDOI{\tempurl}


\bibitem[Yu et~al\mbox{.}(2017)]%
        {YuHuiandCao201705}
\bibfield{author}{\bibinfo{person}{Hui Yu}, \bibinfo{person}{Xi Cao},
  \bibinfo{person}{Zun Liu}, {and} \bibinfo{person}{Yongjun Li}.}
  \bibinfo{year}{2017}\natexlab{}.
\newblock \showarticletitle{Identifying key nodes based on improved structural
  holes in complex networks}.
\newblock \bibinfo{journal}{\emph{Physica A: Statistical Mechanics and its
  Applications}}  \bibinfo{volume}{486} (\bibinfo{date}{05}
  \bibinfo{year}{2017}).
\newblock
\urldef\tempurl%
\url{https://doi.org/10.1016/j.physa.2017.05.028}
\showDOI{\tempurl}


\bibitem[Zareie and Sheikhahmadi(2017)]%
        {Ahierarchicalapproachforinfluential}
\bibfield{author}{\bibinfo{person}{Ahmad Zareie} {and} \bibinfo{person}{Amir
  Sheikhahmadi}.} \bibinfo{year}{2017}\natexlab{}.
\newblock \showarticletitle{A hierarchical approach for influential node
  ranking in complex social networks}.
\newblock \bibinfo{journal}{\emph{Expert Systems with Applications}}
  \bibinfo{volume}{93} (\bibinfo{date}{10} \bibinfo{year}{2017}).
\newblock
\urldef\tempurl%
\url{https://doi.org/10.1016/j.eswa.2017.10.018}
\showDOI{\tempurl}


\bibitem[Zareie et~al\mbox{.}(2017)]%
        {ZAREIE2017485}
\bibfield{author}{\bibinfo{person}{Ahmad Zareie}, \bibinfo{person}{Amir
  Sheikhahmadi}, {and} \bibinfo{person}{Adel Fatemi}.}
  \bibinfo{year}{2017}\natexlab{}.
\newblock \showarticletitle{Influential nodes ranking in complex networks: An
  entropy-based approach}.
\newblock \bibinfo{journal}{\emph{Chaos, Solitons \& Fractals}}
  \bibinfo{volume}{104} (\bibinfo{year}{2017}), \bibinfo{pages}{485--494}.
\newblock
\showISSN{0960-0779}
\urldef\tempurl%
\url{https://doi.org/10.1016/j.chaos.2017.09.010}
\showDOI{\tempurl}


\bibitem[Zareie et~al\mbox{.}(2018)]%
        {ZareieAhmad201811}
\bibfield{author}{\bibinfo{person}{Ahmad Zareie}, \bibinfo{person}{Amir
  Sheikhahmadi}, {and} \bibinfo{person}{Mahdi Jalili}.}
  \bibinfo{year}{2018}\natexlab{}.
\newblock \showarticletitle{Influential node ranking in social networks based
  on neighborhood diversity}.
\newblock \bibinfo{journal}{\emph{Future Generation Computer Systems}}
  \bibinfo{volume}{94} (\bibinfo{date}{11} \bibinfo{year}{2018}).
\newblock
\urldef\tempurl%
\url{https://doi.org/10.1016/j.future.2018.11.023}
\showDOI{\tempurl}


\bibitem[Zeng and Zhang(2013)]%
        {ZENG20131031}
\bibfield{author}{\bibinfo{person}{An Zeng} {and} \bibinfo{person}{Cheng-Jun
  Zhang}.} \bibinfo{year}{2013}\natexlab{}.
\newblock \showarticletitle{Ranking spreaders by decomposing complex networks}.
\newblock \bibinfo{journal}{\emph{Physics Letters A}} \bibinfo{volume}{377},
  \bibinfo{number}{14} (\bibinfo{year}{2013}), \bibinfo{pages}{1031--1035}.
\newblock
\showISSN{0375-9601}
\urldef\tempurl%
\url{https://doi.org/10.1016/j.physleta.2013.02.039}
\showDOI{\tempurl}


\bibitem[Zhang et~al\mbox{.}(2019)]%
        {ZhangJunkai201910}
\bibfield{author}{\bibinfo{person}{Junkai Zhang}, \bibinfo{person}{Bin Wang},
  \bibinfo{person}{Jinfang Sheng}, \bibinfo{person}{Jinying Dai},
  \bibinfo{person}{Jie Hu}, {and} \bibinfo{person}{Long Chen}.}
  \bibinfo{year}{2019}\natexlab{}.
\newblock \showarticletitle{Identifying Influential Nodes in Complex Networks
  Based on Local Effective Distance}.
\newblock \bibinfo{journal}{\emph{Information}}  \bibinfo{volume}{10}
  (\bibinfo{date}{10} \bibinfo{year}{2019}), \bibinfo{pages}{311}.
\newblock
\urldef\tempurl%
\url{https://doi.org/10.3390/info10100311}
\showDOI{\tempurl}


\bibitem[Zhang et~al\mbox{.}(2016a)]%
        {ZhangJian-XiongandChen2016}
\bibfield{author}{\bibinfo{person}{Jian-Xiong Zhang}, \bibinfo{person}{Duanbing
  Chen}, \bibinfo{person}{Qiang Dong}, {and} \bibinfo{person}{Zhi-Dan Zhao}.}
  \bibinfo{year}{2016}\natexlab{a}.
\newblock \showarticletitle{Identifying a set of influential spreaders in
  complex networks}.
\newblock \bibinfo{journal}{\emph{Scientific Reports}}  \bibinfo{volume}{6}
  (\bibinfo{date}{01} \bibinfo{year}{2016}).
\newblock
\urldef\tempurl%
\url{https://doi.org/10.1038/srep27823}
\showDOI{\tempurl}


\bibitem[Zhang et~al\mbox{.}(2016b)]%
        {ZhangJianXiongandChenDuanbingandDong}
\bibfield{author}{\bibinfo{person}{Jian-Xiong Zhang}, \bibinfo{person}{Duanbing
  Chen}, \bibinfo{person}{Qiang Dong}, {and} \bibinfo{person}{Zhi-Dan Zhao}.}
  \bibinfo{year}{2016}\natexlab{b}.
\newblock \showarticletitle{Identifying a set of influential spreaders in
  complex networks}.
\newblock \bibinfo{journal}{\emph{Scientific Reports}}  \bibinfo{volume}{6}
  (\bibinfo{date}{01} \bibinfo{year}{2016}).
\newblock
\urldef\tempurl%
\url{https://doi.org/10.1038/srep27823}
\showDOI{\tempurl}


\bibitem[Zhang et~al\mbox{.}(2007)]%
        {ZhangPengandWang200709}
\bibfield{author}{\bibinfo{person}{Peng Zhang}, \bibinfo{person}{Jinliang
  Wang}, \bibinfo{person}{Xiaojia Li}, \bibinfo{person}{Zengru Di}, {and}
  \bibinfo{person}{Ying Fan}.} \bibinfo{year}{2007}\natexlab{}.
\newblock \showarticletitle{Clustering coefficient and community structure of
  bipartite networks}.
\newblock \bibinfo{journal}{\emph{Physica A: Statistical Mechanics and its
  Applications}}  \bibinfo{volume}{387} (\bibinfo{date}{09}
  \bibinfo{year}{2007}).
\newblock
\urldef\tempurl%
\url{https://doi.org/10.1016/j.physa.2008.09.006}
\showDOI{\tempurl}


\bibitem[Zhang et~al\mbox{.}(2016c)]%
        {ZhangRuishengandYang201608}
\bibfield{author}{\bibinfo{person}{Ruisheng Zhang}, \bibinfo{person}{Zhao
  Yang}, \bibinfo{person}{Rongjing Hu}, \bibinfo{person}{Yongna Yuan},
  \bibinfo{person}{Keqing Li}, {and} \bibinfo{person}{Mengtian Li}.}
  \bibinfo{year}{2016}\natexlab{c}.
\newblock \showarticletitle{Identifying the most influential spreaders in
  complex networks by an Extended Local K-Shell Sum}.
\newblock \bibinfo{journal}{\emph{International Journal of Modern Physics C}}
  \bibinfo{volume}{28} (\bibinfo{date}{08} \bibinfo{year}{2016}).
\newblock
\urldef\tempurl%
\url{https://doi.org/10.1142/S0129183117500140}
\showDOI{\tempurl}


\bibitem[Zhang et~al\mbox{.}(2020)]%
        {ZhangZufanandLiXieliang}
\bibfield{author}{\bibinfo{person}{Zufan Zhang}, \bibinfo{person}{Xieliang Li},
  {and} \bibinfo{person}{Chenquan Gan}.} \bibinfo{year}{2020}\natexlab{}.
\newblock \showarticletitle{Identifying influential nodes in social networks
  via community structure and influence distribution difference}.
\newblock \bibinfo{journal}{\emph{Digital Communications and Networks}}
  \bibinfo{volume}{7} (\bibinfo{date}{05} \bibinfo{year}{2020}).
\newblock
\urldef\tempurl%
\url{https://doi.org/10.1016/j.dcan.2020.04.011}
\showDOI{\tempurl}


\bibitem[Zhao et~al\mbox{.}(2020b)]%
        {ZhaoGouhengandJia202003}
\bibfield{author}{\bibinfo{person}{Gouheng Zhao}, \bibinfo{person}{Peng Jia},
  \bibinfo{person}{Cheng Huang}, \bibinfo{person}{Anmin Zhou}, {and}
  \bibinfo{person}{Yong Fang}.} \bibinfo{year}{2020}\natexlab{b}.
\newblock \showarticletitle{A Machine Learning Based Framework for Identifying
  Influential Nodes in Complex Networks}.
\newblock \bibinfo{journal}{\emph{IEEE Access}}  \bibinfo{volume}{PP}
  (\bibinfo{date}{03} \bibinfo{year}{2020}), \bibinfo{pages}{1--1}.
\newblock
\urldef\tempurl%
\url{https://doi.org/10.1109/ACCESS.2020.2984286}
\showDOI{\tempurl}


\bibitem[Zhao et~al\mbox{.}(2020c)]%
        {ZHAO202018}
\bibfield{author}{\bibinfo{person}{Gouheng Zhao}, \bibinfo{person}{Peng Jia},
  \bibinfo{person}{Anmin Zhou}, {and} \bibinfo{person}{Bing Zhang}.}
  \bibinfo{year}{2020}\natexlab{c}.
\newblock \showarticletitle{InfGCN: Identifying influential nodes in complex
  networks with graph convolutional networks}.
\newblock \bibinfo{journal}{\emph{Neurocomputing}}  \bibinfo{volume}{414}
  (\bibinfo{year}{2020}), \bibinfo{pages}{18--26}.
\newblock
\showISSN{0925-2312}
\urldef\tempurl%
\url{https://doi.org/10.1016/j.neucom.2020.07.028}
\showDOI{\tempurl}


\bibitem[Zhao et~al\mbox{.}(2020d)]%
        {ZhaoJieandSongYutong202003}
\bibfield{author}{\bibinfo{person}{Jie Zhao}, \bibinfo{person}{Yutong Song},
  {and} \bibinfo{person}{Yong Deng}.} \bibinfo{year}{2020}\natexlab{d}.
\newblock \showarticletitle{A Novel Model to Identify the Influential Nodes:
  Evidence Theory Centrality}.
\newblock \bibinfo{journal}{\emph{IEEE Access}}  \bibinfo{volume}{PP}
  (\bibinfo{date}{03} \bibinfo{year}{2020}), \bibinfo{pages}{1--1}.
\newblock
\urldef\tempurl%
\url{https://doi.org/10.1109/ACCESS.2020.2978142}
\showDOI{\tempurl}


\bibitem[Zhao et~al\mbox{.}(2020e)]%
        {ZhaoJieandSong202008}
\bibfield{author}{\bibinfo{person}{Jie Zhao}, \bibinfo{person}{Yutong Song},
  \bibinfo{person}{Fan Liu}, {and} \bibinfo{person}{Yong Deng}.}
  \bibinfo{year}{2020}\natexlab{e}.
\newblock \showarticletitle{The identification of influential nodes based on
  structure similarity}.
\newblock \bibinfo{journal}{\emph{Connection Science}}  \bibinfo{volume}{33}
  (\bibinfo{date}{08} \bibinfo{year}{2020}), \bibinfo{pages}{1--18}.
\newblock
\urldef\tempurl%
\url{https://doi.org/10.1080/09540091.2020.1806203}
\showDOI{\tempurl}


\bibitem[Zhao et~al\mbox{.}(2020f)]%
        {ZhaoJieandWang202004}
\bibfield{author}{\bibinfo{person}{Jie Zhao}, \bibinfo{person}{Yunchuan Wang},
  {and} \bibinfo{person}{Yong Deng}.} \bibinfo{year}{2020}\natexlab{f}.
\newblock \showarticletitle{Identifying influential nodes in complex networks
  from global perspective}.
\newblock \bibinfo{journal}{\emph{Chaos, Solitons \& Fractals}}
  \bibinfo{volume}{133} (\bibinfo{date}{04} \bibinfo{year}{2020}),
  \bibinfo{pages}{109637}.
\newblock
\urldef\tempurl%
\url{https://doi.org/10.1016/j.chaos.2020.109637}
\showDOI{\tempurl}


\bibitem[Zhao et~al\mbox{.}(2020a)]%
        {10.1155/2020/5903798}
\bibfield{author}{\bibinfo{person}{Nan Zhao}, \bibinfo{person}{Jingjing Bao},
  \bibinfo{person}{Nan Chen}, {and} \bibinfo{person}{Hongshu Chen}.}
  \bibinfo{year}{2020}\natexlab{a}.
\newblock \showarticletitle{Ranking Influential Nodes in Complex Networks with
  Information Entropy Method}.
\newblock \bibinfo{journal}{\emph{Complex.}}  \bibinfo{volume}{2020}
  (\bibinfo{date}{jan} \bibinfo{year}{2020}), \bibinfo{numpages}{15}~pages.
\newblock
\showISSN{1076-2787}
\urldef\tempurl%
\url{https://doi.org/10.1155/2020/5903798}
\showDOI{\tempurl}


\bibitem[Zhao et~al\mbox{.}(2014)]%
        {Zhao_Xiang-Yu_and_Huang}
\bibfield{author}{\bibinfo{person}{Xiang-Yu Zhao}, \bibinfo{person}{Bin Huang},
  \bibinfo{person}{Ming Tang}, \bibinfo{person}{Hai-Feng Zhang}, {and}
  \bibinfo{person}{Duanbing Chen}.} \bibinfo{year}{2014}\natexlab{}.
\newblock \showarticletitle{Identifying effective multiple spreaders by
  coloring complex networks}.
\newblock \bibinfo{journal}{\emph{EPL (Europhysics Letters)}}
  \bibinfo{volume}{108} (\bibinfo{date}{10} \bibinfo{year}{2014}).
\newblock
\urldef\tempurl%
\url{https://doi.org/10.1209/0295-5075/108/68005}
\showDOI{\tempurl}


\bibitem[Zhao et~al\mbox{.}(2015)]%
        {ZhaoZhiyingandWang201504}
\bibfield{author}{\bibinfo{person}{Zhiying Zhao}, \bibinfo{person}{Xiaofan
  Wang}, \bibinfo{person}{Wei Zhang}, {and} \bibinfo{person}{Zhiliang Zhu}.}
  \bibinfo{year}{2015}\natexlab{}.
\newblock \showarticletitle{A Community-Based Approach to Identifying
  Influential Spreaders}.
\newblock \bibinfo{journal}{\emph{Entropy}}  \bibinfo{volume}{17}
  (\bibinfo{date}{04} \bibinfo{year}{2015}), \bibinfo{pages}{2228--2252}.
\newblock
\urldef\tempurl%
\url{https://doi.org/10.3390/e17042228}
\showDOI{\tempurl}


\bibitem[Zhao et~al\mbox{.}(2019)]%
        {ZhaoZijuan201912}
\bibfield{author}{\bibinfo{person}{Zi-Juan Zhao}, \bibinfo{person}{Qiang Guo},
  \bibinfo{person}{Kai Yu}, {and} \bibinfo{person}{Jian-Guo Liu}.}
  \bibinfo{year}{2019}\natexlab{}.
\newblock \showarticletitle{Identifying influential nodes for the networks with
  community structure}.
\newblock \bibinfo{journal}{\emph{Physica A: Statistical Mechanics and its
  Applications}}  \bibinfo{volume}{551} (\bibinfo{date}{12}
  \bibinfo{year}{2019}), \bibinfo{pages}{123893}.
\newblock
\urldef\tempurl%
\url{https://doi.org/10.1016/j.physa.2019.123893}
\showDOI{\tempurl}


\bibitem[Zhong et~al\mbox{.}(2015)]%
        {ZHONG20152272}
\bibfield{author}{\bibinfo{person}{Lin-Feng Zhong}, \bibinfo{person}{Jian-Guo
  Liu}, {and} \bibinfo{person}{Ming-Sheng Shang}.}
  \bibinfo{year}{2015}\natexlab{}.
\newblock \showarticletitle{Iterative resource allocation based on propagation
  feature of node for identifying the influential nodes}.
\newblock \bibinfo{journal}{\emph{Physics Letters A}} \bibinfo{volume}{379},
  \bibinfo{number}{38} (\bibinfo{year}{2015}), \bibinfo{pages}{2272--2276}.
\newblock
\showISSN{0375-9601}
\urldef\tempurl%
\url{https://doi.org/10.1016/j.physleta.2015.05.021}
\showDOI{\tempurl}


\bibitem[Zhong et~al\mbox{.}(2018a)]%
        {ZHONG201878}
\bibfield{author}{\bibinfo{person}{Lin-Feng Zhong}, \bibinfo{person}{Quan-Hui
  Liu}, \bibinfo{person}{Wei Wang}, {and} \bibinfo{person}{Shi-Min Cai}.}
  \bibinfo{year}{2018}\natexlab{a}.
\newblock \showarticletitle{Comprehensive influence of local and global
  characteristics on identifying the influential nodes}.
\newblock \bibinfo{journal}{\emph{Physica A: Statistical Mechanics and its
  Applications}}  \bibinfo{volume}{511} (\bibinfo{year}{2018}),
  \bibinfo{pages}{78--84}.
\newblock
\showISSN{0378-4371}
\urldef\tempurl%
\url{https://doi.org/10.1016/j.physa.2018.07.031}
\showDOI{\tempurl}


\bibitem[Zhong et~al\mbox{.}(2018b)]%
        {ZhongLinFengandLiu201807}
\bibfield{author}{\bibinfo{person}{Lin-Feng Zhong}, \bibinfo{person}{Quan-Hui
  Liu}, \bibinfo{person}{Wei Wang}, {and} \bibinfo{person}{Shi-Ming Cai}.}
  \bibinfo{year}{2018}\natexlab{b}.
\newblock \showarticletitle{Comprehensive influence of local and global
  characteristics on identifying the influential nodes}.
\newblock \bibinfo{journal}{\emph{Physica A: Statistical Mechanics and its
  Applications}}  \bibinfo{volume}{511} (\bibinfo{date}{07}
  \bibinfo{year}{2018}).
\newblock
\urldef\tempurl%
\url{https://doi.org/10.1016/j.physa.2018.07.031}
\showDOI{\tempurl}


\bibitem[Zhong et~al\mbox{.}(2018c)]%
        {ZhongLinFengandShang201806}
\bibfield{author}{\bibinfo{person}{Lin-Feng Zhong}, \bibinfo{person}{Ming~Sheng
  Shang}, \bibinfo{person}{Xiao-Long Chen}, {and} \bibinfo{person}{Shi-Ming
  Cai}.} \bibinfo{year}{2018}\natexlab{c}.
\newblock \showarticletitle{Identifying the influential nodes via
  eigen-centrality from the differences and similarities of structure}.
\newblock \bibinfo{journal}{\emph{Physica A: Statistical Mechanics and its
  Applications}}  \bibinfo{volume}{510} (\bibinfo{date}{06}
  \bibinfo{year}{2018}).
\newblock
\urldef\tempurl%
\url{https://doi.org/10.1016/j.physa.2018.06.115}
\showDOI{\tempurl}


\bibitem[Zhou et~al\mbox{.}(2020)]%
        {ZHOU202047}
\bibfield{author}{\bibinfo{person}{Bin Zhou}, \bibinfo{person}{Xin Lu}, {and}
  \bibinfo{person}{Petter Holme}.} \bibinfo{year}{2020}\natexlab{}.
\newblock \showarticletitle{Universal evolution patterns of degree
  assortativity in social networks}.
\newblock \bibinfo{journal}{\emph{Social Networks}}  \bibinfo{volume}{63}
  (\bibinfo{year}{2020}), \bibinfo{pages}{47--55}.
\newblock
\showISSN{0378-8733}
\urldef\tempurl%
\url{https://doi.org/10.1016/j.socnet.2020.04.004}
\showDOI{\tempurl}


\bibitem[Zhu and Wang(2021)]%
        {Zhu_Jingcheng_and_Wang}
\bibfield{author}{\bibinfo{person}{Jingcheng Zhu} {and} \bibinfo{person}{Lunwen
  Wang}.} \bibinfo{year}{2021}\natexlab{}.
\newblock \showarticletitle{Identifying Influential Nodes in Complex Networks
  Based on Node Itself and Neighbor Layer Information}.
\newblock \bibinfo{journal}{\emph{Symmetry}}  \bibinfo{volume}{13}
  (\bibinfo{date}{08} \bibinfo{year}{2021}), \bibinfo{pages}{1570}.
\newblock
\urldef\tempurl%
\url{https://doi.org/10.3390/sym13091570}
\showDOI{\tempurl}


\end{thebibliography}

\end{document}